\documentclass[onecolumn,aps,superscriptaddress,pre]{revtex4}

\usepackage{amsmath,amssymb,graphicx}
\usepackage{subfigure}
\usepackage{algorithmic}
\usepackage{enumerate}
\usepackage{times}
\usepackage{color}
\usepackage{soul}
\definecolor{yblue}{rgb}{0.06, 0.3, 0.57}
\usepackage[pdftex]{hyperref}
\hypersetup{colorlinks=true,linkcolor=blue,citecolor=blue,urlcolor=blue}

\newcommand{\state}[1]{$\mathcal{S}_{#1}$}

\begin{document}

\title{Dark-dark soliton breathing patterns in multi-component Bose-Einstein condensates}

\author{Wenlong Wang}
\email{wenlongcmp@scu.edu.cn}
\affiliation{College of Physics, Sichuan University, Chengdu 610065, China}

\author{Li-Chen Zhao}
\affiliation{School of Physics, Northwest University, Xi'an 710069, China}
\affiliation{Shaanxi Key Laboratory for Theoretical Physics Frontiers, Xi'an 710069, China}

\author{Efstathios G. Charalampidis}
\email{echarala@calpoly.edu}
\affiliation{Mathematics Department, California Polytechnic State University, 
San Luis Obispo, CA 93407-0403, USA}

\author{Panayotis G. Kevrekidis}
\email{kevrekid@math.umass.edu}
\affiliation{Department of Mathematics and Statistics, University of Massachusetts,
Amherst, Massachusetts 01003-4515, USA}
\affiliation{Mathematical Institute, University of Oxford, Oxford, UK}

\begin{abstract}
In this work, we explore systematically various
SO(2)-rotation-induced
multiple dark-dark 
soliton breathing patterns obtained from stationary and spectrally stable multiple dark-bright and dark-dark waveforms in trapped one-dimensional, two-component atomic 
 Bose-Einstein condensates (BECs). The stationary
 states stem from the associated
linear limits (as the eigenfunctions of the quantum harmonic
oscillator problem) and  are 
parametrically continued to the nonlinear regimes by varying the respective chemical 
potentials, i.e., from the low-density linear limits to the high-density Thomas-Fermi 
regimes. We perform a Bogolyubov-de Gennes (BdG) spectral stability
analysis to
identify 
stable parametric regimes of these states. Upon SO(2)-rotation, the
stable steady-states,
one-, two-, three-, four-, and many 
dark-dark soliton breathing
patterns are observed
in the
numerical simulations. Furthermore, analytic solutions up to three dark-bright solitons in the
homogeneous setting, and three-component systems are also investigated.
\end{abstract}

\maketitle

\section{Introduction}
Bose-Einstein condensates (BECs) have attracted a significant amount
of
attention over more than two decades for 
investigating macroscopic quantum phenomena~\cite{becbook1,becbook2}.
One major theme of research concerns (effectively) nonlinear
coherent structure solutions in the form of solitary 
waves that are supported by these quantum gases~\cite{Panos:book},
which share many similarities with nonlinear 
optics \cite{DSoptics}. A large variety of solitary waves has been studied in the context of 
BECs, ranging from bright solitons in attractive condensates~\cite{tomio} to dark solitons~\cite{Dimitri:DS}, 
vortices \cite{Alexander2001}, and vortical filaments as well as
rings~\cite{fetter2,komineas,PhysRevE.85.036306} 
in repulsive condensates.

One important extension of these studies is the investigation of 
multicomponent condensates supporting, e.g., dark-bright (DB),
dark-dark (DD), dark-antidark (DAD) structures 
in repulsive condensates~\cite{DBS1,DBS2, PS:DB,PS:DBDD,Dong:MDB}; see, e.g.,~\cite{revip} for
a relatively recent review summarizing some of the early work
on the subject both in atomic physics, as well as in nonlinear
optics.
Note that the bright {soliton} 
cannot be sustained on its own
in a single repulsive condensate, {but} exists as a result 
of the effective trapping of the dark
soliton in the other component. It is also relevant to mention
that the study of such structures has motivated extensions thereof
also in higher dimensions~\cite{Wang:DBS,Wang:AI2,Wang:RDS}.
In recent years, there has been a significant number of
further efforts to extend this multi-component understanding
to a variety of more complex settings, including, e.g., the one
of three-component condensates~\cite{engels18}, that of
magnetic solitons in both binary~\cite{string} and even
spinor~\cite{rueda}
BECs, and very recently the examination of multiple DAD states
in two-component systems~\cite{engels20}. 


Our primary focus herein will be more concretely on the two-component setting.
In this case and when only incoherent coupling between the
components is involved,
the system  trivially supports
the U$(1) \times$ U$(1)$ symmetry. In the special Manakov case, in 
which all the intra- and inter-component interaction strengths are equal, there is an additional SU$(2)$ 
symmetry~\cite{Park:Opticalsolitons}; see the next section for
details. One particularly
interesting result 
is that this SU$(2)$-symmetry can induce the formation of the
so-called
dark-dark breathing or beating dynamics upon rotating stationary and
stable
dark-bright
soliton solutions~\cite{Yan:DD}.
This rotation has been exploited to produce single DD states
from
corresponding single DB ones, 
and these DD states have been studied in various (both one- and
higher-dimensional)
settings~\cite{Yan:DD,Wang:SO2,Lichen:DD,Wang:DBS,Wang:RDS}. Nevertheless,
the methodology has not been
extended to multiple-wave patterns and 
the 
{DB} soliton
crystal states that can also be realized~\cite{Wang:OD}.
Note that in the present work, we are principally interested in stable
patterns,
{looking for} stable dark-bright solitons, 
although the symmetry is not limited to stable structures or even
stationary states. 

Given the above state of the field, 
the main purpose of the present work is to offer a systematic study of multiple 
{DB} solitons or more precisely 
{DB} and 
{DD} mixtures, and their associated stable multiple 
DD soliton breathing patterns via an 
SU$(2)$ rotation.
Given the recent experimental developments enabling both the
sequential and alternating
seeding
of dark and antidark structures in the two components~\cite{engels20}, this possibility
is  especially timely and interesting. 
We focus on the case of a two-component condensate in 1D confined in a
harmonic trap. A key feature of our study is that we explore these structures systematically from the low-density linear
limits to the high-density Thomas-Fermi (TF) regimes, and their Bogolyubov-de Gennes (BdG) spectra are computed in the 
realm of spectral stability analysis. These computations shed light on potential stable parametric regimes in the chemical
potentials in which bound-state modes are long lived ones (and observed in our simulations). Such a methodology can be utilized 
to construct a whole series of topologically distinct stationary states. To that end, a component with $n>0$ solitons stemming
from the quantum harmonic oscillator eigenfunction $|n\rangle$ is progressively coupled to $m=0, 1,\dotsc, n-1$ solitons in 
the other component stemming from the state $|m\rangle$. These states are therefore expected to exist as the two components
decouple in the low-density linear limits. We assume (without loss of generality) $n>m>0$ and refer to the composite structure 
as state $\mathcal{S}_{nm}$, where $\mathcal{S}$ stands for both state
and soliton. For
each integer $n$, there is a total of 
$n$ \textit{distinct}
stationary states; thus, $n$ also
corresponds to the number of distinct 
DD breathing patterns.
Note that this enumeration accounts for the (definite parity)
states with $m=0, \dots,
n-1$ relevant in the vicinity of the linear limit.
In principle, this
does not preclude the potential of other (asymmetric) states to arise
in regimes of high nonlinearity, without persisting all the way to the
linear
limit. 
Therefore, the number of patterns grows rapidly for these composite structures. For 
example, up to $n=4$, there are already remarkably $10$ breathing
patterns; up to arbitrary $n$,  this number is $n (n+1)/2$. In the specific case of $n=2$, our 
procedure automatically reproduces both the in-phase (coupling with $m=0$) and the out-of-phase 
(coupling with $m=1$) 
DB solitons, and upon rotation their corresponding 
DD breathing patterns of~\cite{Yan:DD}.

It is straightforward to see that in $\mathcal{S}_{n0}$ the bright
solitary waves are all in phase as the second component is uniform in
phase, while in $\mathcal{S}_{n,n-1}$ (here a comma is added for clarity) the bright ones are fully out of phase as the roots of
neighbouring orthogonal polynomials alternate \cite{Numanalysis}. Interestingly, the fully in-phase 
DB will produce, upon rotation, 
an out-of-phase DD 
breathing pattern, as each of the DB solitary waves converts into a DD one.
As $m$ grows from $0$ to 
$n-1$, the number of dark solitons in the second component increases
by one successively, and the
resulting rotated patterns will convert each of the DBs into a DD,
while collocated zero crossings will be preserved under the transformation.
The breathing patterns are, in fact, reminiscent of a
1D mass-spring system with fixed boundary 
conditions, and for $n$ masses, there are $n$ normal modes increasingly out of phase.

In this work, we explore all the distinct states up to $n=4$. For higher $n$, the computation gets increasingly tedious as well 
as more expensive. One reason is that the number of combinations grows with $n$ as mentioned above, and there is an additional 
much more severe factor from the progressively larger number of
unstable modes
of the states.
This, in turn, necessitates
much higher densities or chemical potentials in order to stabilize the
configurations compared with the low-lying structures. In order to reach large chemical potentials, 
both a larger domain (to ensure that the patterns identified are located
comfortably
within the condensate) and a finer spacing (to accurately
resolve the solitonic structures) are required to achieve high accuracy in numerical computations. For higher $n$, we examine only
the state $\mathcal{S}_{n0}$ which typically has a wider region of stability (in chemical potentials) among the different values of 
$m$. In this work, we have explored the cases with $n=5, 6, ...,$ up to $10$, thus forming a 
DB ``mini-lattice''. In fact, our 
results involve quite substantial computations, despite our work
being
restricted to 1D: for example, to stabilize the $\mathcal{S}_{10,0}$ 
structure, we have to reach chemical potentials on the order of $100$.

In addition to breathing patterns in a harmonic trap, we
discuss the homogeneous setting with up to three 
soliton structures (and the states that emerge from their rotation);
finally, we extend our considerations to three-component systems. In
the former case,
we are interested in exact solutions 
of bound DB 
solitons and the corresponding DD 
breathing waveforms. In the latter case, the number of stationary
states
is even higher due to the different combinations of the pertinent
eigenstates. To this end, we introduce in this case
the state symbolism $\mathcal{S}_{mnp}$ 
with $m>n>p>0$ which stems itself from the coupling of the harmonic oscillator states $|m\rangle$, $|n\rangle$, and $|p\rangle$. We shall 
not explore all of these structures in detail in this work, but rather our goal is to illustrate prototypical examples involving them and 
demonstrate the applicability of our current approach in tracing states from the linear limits. In the three-component case, we will 
explore 
SU$(3)$ rotated breathing patterns, again using stable
stationary solitonic structures as a starting point for performing the
corresponding rotations. 


Our presentation is organized as follows. In Sec.~\ref{setup}, we introduce the model, 
the SU$(2)$ (and SO$(2)$) symmetry and the various numerical 
methods employed in this work. Next, we present our numerical and analytical results in Sec.~\ref{results}. Finally, our conclusions 
and a number of open problems for future consideration are given in Sec.~\ref{conclusion}.

\section{Models and methods}
\label{setup}

We first present the mean-field Gross-Pitaevskii equation and the SU$(2)$ symmetry for a two-component condensate 
at the Manakov limit. Then, we discuss
our methodology for constructing stationary solitons from the linear limits, 
and the numerical methods employed in the nonlinear realm
for identifying stationary states, performing stability analysis, and dynamics. 
Finally, we briefly describe the analytical method and the generalization to three-component systems.

\subsection{Computational setup}

In the framework of mean-field theory, and for sufficiently low temperatures, the dynamics of a strongly
transversely confined 1D two-component repulsive BEC in a time-independent trap $V=V(x)$, is described by 
the following coupled dimensionless Gross-Pitaevskii equation (GPE)~\cite{Panos:book}:
\begin{subequations}
\begin{align}
i \frac{\partial \psi_1}{\partial t} &= -\frac{1}{2} \frac{\partial^{2}\psi_1}{\partial x^{2}}+%
V \psi_1 +(g_{11}| \psi_1 |^2+ g_{12}| \psi_2 |^2) \psi_1,  %
\label{GPE_1} \\
i \frac{\partial \psi_2}{\partial t} &= -\frac{1}{2} \frac{\partial^{2}\psi_2}{\partial x^{2}}+%
V \psi_2 +(g_{21}| \psi_1 |^2 + g_{22} |\psi_2|^2) \psi_2,
\label{GPE_2}
\end{align}
\end{subequations}
where $\psi_1=\psi_{1}(x,t)$ and $\psi_2=\psi_2(x,t)$ are two complex scalar macroscopic wavefunctions. 
In order to study the SU$(2)$-induced breathing patterns, we consider mainly in this work the Manakov 
limit $g_{11}=g_{12}=g_{21}=g_{22}=1$, although effects of weak
deviations
are also considered in our 
subsequent discussion.
Such effects are relevant for the weak deviations from equal
coefficients
that are encountered, e.g., in the study of hyperfine states of
$^{87}$Rb~\cite{becbook1,becbook2,Panos:book}.
The condensates, unless otherwise specified, are confined in a harmonic magnetic
trap of the form: 
\begin{equation}
V=\frac{1}{2} \omega^2 x^2,
\label{potential}
\end{equation}
where the trapping frequency $\omega$ is set (without loss of
generality) to $\omega=1$.
Stationary states with chemical potentials $\mu_1$ and $\mu_2$ for the first and second
components, respectively, are constructed by considering the Ans\"atze:
\begin{align}
\psi_1(x,t) &= \psi^0_1(x)e^{-i\mu_1t}, \nonumber \\
\psi_2(x,t) &= \psi^0_2(x)e^{-i\mu_2t},
\end{align}
which lead to the stationary equations:
\begin{eqnarray}
\label{SS1}
-\frac{1}{2} \frac{\partial^{2}\psi^0_1}{\partial x^{2}}+%
V \psi^0_1 +(| \psi^0_1 |^2+| \psi^0_2 |^2) \psi^0_1 &=& \mu_1 \psi^0_1, \nonumber \\
-\frac{1}{2}  \frac{\partial^{2}\psi^0_2}{\partial x^{2}}+%
V \psi^0_2 +(| \psi^0_1 |^2+| \psi^0_2 |^2) \psi^0_2 &=& \mu_2 \psi^0_2.
\end{eqnarray}

The system described by Eqs.~\eqref{GPE_1}-\eqref{GPE_2} admits the U$(1) \times$ U$(1)$ symmetry, 
and additionally the SU$(2)$ symmetry in the Manakov case (where all interaction coefficients
are set to unity). Indeed, if $(\psi_1,\psi_2)^T$ is a solution to the system~\eqref{GPE_1}-\eqref{GPE_2}, 
then $(\psi_1 \exp(i\theta_1),\psi_2\exp(i\theta_2))^T$ also is, where $\theta_1$ and $\theta_2$ 
are two real constants. In the Manakov case, it is straightforward to show that
\begin{eqnarray}
\begin{pmatrix}
\psi_1' \\
\psi_2'
\end{pmatrix}
=
U
\begin{pmatrix}
\psi_1 \\
\psi_2
\end{pmatrix}
=
\begin{pmatrix}
&\alpha &-\beta^{\ast} \\
&\beta &\alpha^{\ast}
\end{pmatrix}
\begin{pmatrix}
\psi_1 \\
\psi_2
\end{pmatrix},
\end{eqnarray}
is also a solution, where $UU^{\dagger}=\mathbb{I}$, $|\alpha|^2+|\beta|^2=1$, and a star ($\ast$) denotes 
complex conjugation. Note that the total density profile is invariant upon rotation, i.e., 
$|\psi_1|^2+|\psi_2|^2 = |\psi_1'|^2+|\psi_2'|^2$. In this work, we
explore the subset
of SO$(2)$ rotations: 
\begin{eqnarray}
\begin{pmatrix}
\psi_1' \\
\psi_2'
\end{pmatrix}
=
\begin{pmatrix}
&\cos \delta &\sin \delta \\
&-\sin \delta &\cos \delta
\end{pmatrix}
\begin{pmatrix}
\psi_1 \\
\psi_2
\end{pmatrix},
\end{eqnarray}
and focus on the most symmetric case using $\delta = \pi/4$.

For the two-component case, we identify stationary states by using a finite element method for the spatial discretization 
and employing Newton's method for the underlying root-finding
problem. The linear harmonic oscillator states (which are suitable in
the
low density limit where the cubic nonlinear terms can be neglected) are used as 
initial guesses near the respective linear limits. The
obtained solutions (upon convergence of Newton's method in this weakly
nonlinear regime) are parametrically 
continued to large chemical potentials by performing a sequential continuation. Since our goal in the present work is to 
identify stable stationary states, we systematically compute the
BdG stability spectrum (see,
e.g.,~\cite{Panos:book}
for a discussion thereof for the multi-component system) along the 
$(\mu_1,\mu_2)$ parametric continuation line considered and
select a stable solution which will be rotated subsequently; the
interested reader  can also find details of the BdG stability matrix
in~\cite{Wang:DBS}. The real and imaginary parts of the eigenvalues 
$\lambda$ of the spectrum denote unstable and stable modes,
respectively. Our dynamics  of either the original stationary states
or of
the rotated (and expected to be breathing) ones is performed by using the standard 
fourth-order Runge-Kutta method.

The analytical multiple DB 
soliton solutions and the rotation
thereof for the homogeneous setting~\cite{BoxP} are
also used in order to produce breathing states.
In this 
work, we discuss the two and  three DB 
soliton states, and then
their
corresponding rotated breathing patterns. It is worth noting 
that these solutions generally cannot be tuned to be fully stationary, despite the fact that they can be approximately stationary 
when multiple solitons are well separated.
This can also be understood intuitively as in the trapped case
discussed above the stationarity {stems} 
from the interplay between
the pairwise interaction of the DB structures and the restoring effect
of the trap on each of the waves~\cite{DBS1}. For homogeneous
settings,
the absence of the latter does not allow an equilibrium configuration
given the absence of a counterbalance for the former.
Nevertheless, the SO$(2)$ rotation and symmetry is not limited to stationary states, 
and applies to these dynamic cases as well. Consequently, several time
scales can manifest themselves in the dynamics, in contrast to 
the periodic solutions rotated from stationary states.

The computational setup for the three-component system is similar to that of the two-component case. The pertinent equation 
of motion and the corresponding BdG stability matrix will be presented in Sec.~\ref{TC} for 
{completeness}.
In this system, there 
are three chemical potentials, extending
from the linear limits to the TF regimes in the $(\mu_1, \mu_2,
\mu_3)$
parameter space. 
In this work, we investigate the states \state{210} and \state{310}, including their existence, stability, and SU$(3)$-induced 
breathing dynamics.
Here, it is important to comment on the nature of the corresponding
model. It is well-known that the spinor condensate mean-field
model~\cite{kawueda,stampueda} that has
recently been explored also experimentally
for various solitonic configurations~\cite{engels18,rueda} is
nontrivially different from the Manakov model. In particular, the
latter
contains only the spin-independent part of the hyperfine state
interactions,
while the former contains also the spin-dependent part coupling
the phases of the different components~\cite{kawueda,stampueda}.
Here, motivated also by multi-component nonlinear optical
problems~\cite{Park:Opticalsolitons}, we restrict our considerations
to the Manakov case, however, we note that a more detailed
consideration
of the spin-dependent effect on these states  would be of
interest in its own right.


\subsection{Constructing irreducible topologically distinct stationary states from the linear limits}

Having discussed the computational techniques, we now focus on the construction of topologically distinct 
stationary states from their respective linear limits. The simplest
single DB 
soliton has been 
extensively studied and produces the DD 
breathing state \cite{Yan:DD,Wang:SO2} upon rotation. 
This soliton in the associated linear limit involves the coupling of the first harmonic oscillator 
excited state $|1\rangle$ with the ground state $|0\rangle$. By contrast, there are two cases for two 
DB 
solitons~\cite{OS99}~: (a) the in-phase case where the bright peaks have the same phase and (b) the 
out-of-phase case where the bright peaks have the opposite phase. These solitons again have their respective 
linear limits. These involve the coupling of the second excited
state $|2\rangle$
in the first component with the $|0\rangle$ and the 
$|1\rangle$ state in the second component, respectively~\cite{Wang:OD}. 
From these linear limits, the states can be continued to high 
density regimes in the $(\mu_1, \mu_2)$ parameter space. In our work, we take a simple linear trajectory 
from the linear limit to a final typical high-density regime. These states are conveniently labelled as 
$\mathcal{S}_{10}$, $\mathcal{S}_{20}$, and $\mathcal{S}_{21}$ in the
above notation, respectively.

These considerations can be generalized to any pair of harmonic oscillator states. Specifically, the state 
$|n\rangle$ can be coupled successively with the $|m\rangle$ state, thus forming the $\mathcal{S}_{nm}$ state 
with $m=0, 1, 2, ..., n-1$. It should be noted that not all of these
structures are DB 
solitons. 
For example, the state $\mathcal{S}_{31}$ has three dark solitons in the first component but with only two 
out-of-phase bright peaks at the sides, in the second
component. Between these peaks, naturally per the anti-symmetric
nature
of the $m=1$ state lies 
a dark solitonic structure in the second component. Therefore, 
the state $\mathcal{S}_{31}$ is a stationary state concaternating a
DB 
wave on the one end, with a DD 
one in the  middle and a DB 
structure
on the other end. It is noted in
passing that the DD 
structure is expected to exist whenever both $n$ and $m$ are odd. Finally, for 
each integer $n$, there is a total of $n$ distinct stationary states
and corresponding breathing patterns {\it stemming from the linear
  limit};
this is noted because in principle states that do not terminate at the
linear
limit may exist in the highly nonlinear regime.

We have omitted the structures stemming from the same linear states, i.e., \state{mn} with $m=n$. While these 
states are topologically distinct as well, there is a stringent constraint that the two fields must have the 
same chemical potentials $\mu_1=\mu_2$. In fact, one such state can be viewed as a splitting of the corresponding 
single-component state. If $\psi_{1D}$ is a stationary state of the one-component system, 
then $(\cos(\delta)\psi_{1D}, \sin(\delta)\psi_{1D})^T$ is a solution of the two-component system with same 
interaction strengths. Therefore, 
we focus on states of distinct quantum numbers, both for the
two-component
but also for the following three-component system. In a sense, we
investigate all the \textit{irreducible} topologically distinct states.
The construction can be further generalized to the three-component system, where state \state{mnp} is expected to be 
formed by coupling the harmonic oscillator states, namely $|m\rangle$, $|n\rangle$, and $|p\rangle$, where $m>n>p>0$. 
In this work, for proof-of-principle purposes, we only
explore 
two {specific yet} typical low-lying states \state{210} and
\state{310}, focusing
on their 
existence, stability, and the SU$(3)$-induced breathing patterns.

\section{Results}
\label{results}

\subsection{Multiple dark-dark breathing patterns in two components}
\begin{figure*}
\subfigure[]{\includegraphics[width=0.33\textwidth]{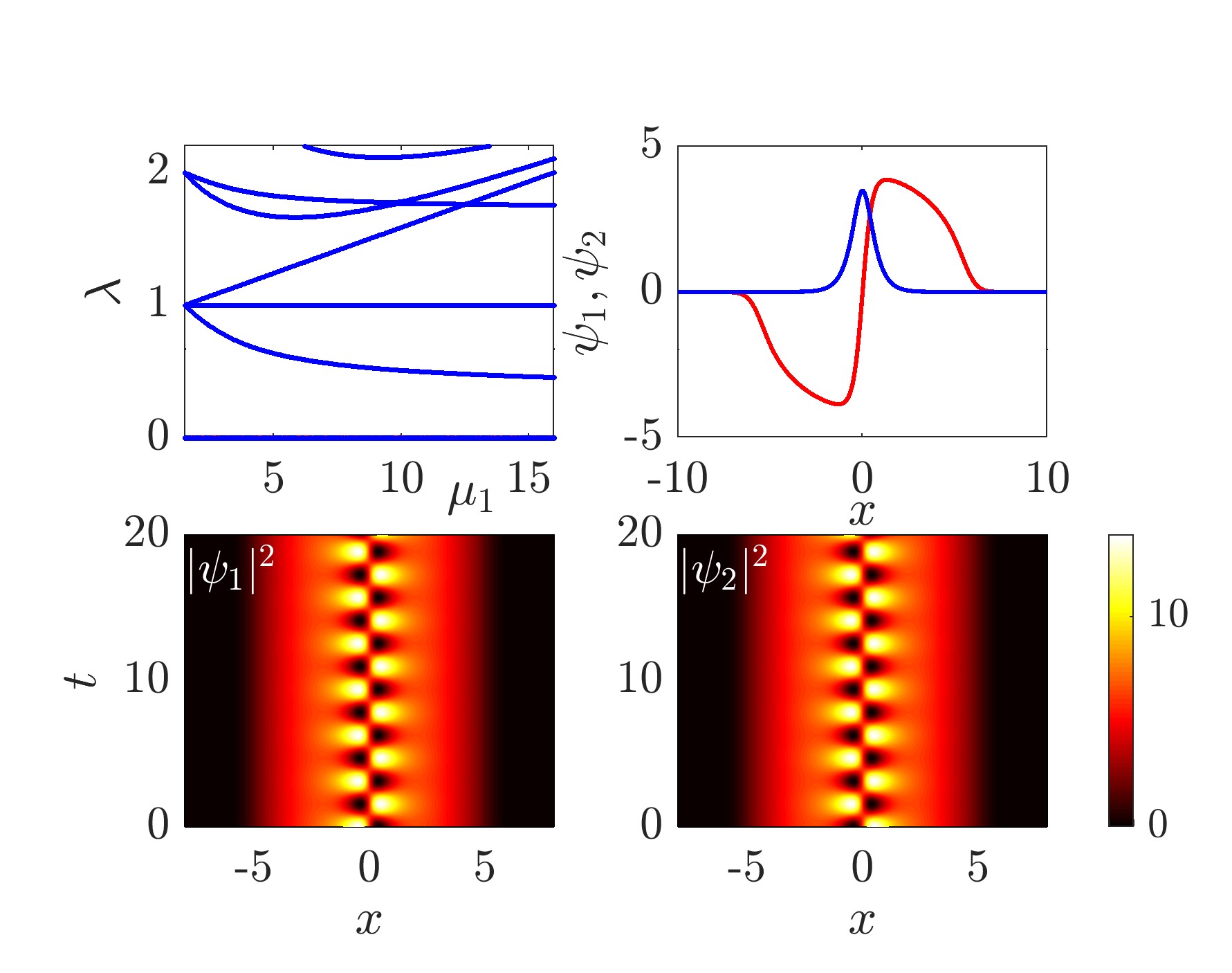}}
\subfigure[]{\includegraphics[width=0.33\textwidth]{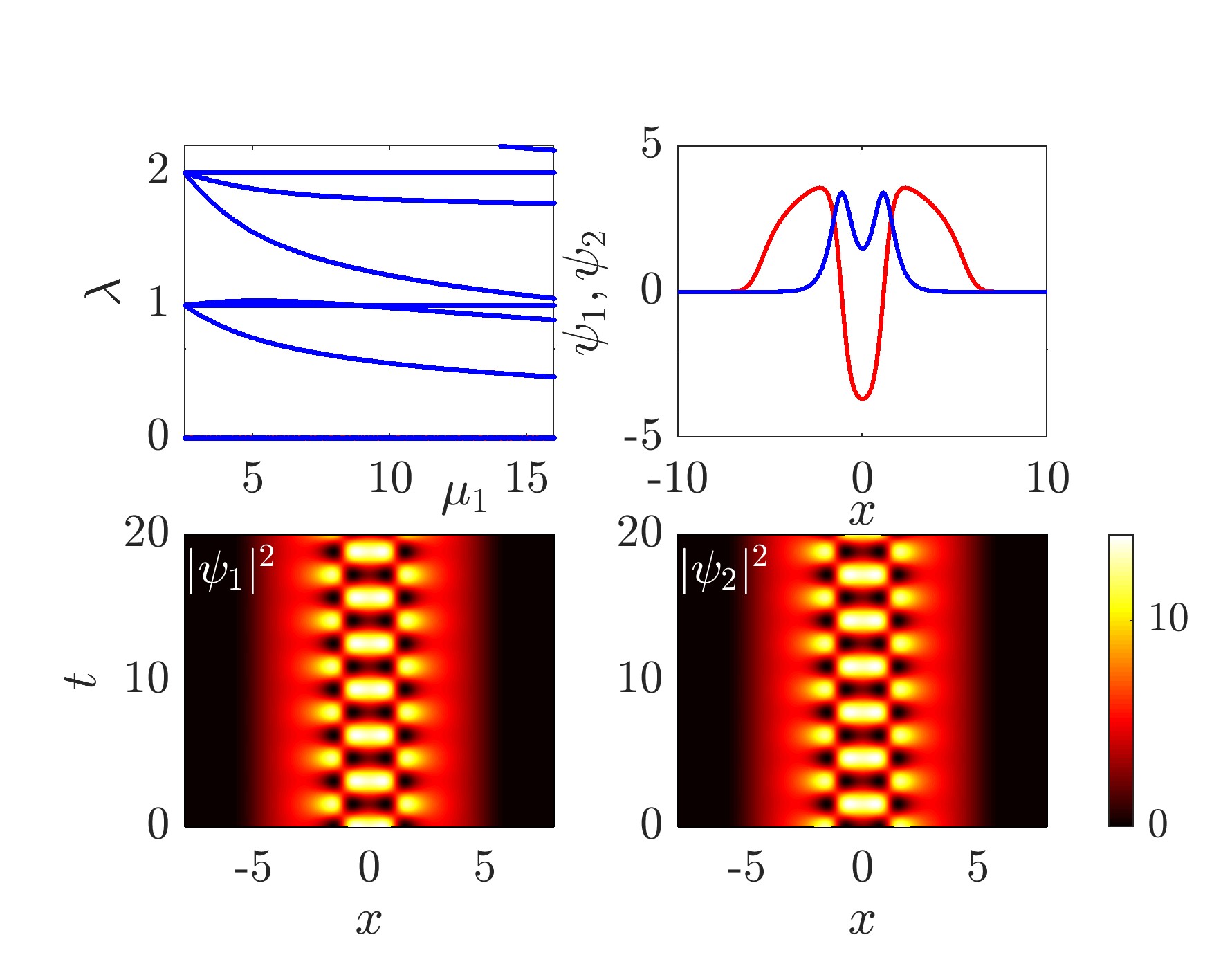}}
\subfigure[]{\includegraphics[width=0.33\textwidth]{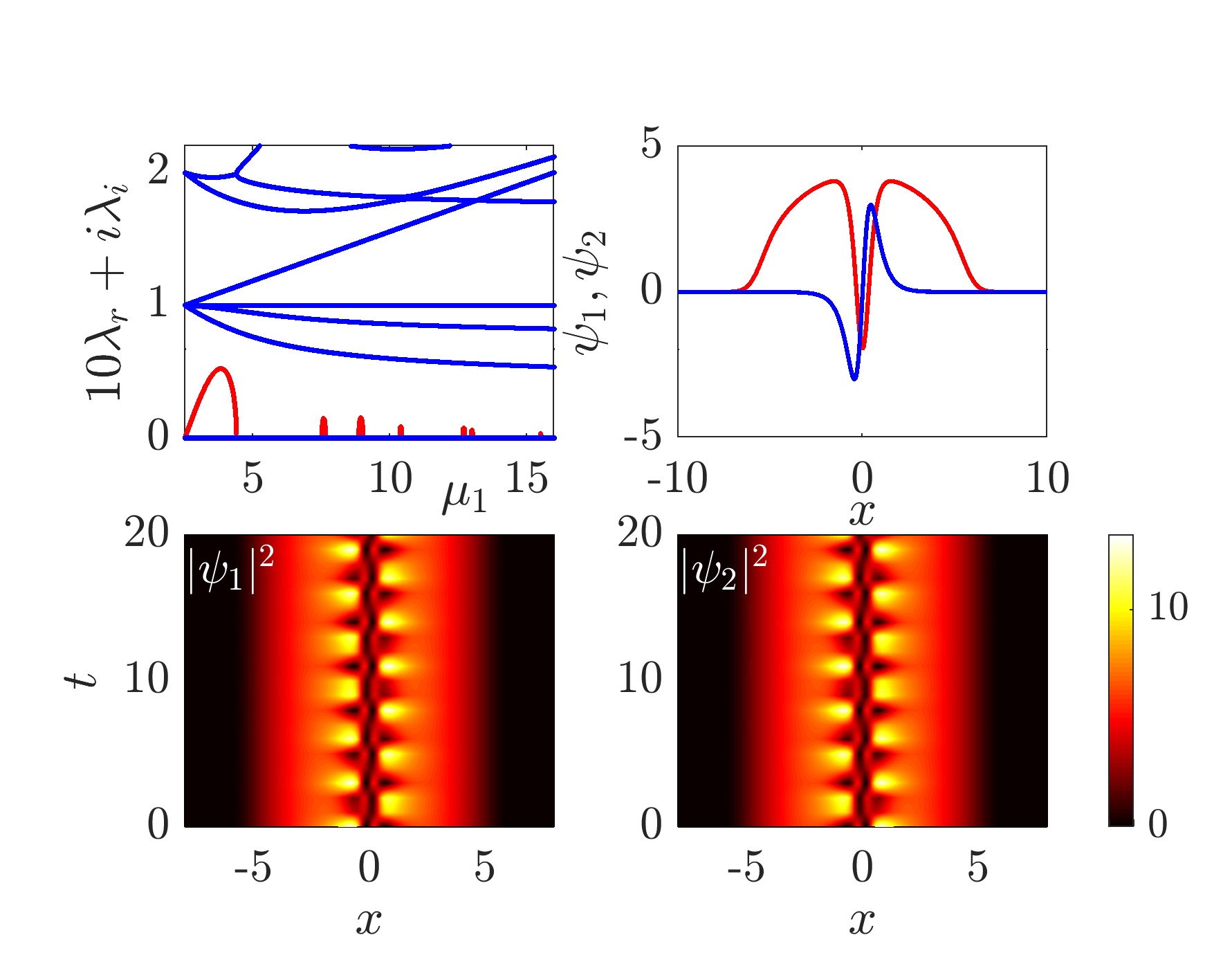}}
\caption{
(Color online) %
\textit{Left panel}: The BdG stability spectrum of the DB 
soliton along a linear trajectory 
from the linear limit $(1.5,0.5$) to a typical large-density limit $(16,14)$ in the $(\mu_1,\mu_2)$ 
parameter space. Red and blue points denote unstable and stable modes, respectively. The stationary 
DB profile at the end of the parametric line are illustrated, along with the induced DD 
oscillating patterns.
I.e., the DB pattern is SO$(2)$-rotated 
and fed into the time evolution
dynamics in order to observe this dynamical phenomenology.
Here and in the contour plots
that will follow  throughout the manuscript, the density
($|\psi_i|^2$, $i=1, 2$) will be shown for the different components as a function
of space and time.
The DB, as well as its rotated variant are fully stable and robust. \textit{Middle and right panels}: 
Same as the left panel, 
but for the $\mathcal{S}_{2m}$ family from the linear limit $(2.5,m+0.5$) to a typical large-density 
limit $(16,14)$. The bright solitons are in phase for $m=0$ and out of
phase for $m=1$, the in-phase two-DB state
is fully stable, but the out-of-phase one has several (here seven) weak instability peaks.
Note that the
real part of the eigenvalues for $\mathcal{S}_{21}$ is enlarged by a factor of $10$ for 
{visualization purposes}, i.e., the maximum growth rate is approximately $0.5/10=0.05$. All of the breathing patterns in this work 
are integrated and found to be robust up to $t=1000$. 
}
\label{DB10}
\end{figure*}

The single DB 
soliton appears to be very robust and is found to be fully stable over the parameters studied, 
as illustrated in the left panel of Fig.~\ref{DB10}. The same is true for the two
DB 
solitons in phase (see the middle panel of the figure).
On the other hand, the
two DB 
solitons out of phase encounter a series of instabilities, a total of seven unstable peaks along the
parametric line.
These instabilities are in line with what is known about both multiple
dark
solitons (in one-component condensates)~\cite{Panos:book} and
also about multiple DB 
and even DAD 
solitary waves
in two-component condensates; for a recent discussion, see,
e.g.,~\cite{engels20}. In particular, a so-called negative Krein (or
energy)
signature mode associated with the out-of-phase vibration of the
two DB solitary waves becomes resonant with modes of the background
cloud sequentially. The first of these resonances in the vicinity of
$\lambda_i=2$ can be observed in the right panel of Fig.~\ref{DB10}
(this also corresponds to the largest instability (red) ``bubble'').
However, most of the peaks are rather narrow and all of the peaks are
rather weak, i.e., they correspond to low growth rates of the
associated
instability. Note that the real
part of the eigenvalues is enlarged by a factor of $10$ for ease of visualization, i.e., the maximum growth rate is only
about $0.5/10=0.05$.
Therefore, there are wide intervals of stability
for these low-lying states. It is interesting that the DD 
breathing patterns involve the conversion of each of the DB 
structures into a DD, 
creating new phase alternations
(e.g. in the bright component) as a consequence of the emergence of
the
DD 
states.

\begin{figure*}
\subfigure[]{\includegraphics[width=0.33\textwidth]{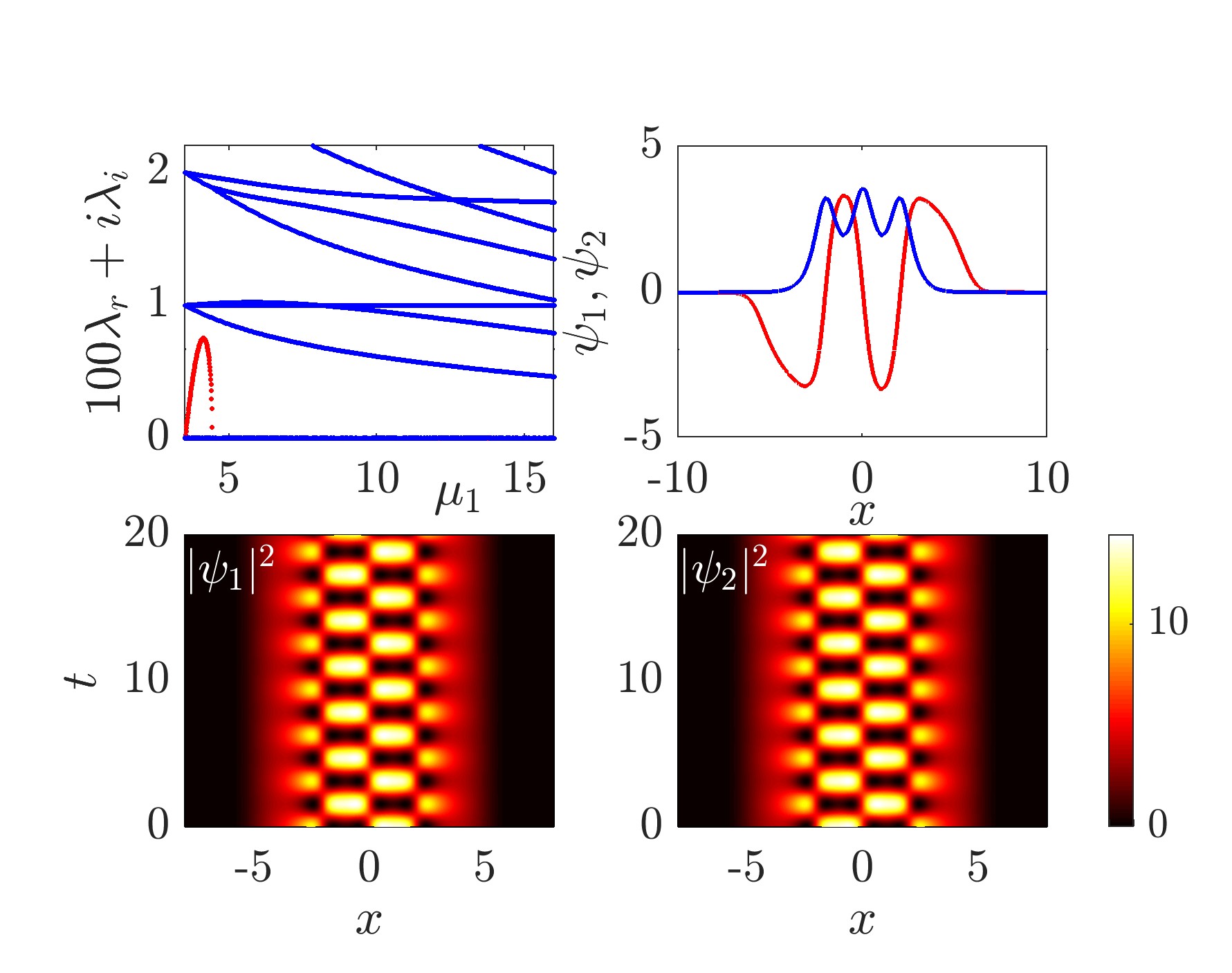}}
\subfigure[]{\includegraphics[width=0.33\textwidth]{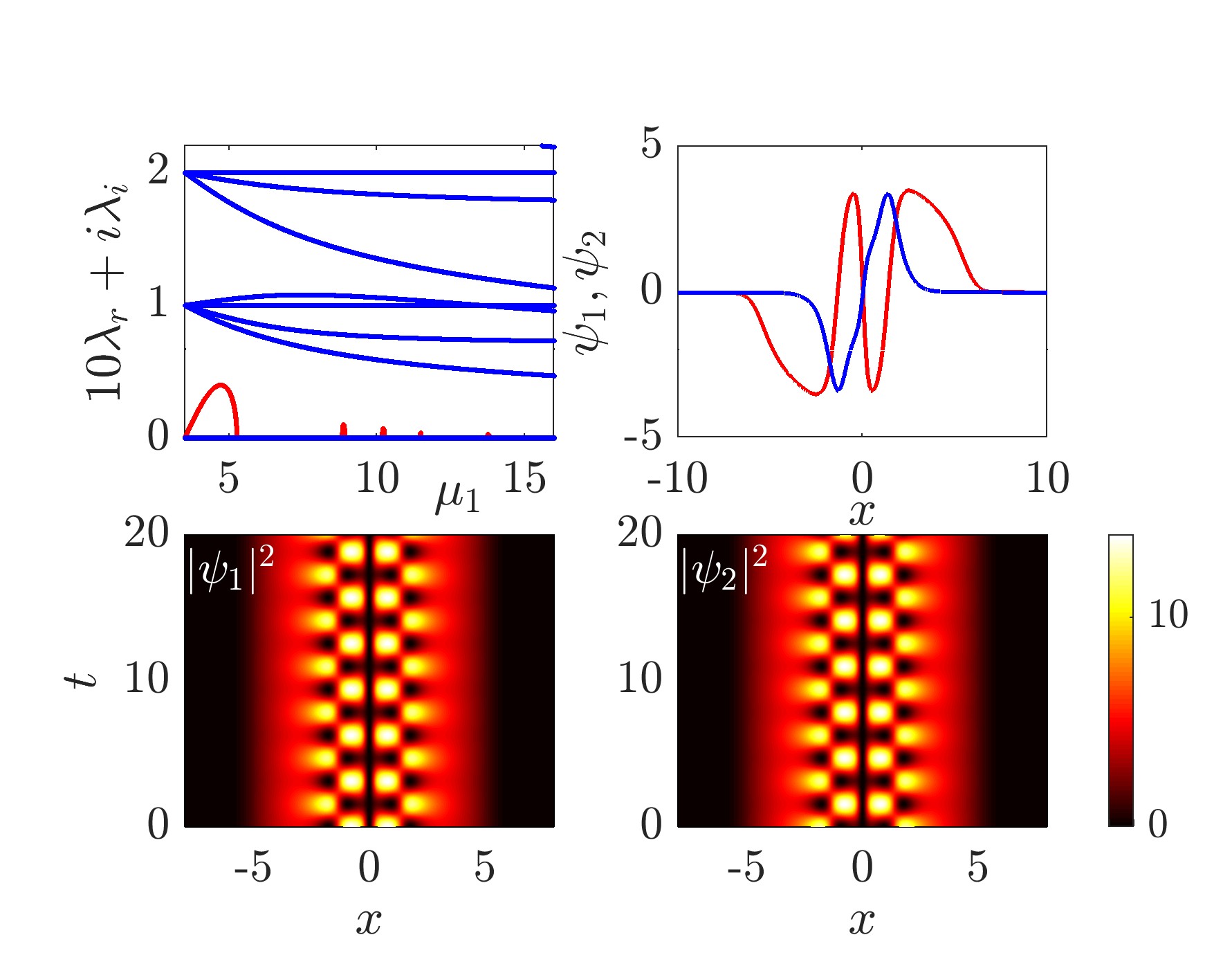}}
\subfigure[]{\includegraphics[width=0.33\textwidth]{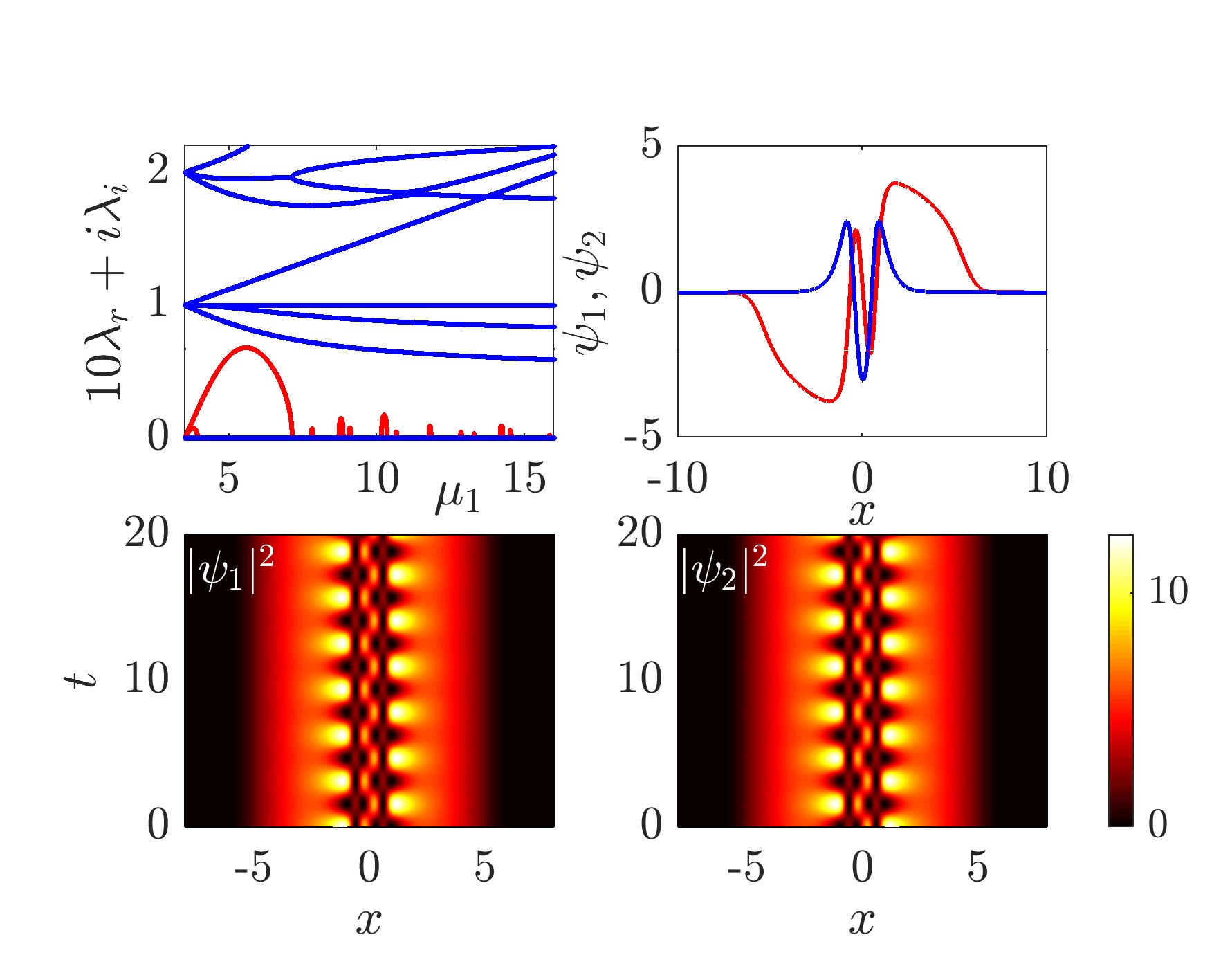}}
\caption{
(Color online) %
Same as Fig.~\ref{DB10}, but for the $\mathcal{S}_{3m}$ family from the linear limits $(3.5,m+0.5$) to 
a typical large-density limit $(16,14)$ in the $(\mu_1,\mu_2)$ parameter space. Here, even the in-phase 
state has an unstable mode, although it is rather weak. The states get
progressively
more unstable as $m$ increases, 
i.e., as the number of dark soliton increases in the second component. Note that the second state 
$\mathcal{S}_{31}$ consists of a DB 
(left end), DD 
(middle), and
DB 
(right end) structure.
}
\label{DB30}
\end{figure*}

\begin{figure}
\subfigure[]{\includegraphics[width=0.495\columnwidth]{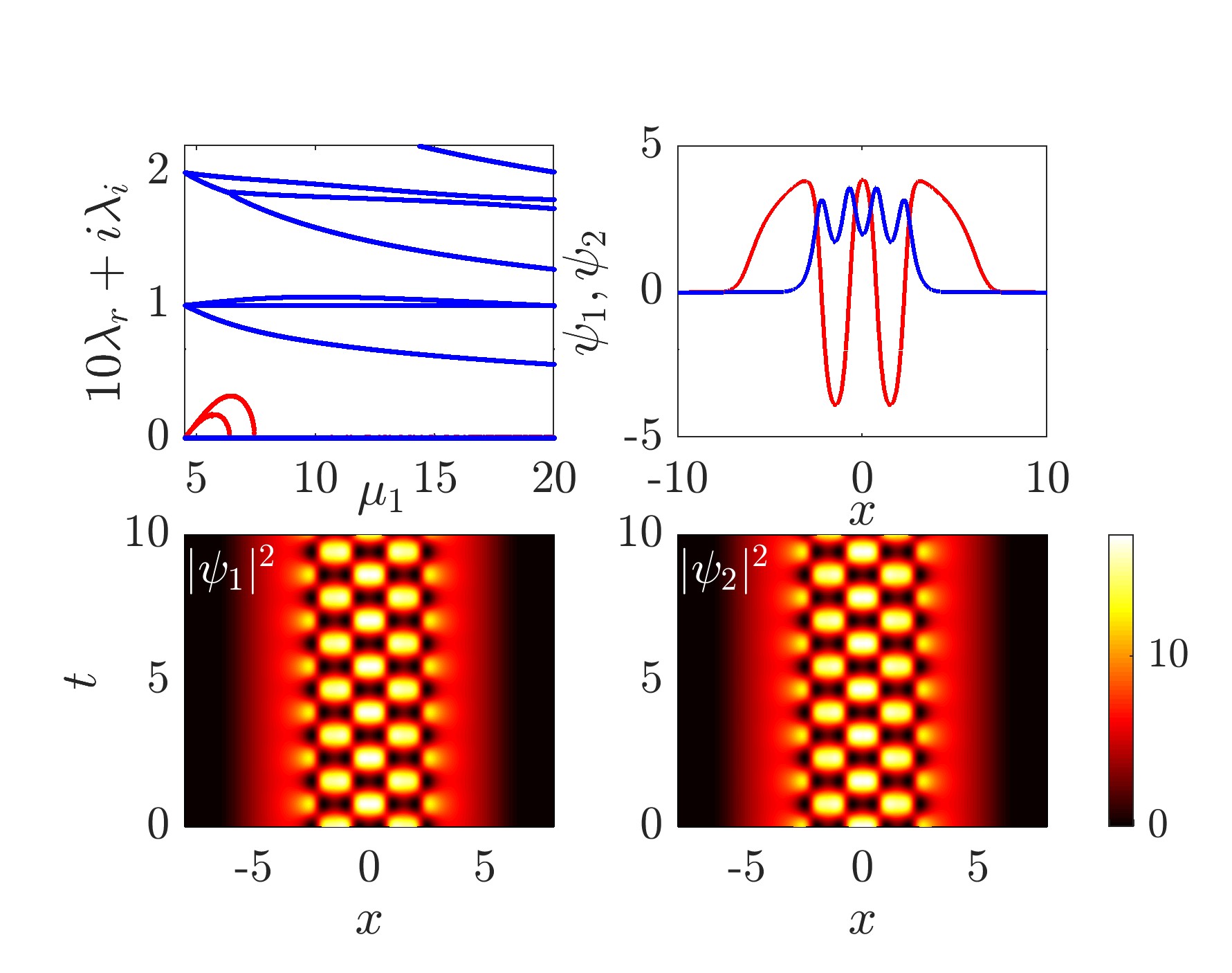}}
\subfigure[]{\includegraphics[width=0.495\columnwidth]{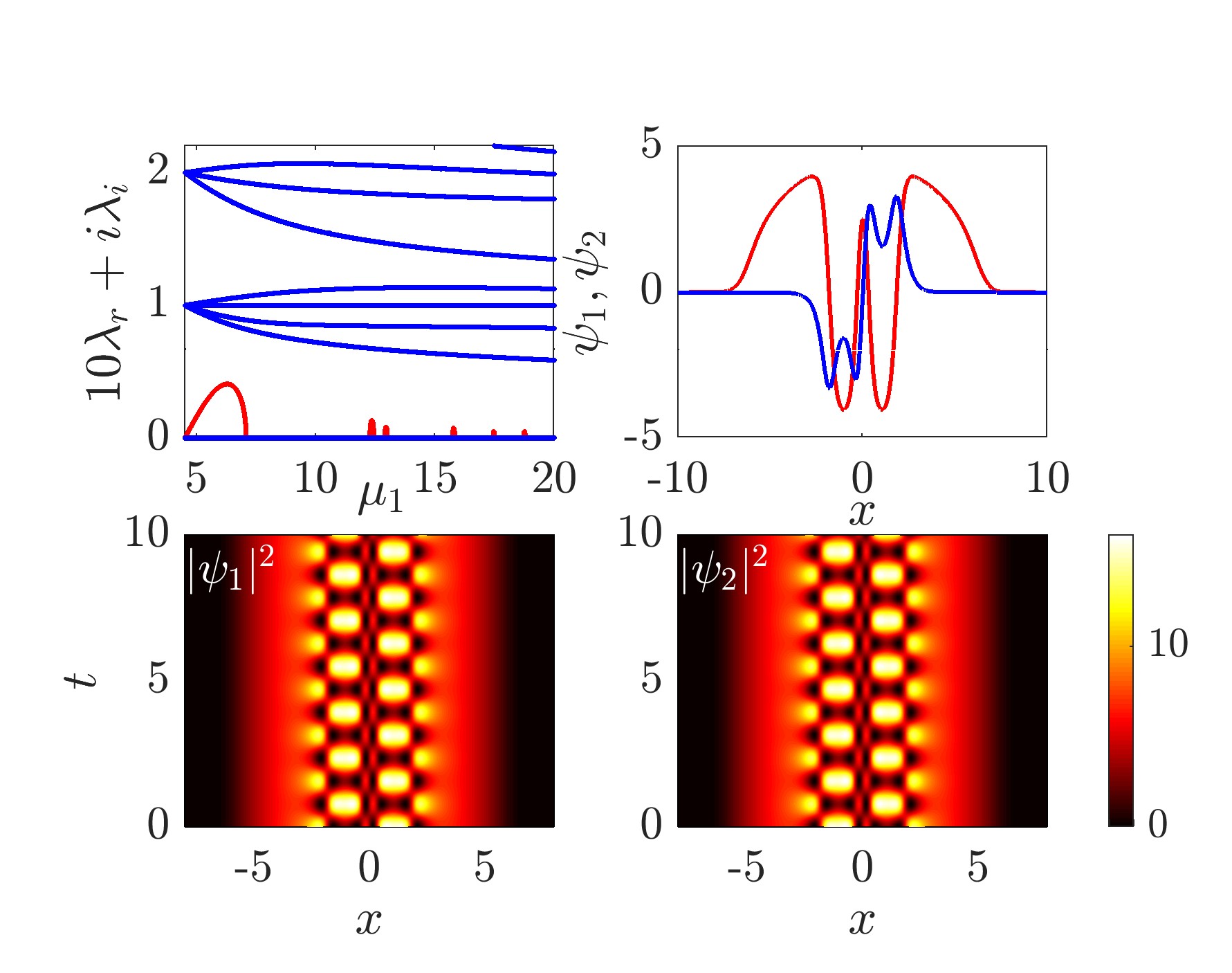}}
\subfigure[]{\includegraphics[width=0.495\columnwidth]{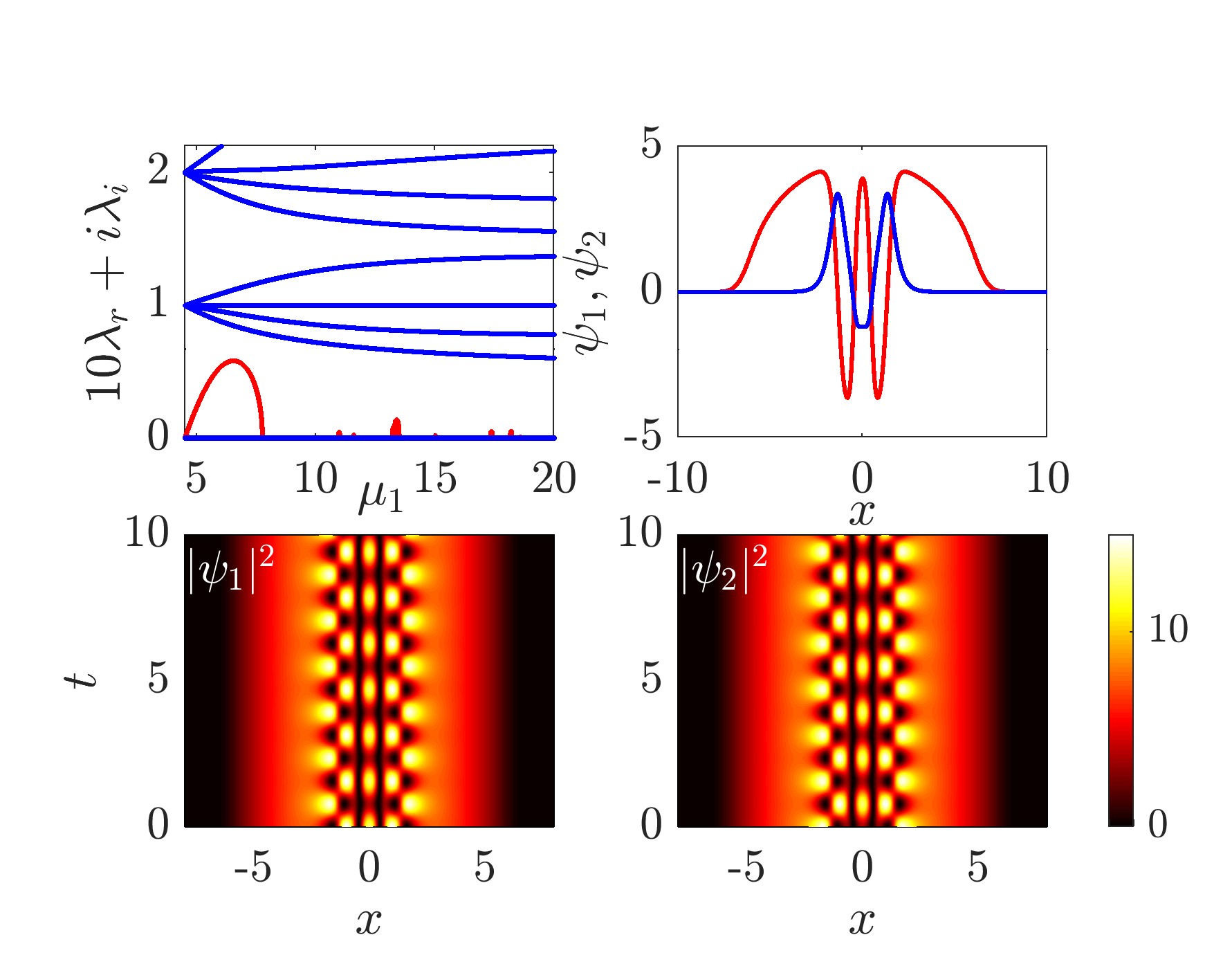}}
\subfigure[]{\includegraphics[width=0.495\columnwidth]{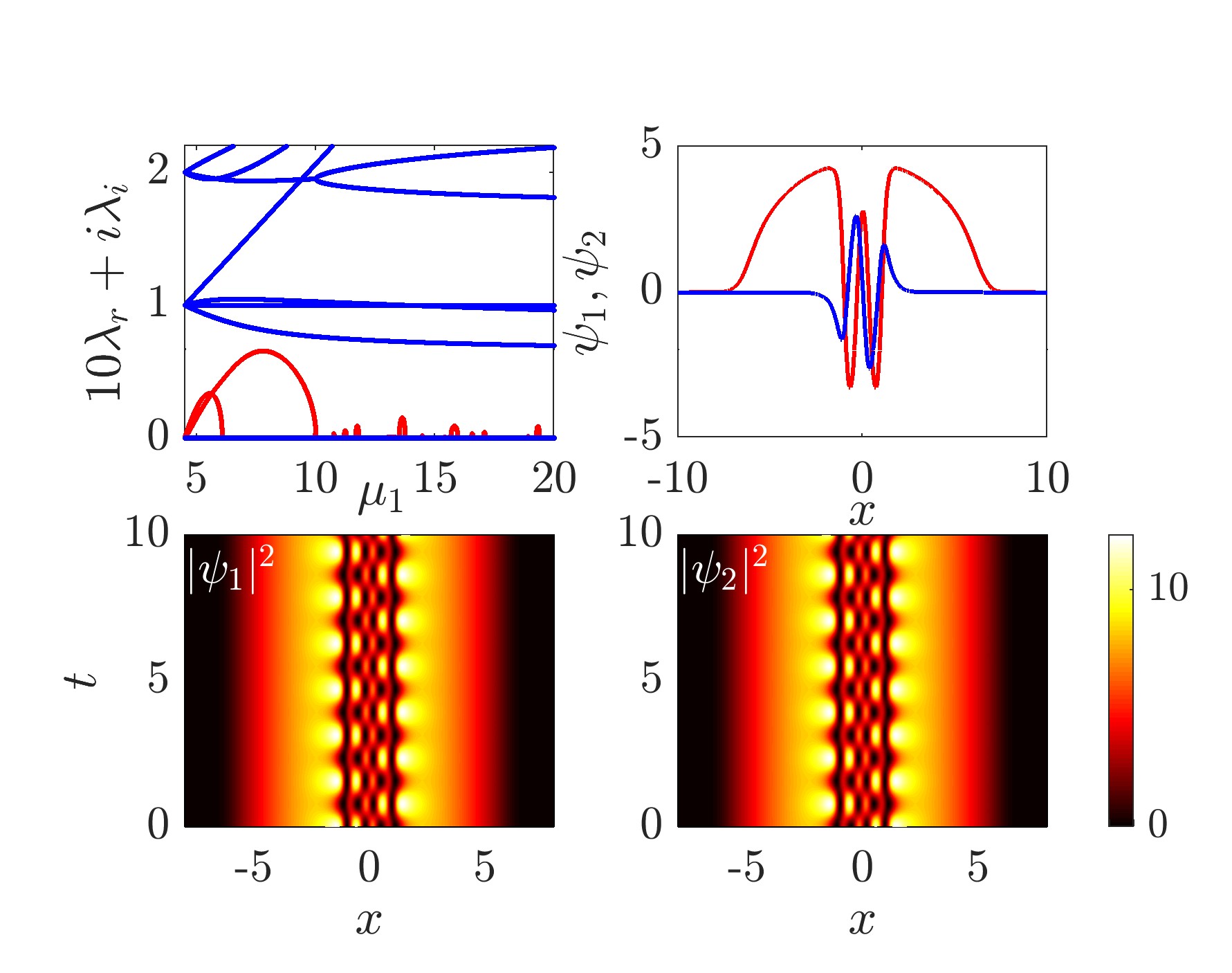}}
\caption{
(Color online) %
Same as {Fig.~\ref{DB10}}, 
but for the $\mathcal{S}_{4m}$ family from
the linear limits $(4.5,m+0.5$) and up to a 
typical large-density limit $(20,16)$ in the $(\mu_1,\mu_2)$ parameter
space. The top left quartet of panels concerns the $\mathcal{S}_{40}$
state,
the top right the $\mathcal{S}_{41}$ configuration, while,
respectively,
the bottom left and right constitute the $\mathcal{S}_{42}$ and
$\mathcal{S}_{43}$ states. In this case too, the initial conditions obtained as a result
of the
SO$(2)$ rotations lead to robust breathing states.
}
\label{DB40}
\end{figure}

Next, we {focus on} 
the $\mathcal{S}_{3m}$ family as shown in
Fig.~\ref{DB30}. In this family, all the states  
considered bear unstable modes; in fact all the states have {\it at
  least}
3 potentially unstable modes because of $n=3$. Furthermore,  the number of
potential instabilities grows {with $m$}. 
In this context, it is reasonable to expect that for states
$\mathcal{S}_{nm}$, the maximal number of potentially
unstable eigendirections is $n+m$.
However, it is important to emphasize that there are again wide ranges of stability.
The associated
instability bubbles will bear quite small growth rates, especially so
in the exception of the first one associated with resonances emerging
for small chemical potentials (particle numbers) right off of the
linear
limit.
From a structural perspective, we can observe in the corresponding
configurations that each of the DBs is converted, as a result
of the transformation, into a DD structure. 
{On the other hand,} a DD remains a DD
when a collocated DD state exists, as in the case of
$\mathcal{S}_{31}$ {where} the relevant zero crossing will be preserved
even after the SO$(2)$ rotation. 
Similar features to the above ones can be detected
for the $\mathcal{S}_{4m}$ family as shown in Fig.~\ref{DB40}. Here, again the
\state{40} state is the one that features the smallest number of
instability bubbles, although it is relevant to note that off of the
linear limit both the $\mathcal{S}_{40}$ and the $\mathcal{S}_{43}$
state
feature two such bubbles (while  $\mathcal{S}_{41}$ and
$\mathcal{S}_{42}$
have only one associated instability).
Nevertheless, all selected states at the DB level when dynamically
robust,
upon their SO$(2)$ rotation yield a number of stable internal
vibrations as manifested in the corresponding dynamical evolutions
in the bottom sets of space-time contours within the
Figs.~\ref{DB30} and~\ref{DB40}.

\begin{figure*}
\subfigure[]{\includegraphics[width=0.33\textwidth]{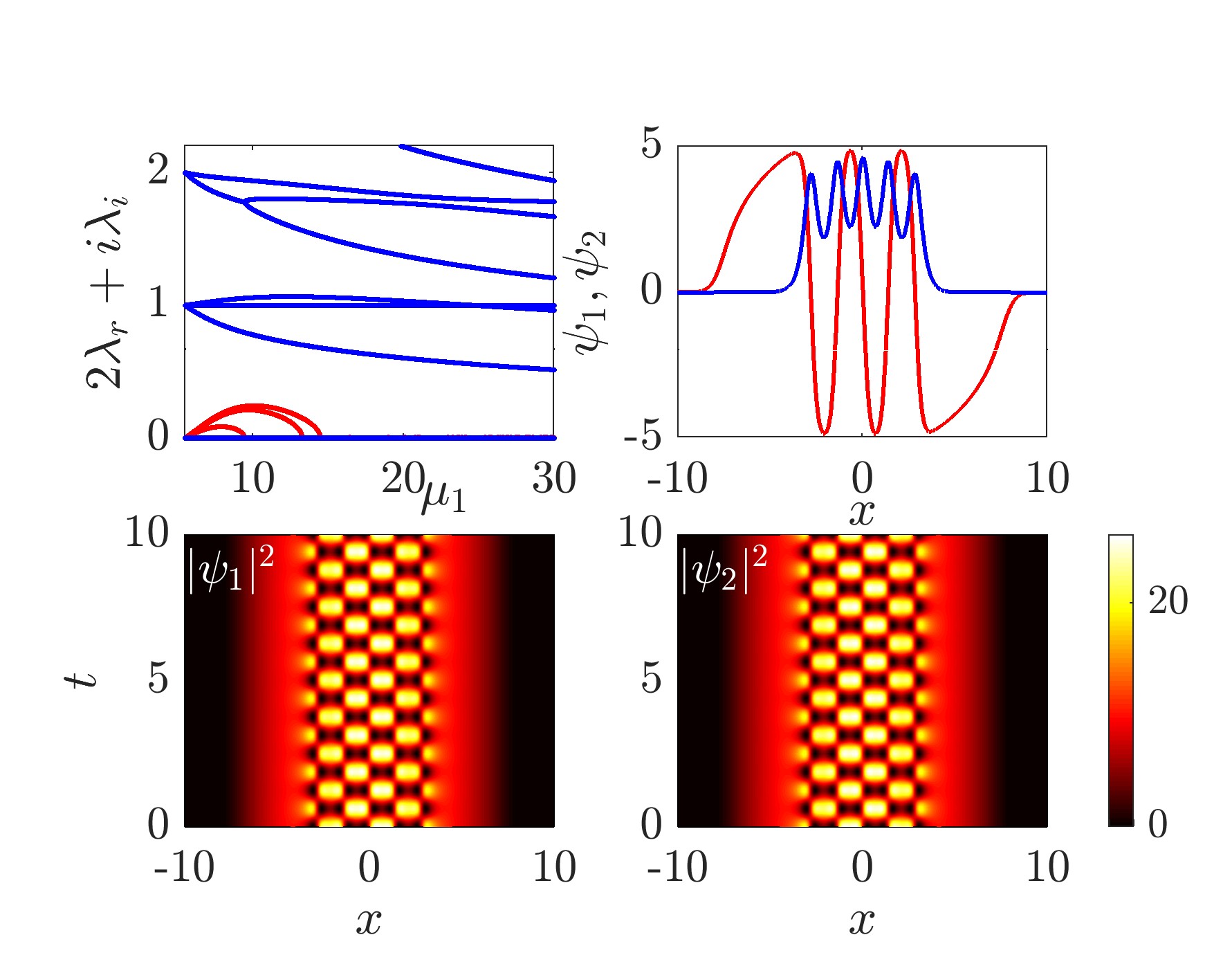}}
\subfigure[]{\includegraphics[width=0.33\textwidth]{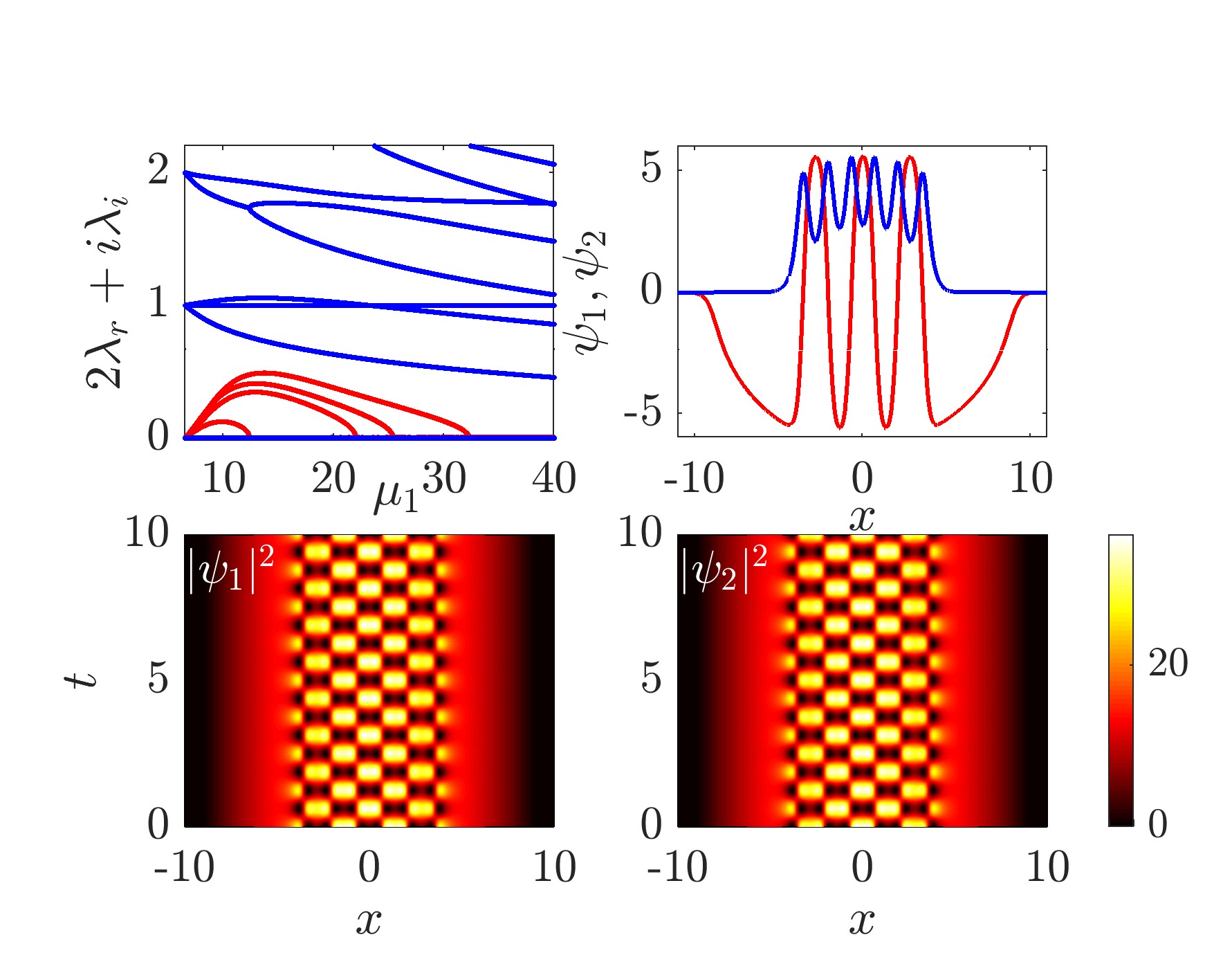}}
\subfigure[]{\includegraphics[width=0.33\textwidth]{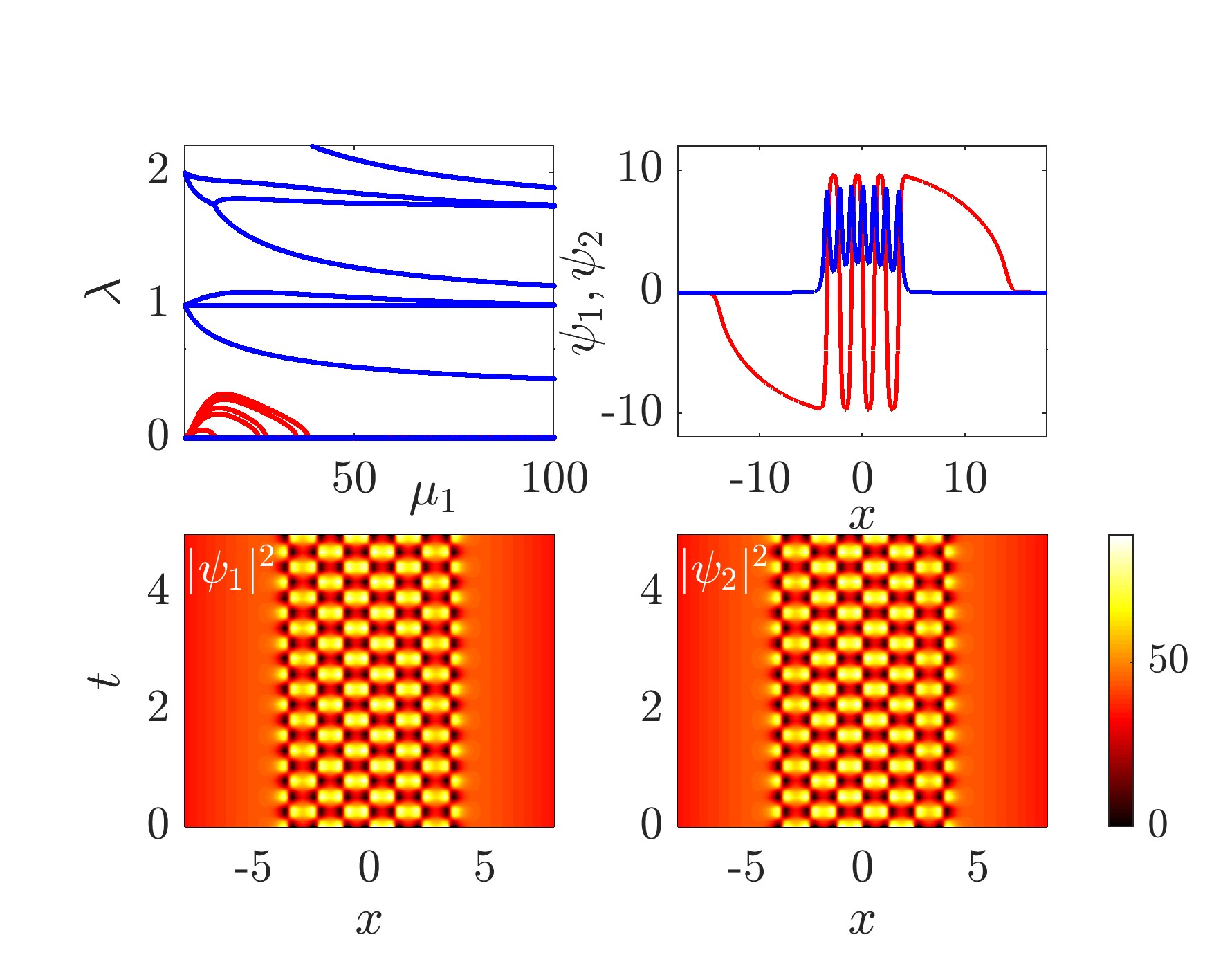}}
\caption{
(Color online) %
Same as Fig.~\ref{DB10}, but for the $\mathcal{S}_{n0}$ states from the linear limits $(n+0.5,0.5$) to typical 
large-density limits. The final chemical potentials are $(30,25)$, $(40,35)$, $(100,88)$ for $n=5, 6$, and $7$, 
respectively. Note that the number of unstable modes increases by $1$ as $n$ increases by $1$.
}
\label{DB50}
\end{figure*}

\begin{figure*}
\subfigure[]{\includegraphics[width=0.33\textwidth]{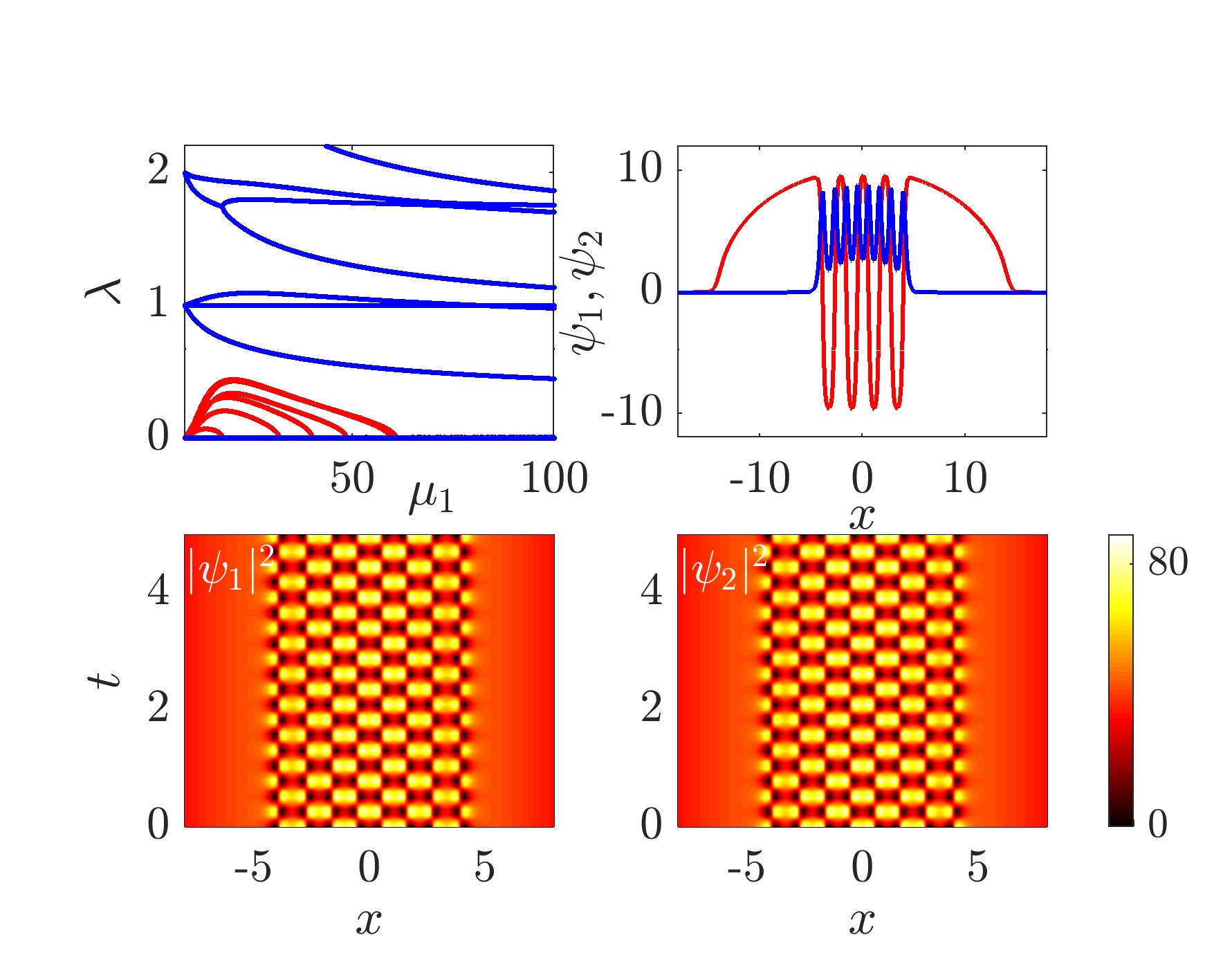}}
\subfigure[]{\includegraphics[width=0.33\textwidth]{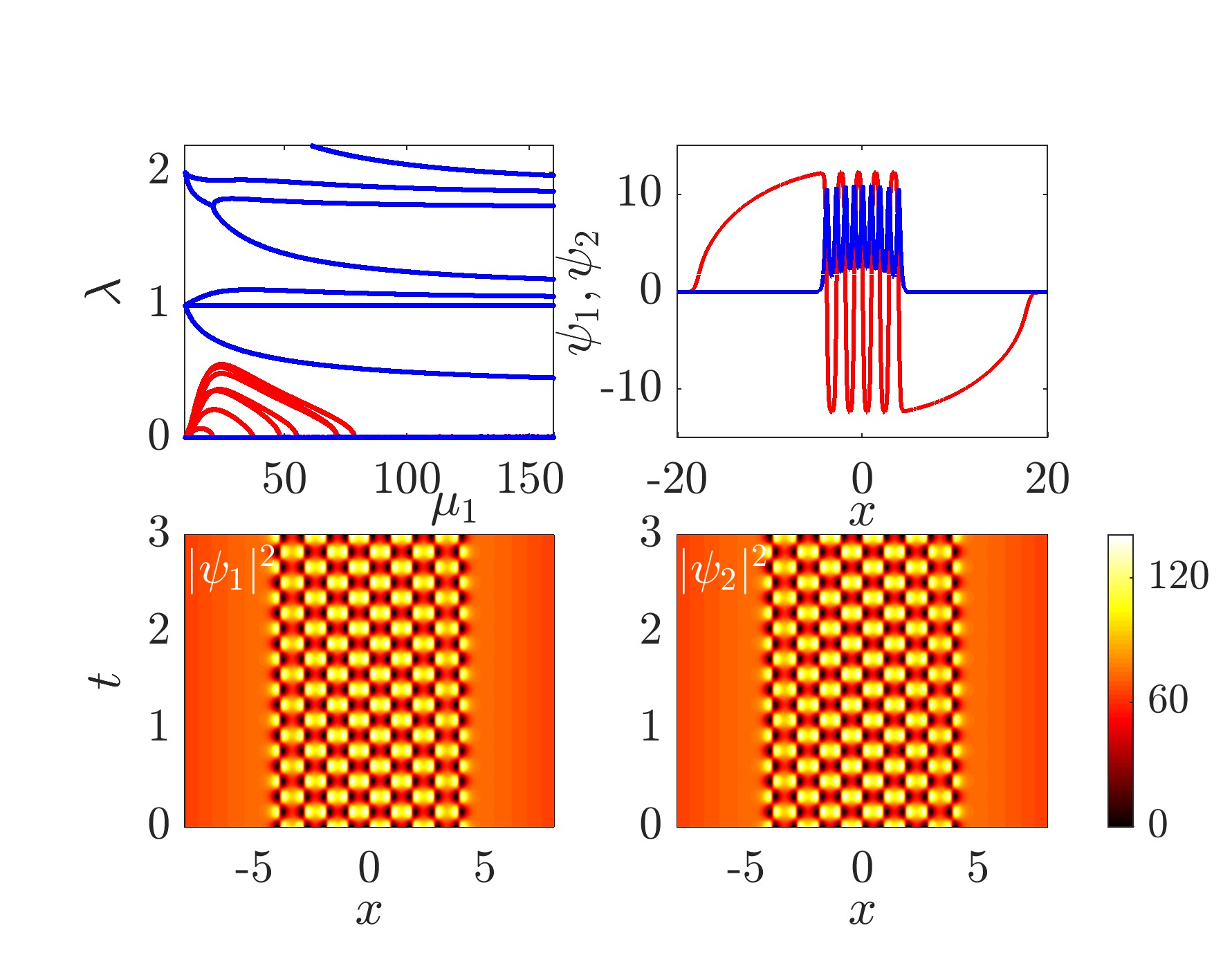}}
\subfigure[]{\includegraphics[width=0.33\textwidth]{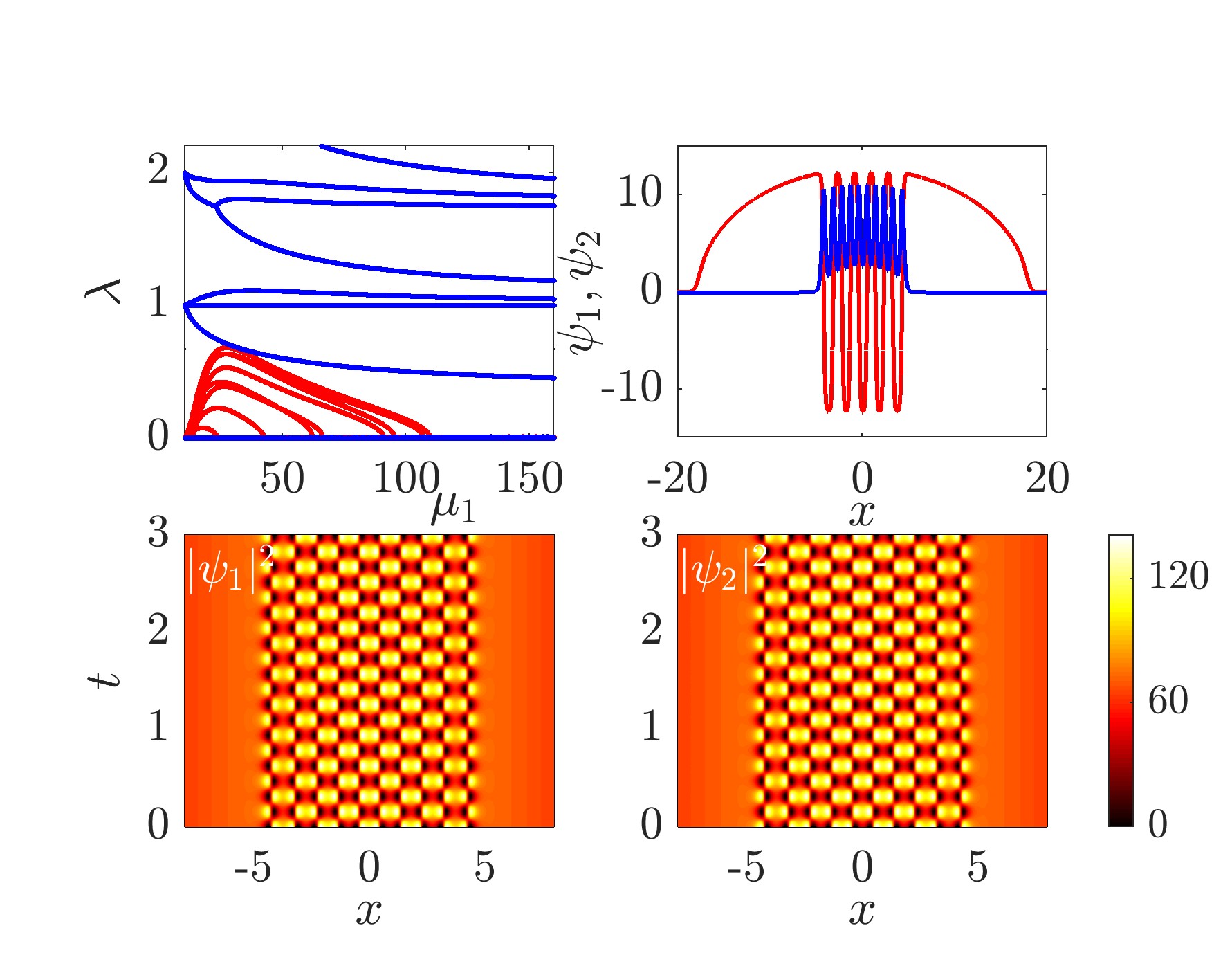}}
\caption{
(Color online) %
Same as Fig.~\ref{DB10}, but for the $\mathcal{S}_{80}$, $\mathcal{S}_{90}$, and $\mathcal{S}_{10,0}$ states
from the linear limits to $(100,88)$, $(160,140)$, $(160,140)$, respectively. The number of unstable modes 
continues to grow by one as the number of dark soliton grows by one, upon examining the closely spaced unstable
modes. Note that $10$ {DB} 
solitons require as large as $\mu_1 \approx 100$ to be fully stabilized.
}
\label{DB80}
\end{figure*}

Motivated by this observation, we next only look at $\mathcal{S}_{n0}$ states for $n=5, 6, ..., 10$. The 
results are presented in Figs.~\ref{DB50}-\ref{DB80}. Naturally,
per the above observations, and in line with the results
of~\cite{Panos:book},
the number of unstable modes, stemming 
from the linear limits, increases by one whenever a dark soliton is added to the first component. This trend 
makes it challenging to stabilize multiple dark-bright solitons. Indeed, for the \state{10,0} state, we need 
chemical potentials of the order of $100$ to fully stabilize this
structure.
However, for these sufficiently high values of the chemical potential,
our direct numerical simulations confirm the presence of breathing
rotated states with a large number of DD 
structures which lead
to the corresponding internal vibrations and the associated breathing
patterns. Since such initial conditions have been realized
in the recent experiments of~\cite{engels20}, it should, in principle,
be possible to visualize and resolve the relevant dynamics.

\begin{figure*}
\subfigure[]{\includegraphics[width=0.33\textwidth]{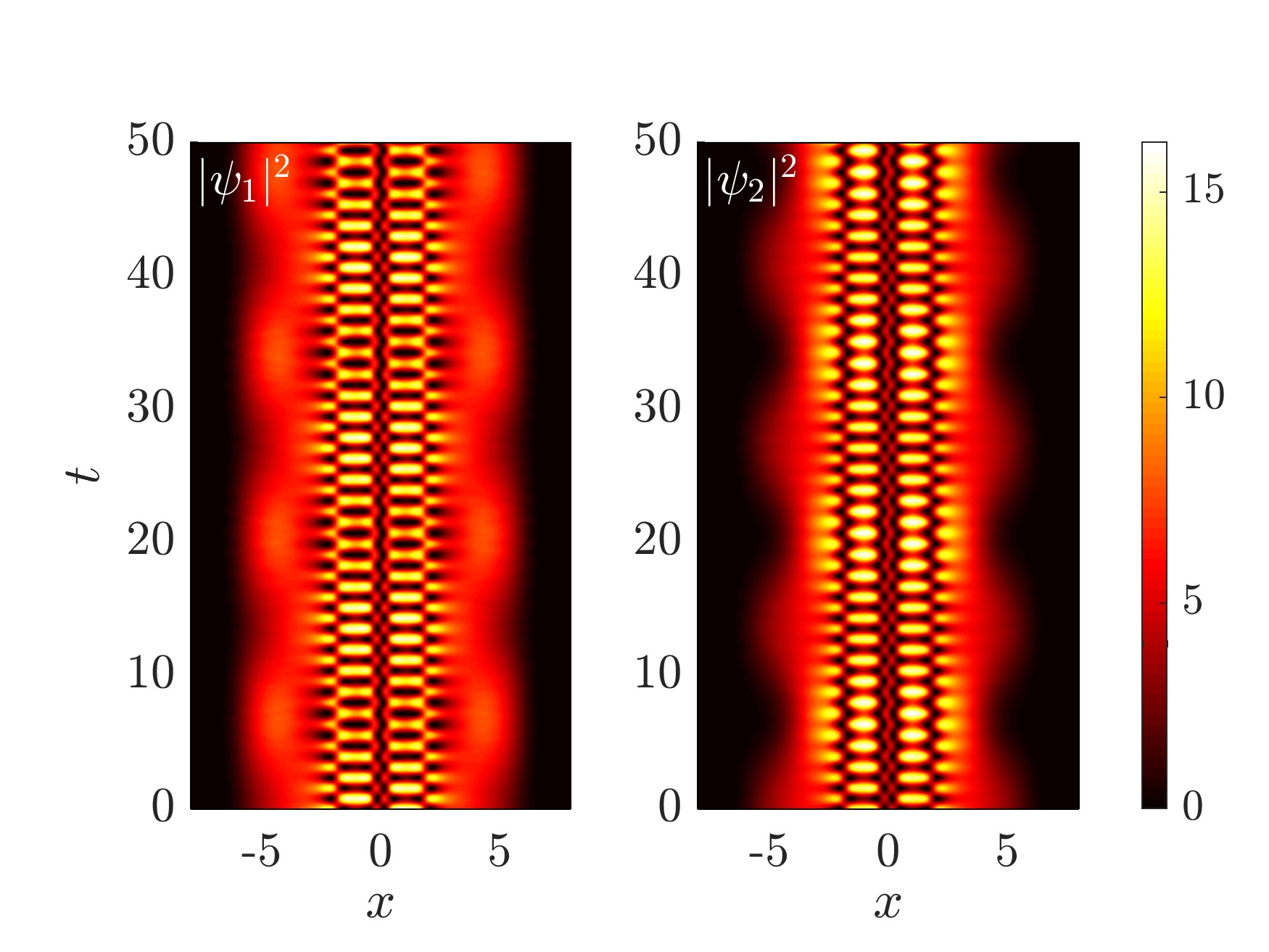}}
\subfigure[]{\includegraphics[width=0.33\textwidth]{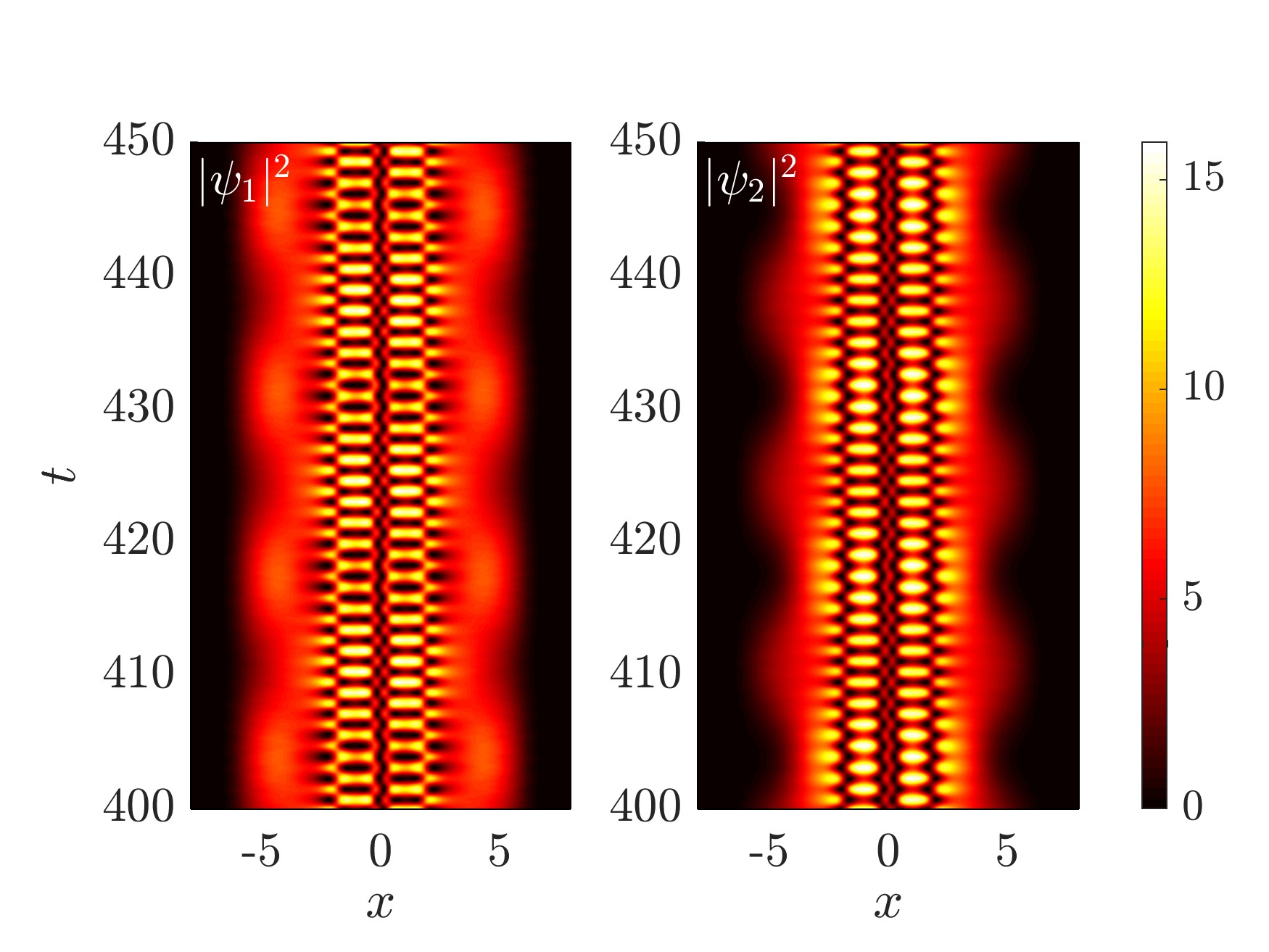}}
\subfigure[]{\includegraphics[width=0.33\textwidth]{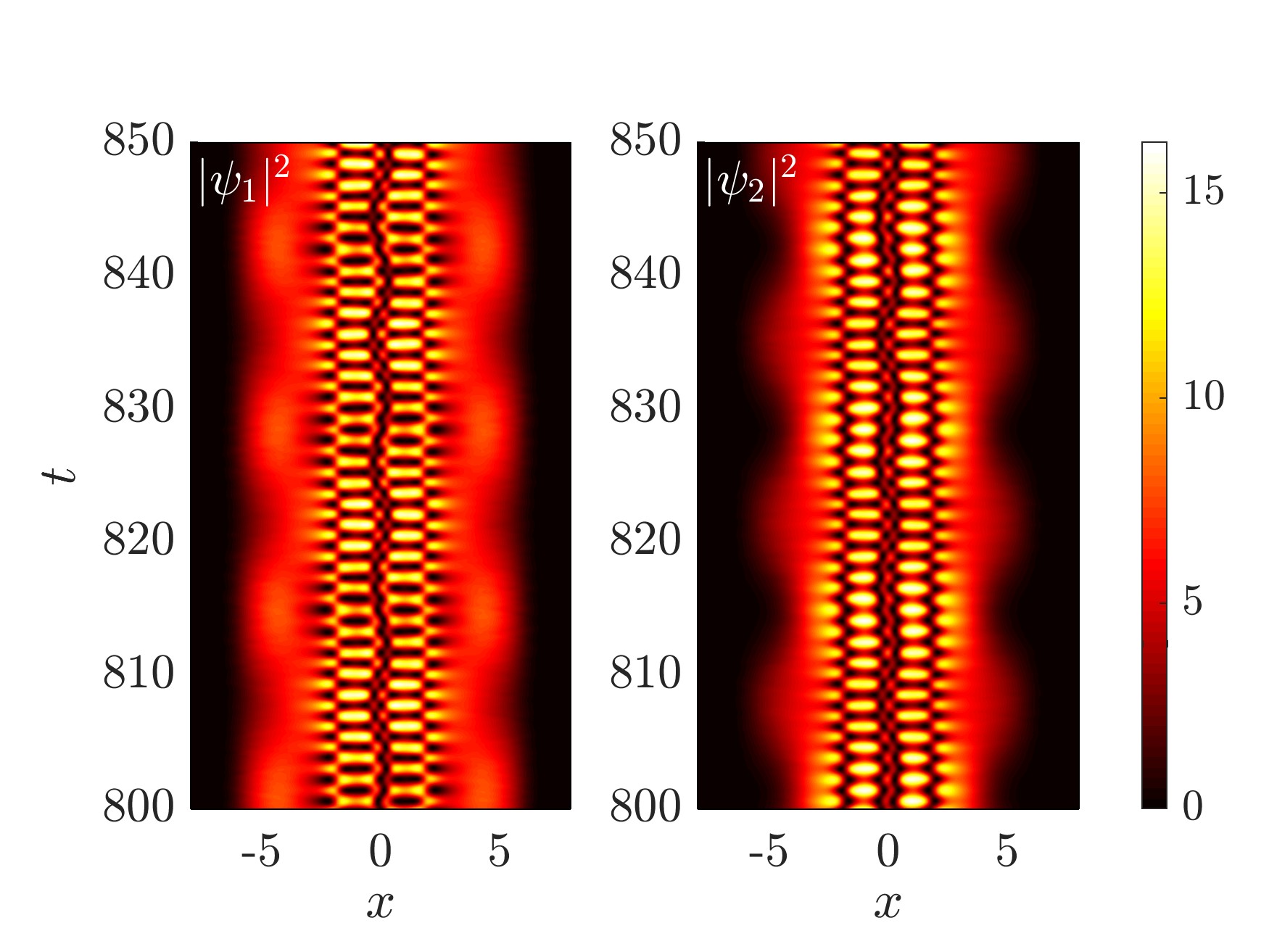}}
\caption{
(Color online) %
Weak disorder for few solitons caused by slight deviations from the perfectly symmetric Manakov limit using 
the experimentally relevant values $g_{11}=1.03$, $g_{22}=0.97$, and $g_{12}=g_{21}=1$ \cite{Dong:MDB}. Here,
the breathing patterns emanating from the $\mathcal{S}_{41}$ state are illustrated, where the stable dynamics 
is suddenly subjected to the above interaction parameters starting from $t=0$. Note that the ground state 
breathing mode is immediately excited, and the two condensates breathe in a correlated manner; see the 
{boundary undulation of the condensates.} 
The DD 
soliton breathing mode is finally also excited (around $t=700$) and the breathing patterns become 
distorted. Nevertheless, the DD 
soliton breathing patterns 
{remain} robust for several hundred periods 
before getting disordered. }
\label{disorder}
\end{figure*}

\begin{figure*}
\subfigure[]{\includegraphics[width=0.33\textwidth]{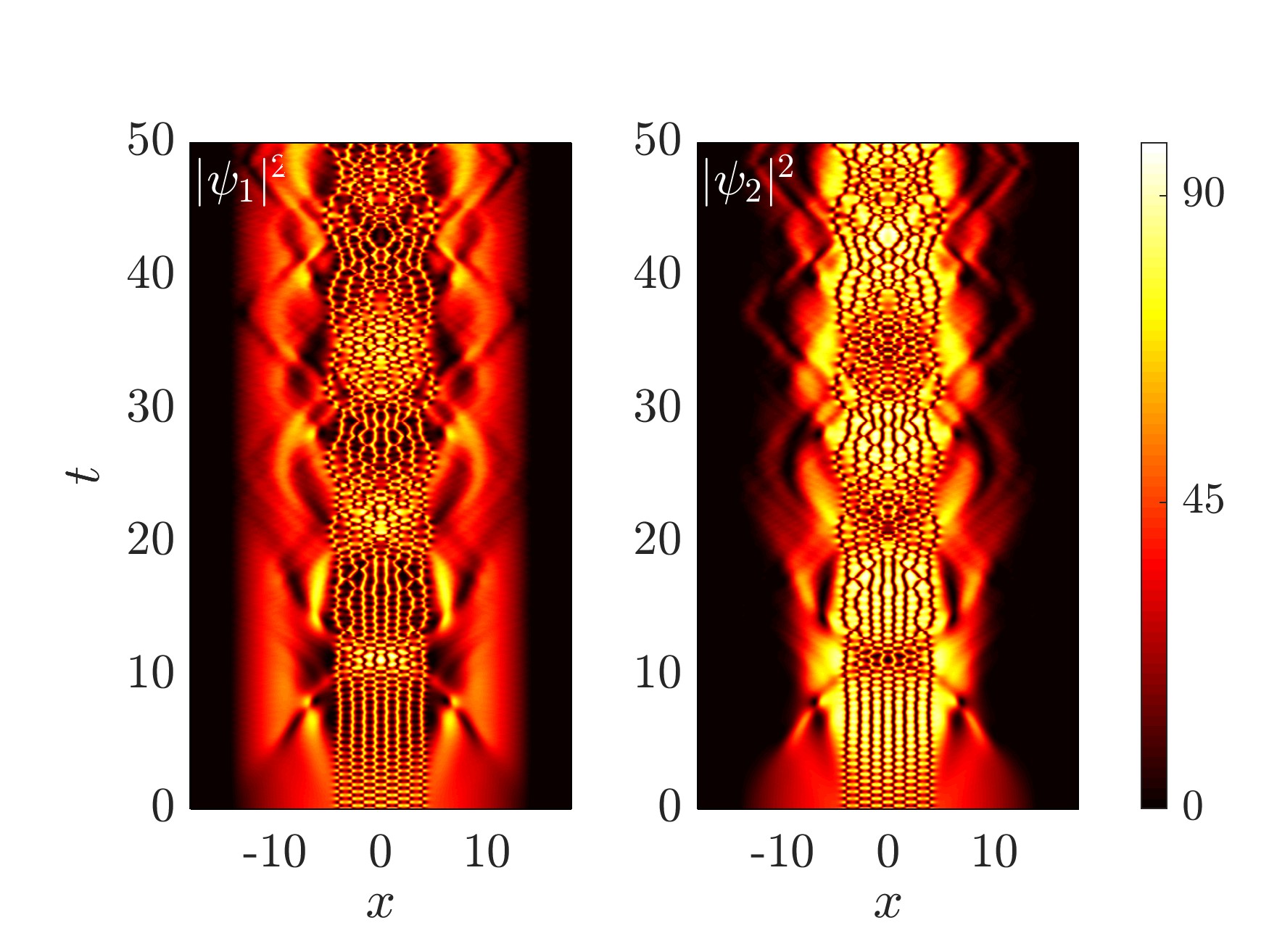}}
\subfigure[]{\includegraphics[width=0.33\textwidth]{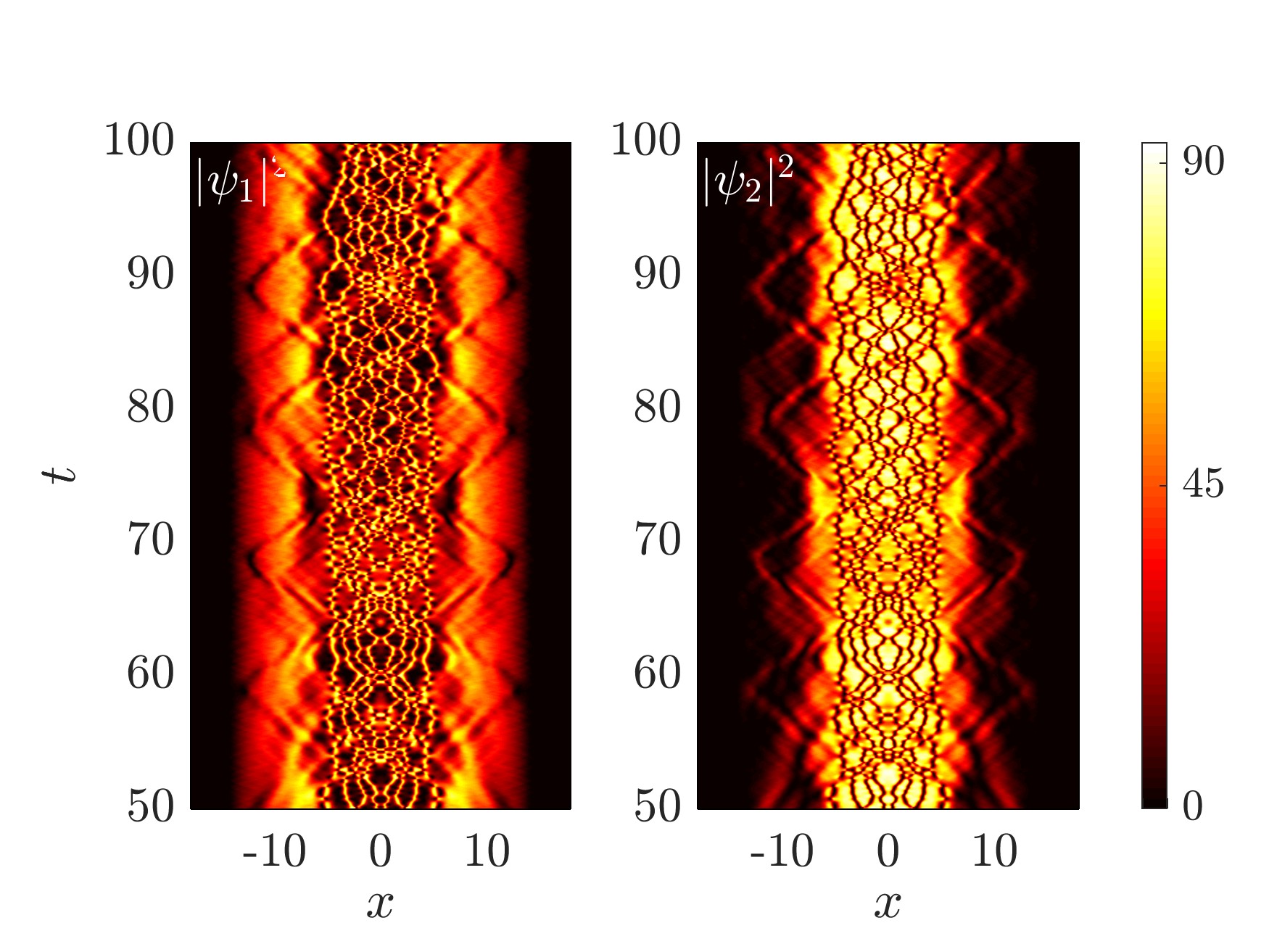}}
\subfigure[]{\includegraphics[width=0.33\textwidth]{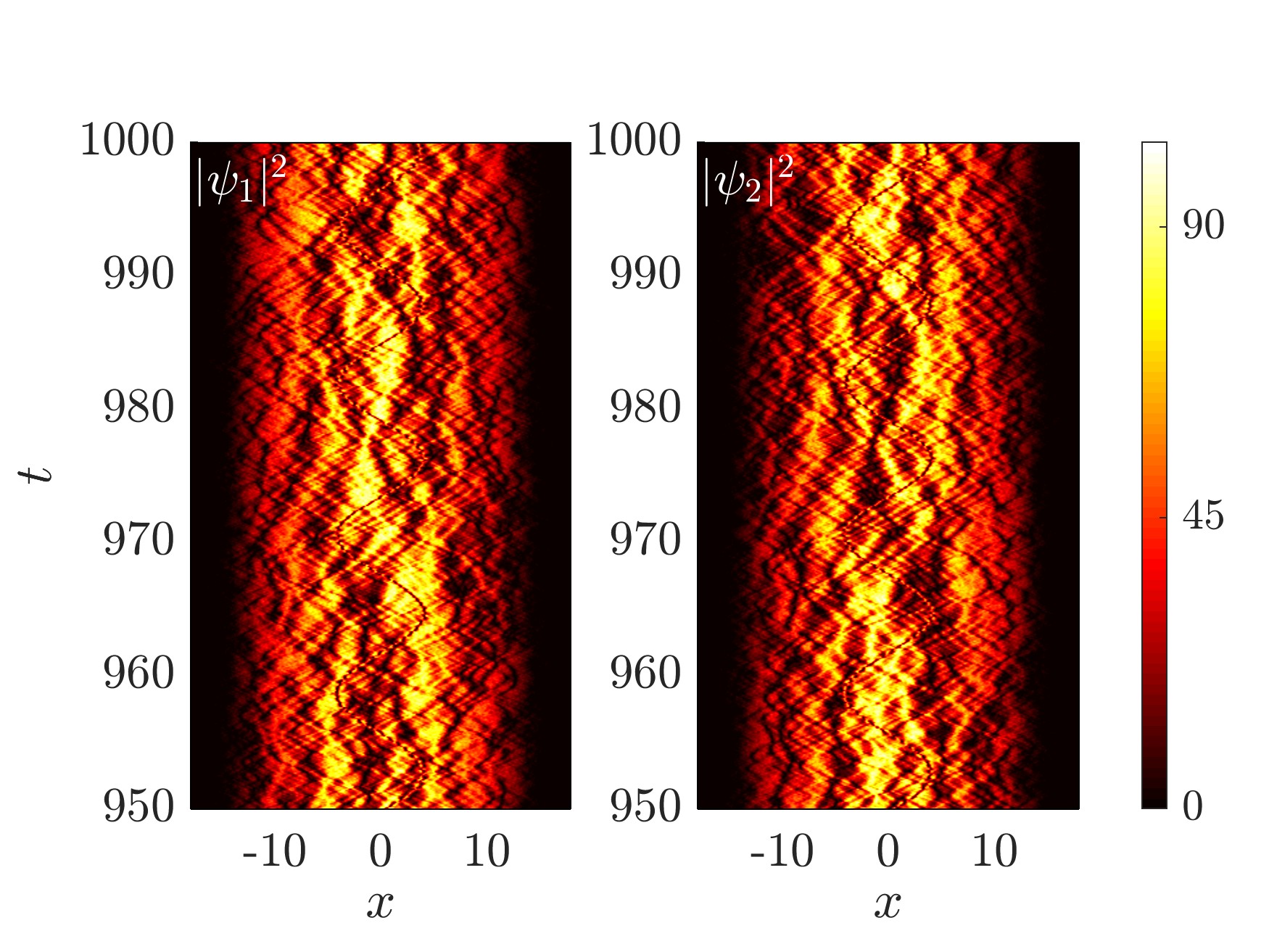}}
\caption{
(Color online) %
Same as Fig.~\ref{disorder} but for breathing patterns emanating from the $\mathcal{S}_{80}$ 
state. From about $t=10$ onwards, the dark soliton ``lattice'' gradually undergoes a transition 
towards a ``gaseous'' state. In this state,
the dark solitons 
frequently collide, thus generating dark bands in the density
profiles.
In addition, the background 
becomes highly excited and fragmented. See the text for more details.
}
\label{disorder2}
\end{figure*}

Finally, we study the effects of weak deviations from the perfectly symmetric Manakov limit using the 
experimentally relevant values $g_{11}=1.03$, $g_{22}=0.97$, and $g_{12}=g_{21}=1$~\cite{Hall:BEC2,Dong:MDB}. 
In particular, Fig.~\ref{disorder} illustrates the breathing patterns corresponding to the $\mathcal{S}_{41}$ state, 
where the stable dynamics is suddenly subjected to the above
interaction parameters starting from $t=0$ (i.e., a quench to the
above values of the interaction coefficients). Note that
the ground state breathing mode is immediately excited, and the two
condensates 
{breathe}
in a correlated (out of phase) manner; 
see the relevant condensate boundaries.
A DD 
soliton breathing mode is finally also excited (around $t=700$)
and the breathing patterns become distorted. Nevertheless, the soliton breathing patterns remain robust 
for several hundred periods before getting disordered. Similar behaviour is found for other patterns, where some patterns
persist for somewhat shorter periods (e.g. patterns resulting from $\mathcal{S}_{21}$ and $\mathcal{S}_{30}$) and 
others remain robust for a much larger number of periods (e.g. patterns resulting from $\mathcal{S}_{31}$ and 
$\mathcal{S}_{40}$).

Strong disorder can manifest quickly and
in a pronounced manner for many solitons for the same parameters. A typical time
evolution for the state $\mathcal{S}_{80}$ is shown in Fig.~\ref{disorder2}. In addition to the aforementioned weak 
disorder, the dark ``lattice'' in each component can quickly evolve
from a more ``crystalline'' 
into a ``gaseous'' state (in line with the terminology
of~\cite{Wang:OD}),
where the synchronization 
of the DD 
soliton vibrations is gradually lost. The states
then become so disordered that there is no clearly discernible
stationary or periodic pattern.
Indeed,
dark solitons in the two components 
frequently collide forming some ``dark bands'' in the density profile. The backgrounds of both states are also
highly excited with this phenomenology persisting up to the time
horizon of
the very long
evolution simulations shown in Fig.~\ref{disorder2}.

\subsection{Multiple dark-dark soliton breathing patterns in a homogeneous setting}

The model with no external trap is an integrable Manakov model \cite{Manokov74}, and various types
of solitons have been deduced using the traditional inverse scattering method, B\"{a}cklund transformation
method, and Hirota bilinear method
\cite{DT,Dressingmethod,Hirota,Lakshman,Lichen:DT}, such as bright-bright (BB),
DB
and 
DD 
solitons. The DB 
soliton
solutions have been extensively investigated \cite{DBS1,Rajendran_2009,DDDB,DBcounterflow,Dong:MDB,DBtunneling,DBcollisions}.
Recently, a modified Darboux transformation method focusing towards dark and
DB 
solitary waves in repulsively interacting BECs was developed \cite{Lichen:DT}.
Upon use this method to identify the multi-DB 
soliton solutions, here we focus on the breathing
variants thereof arising through SO$(2)$ rotations. The
analytical expressions are similar to
the ones developed in the above mentioned earlier works,
therefore we do not present them in detail.
\begin{figure*}[htbp]
\begin{center}
\includegraphics[height=70mm,width=165mm]{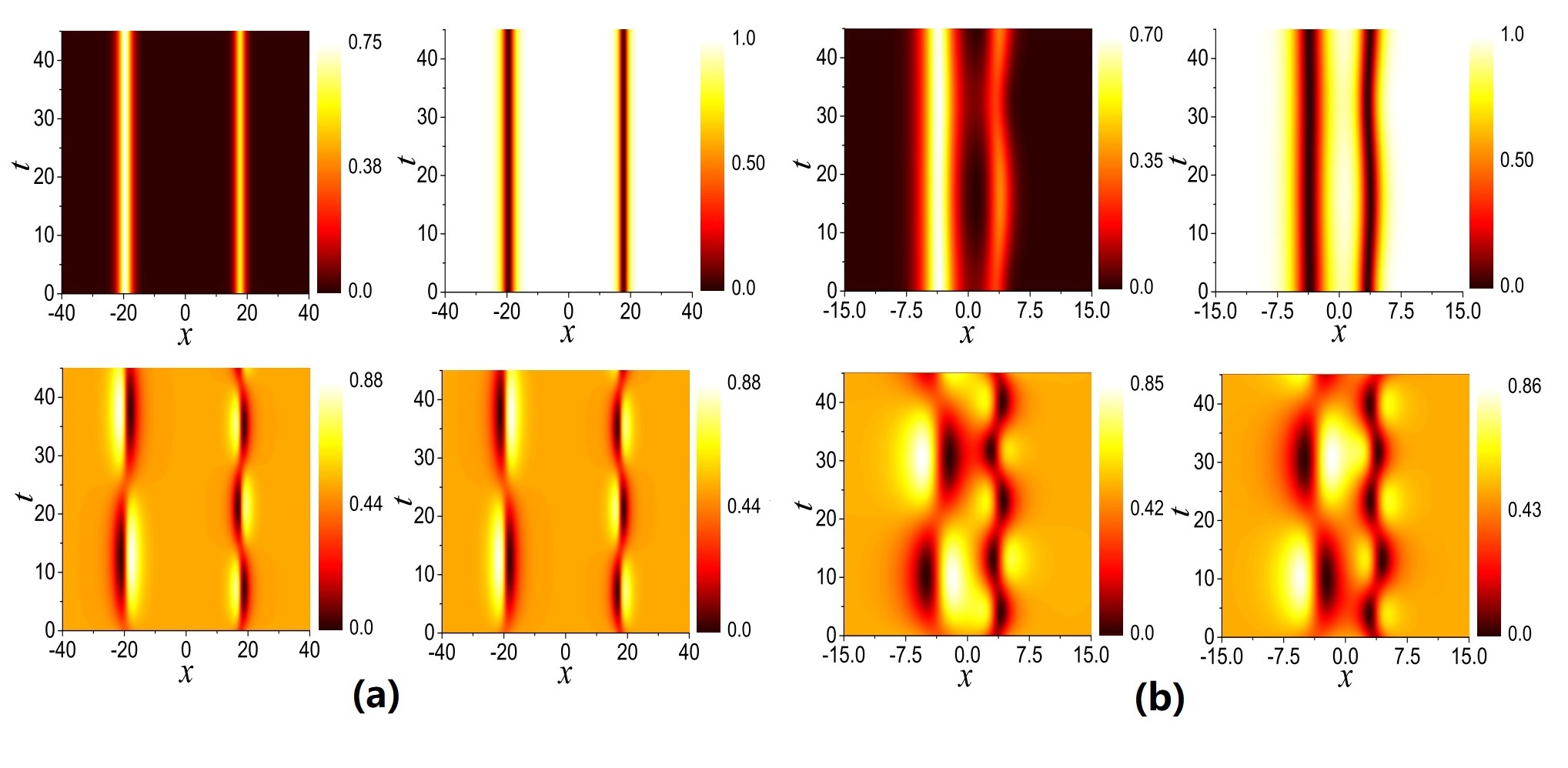}
\end{center}
\caption{
(Color online)
Space-time density evolutions of two DB 
solitons and 
{DD} breathing patterns for well-separated (a) and closely initialized 
(b) cases. The top panels show the bright-soliton component and the dark-soliton 
component, and the lower panels show the rotated {DD} 
breathing patterns. 
The well-separated solitons yield effectively isolated beating dark-dark solitons, 
while the nonlinear interaction between solitons changes the beating patterns
significantly.}
\label{Figdb2}
\end{figure*}

\begin{figure*}[htbp]
\begin{center}
\includegraphics[height=70mm,width=165mm]{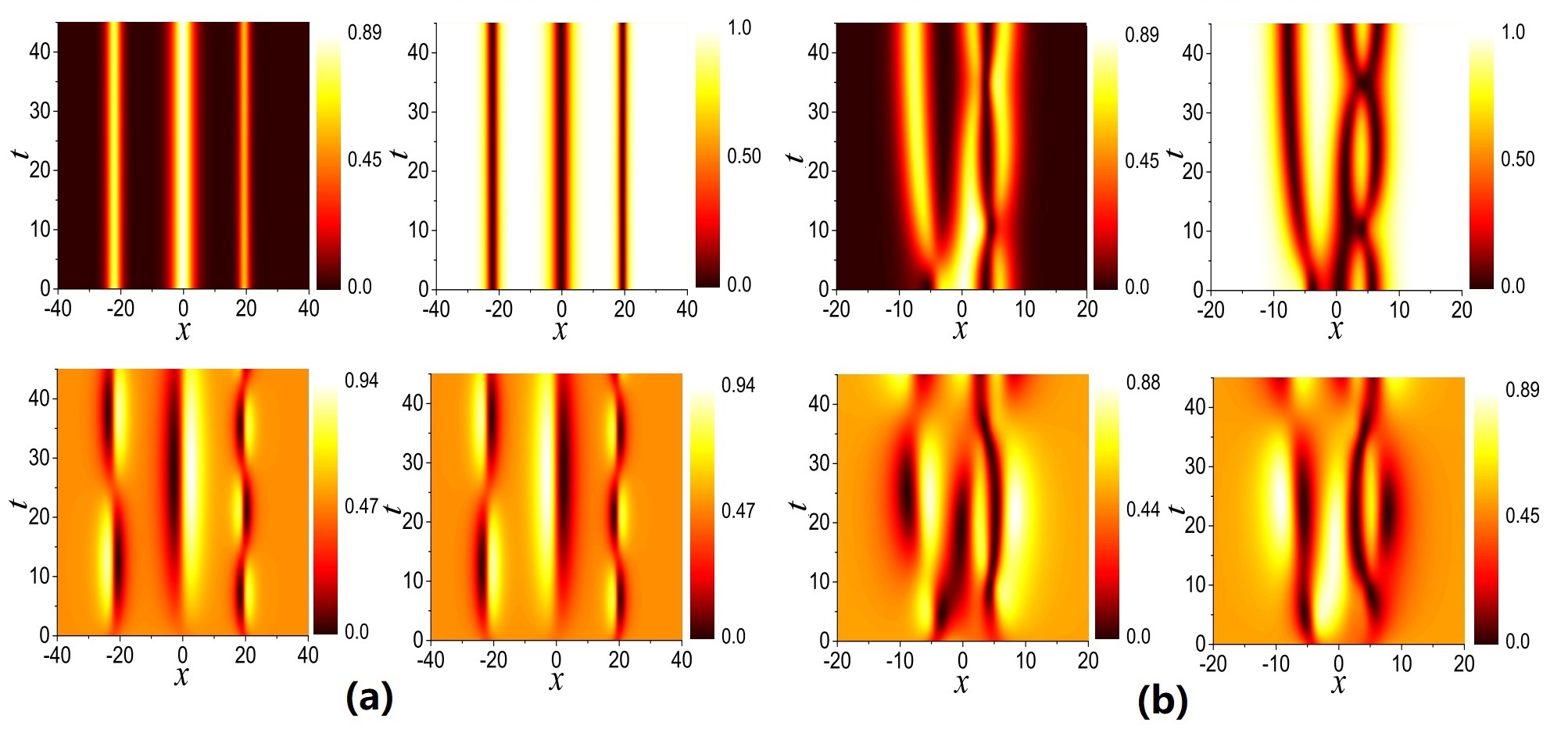}
\end{center}
\caption{
(Color online)
Similar to the previous figure but now for the 3 dark-bright soliton case, once again when well-separated (a) and nonlinearly interacting (b).}\label{Figdb3}
\end{figure*}

Two typical quasi-static solutions along with the
symmetric SO(2) rotated solutions are depicted in 
Fig.~\ref{Figdb2} {where the panel (a) of the figure} 
shows the evolution of two well-separated DB 
solitons; 
essentially these waves are sufficiently far away from each other 
and, hence, do not feel the presence of each other over
the time scale of the simulation. As a result, over the horizon
of the simulation shown in panel (a) of Fig.~\ref{Figdb2}, the
internal beating of each of the two DD solitary waves occurs with
different frequencies.
When the solitons are initialized closer, the interaction between
them changes the beating 
patterns as illustrated in Fig.~\ref{Figdb2}(b). Similarly, the cases for three DB 
solutions and the rotated dynamics are shown in Fig.~\ref{Figdb3}. 
Fig.~\ref{Figdb3}(a) shows the evolution
of three well-separated DB 
solitons, again with very distinct breathing frequencies.
As the initial DBs are brought closer,
the beating patterns again become strongly affected by the
interaction between solitons;
cf. Fig.~\ref{Figdb2}(b). These beating
patterns are found to be stable against weak perturbations.
The resulting pattern while highly dynamical remains spatially
localized,
while this would no longer be true due to modulational instability of
the density
background in the attractive case~\cite{Lichen:MI2}.

\subsection{Three-component dark-dark-dark breathing patterns}
\label{TC}
In the last section of our present work, we turn our focus to the 
three-component case. In particular, the GPEs can be generalized to
the following system:
\begin{subequations}
\begin{align}
i\frac{\partial \psi_{1}}{\partial t} &= -\frac{1}{2}\frac{\partial^{2}\psi_{1}}{\partial x^{2}} %
+\left(g_{11}|\psi_{1}|^{2}+g_{12}|\psi_{2}|^{2}+g_{13}|\psi_{3}|^{2}\right)\psi_{1} %
+V(x)\psi_{1}, \label{three_comp_GPE_orig_1}  \\
i\frac{\partial \psi_{2}}{\partial t} &= -\frac{1}{2}\frac{\partial^{2}\psi_{2}}{\partial x^{2}} %
+\left(g_{21}|\psi_{1}|^{2}+g_{22}|\psi_{2}|^{2}+g_{23}|\psi_{3}|^{2}\right)\psi_{2} %
+V(x)\psi_{2},\label{three_comp_GPE_orig_2} \\
i\frac{\partial \psi_{3}}{\partial t} &= -\frac{1}{2}\frac{\partial^{2}\psi_{3}}{\partial x^{2}} %
+\left(g_{31}|\psi_{1}|^{2}+g_{32}|\psi_{2}|^{2}+g_{33}|\psi_{3}|^{2}\right)\psi_{3} %
+V(x)\psi_{3}, \label{three_comp_GPE_orig_3}
\end{align}
\end{subequations}
where $\psi_{j}(x,t)$ ($j=1,2,3$) are similarly the macroscopic wavefunctions and 
$g_{ij}$ ($i,j=1,2,3$) are the interaction coefficients with $g_{21}\equiv g_{12}$, $g_{31}\equiv g_{13}$, 
$g_{32}\equiv g_{23}$. Note that we will explore the Manakov case
herein corresponding to $g_{ij}=1$.
As discussed in the introduction, in addition to its mathematical
interest,
this scenario has been touched upon in nonlinear optical
multi-component
settings; a corresponding BEC framework would also need to incorporate
the spin-dependent aspect of interactions within the spinor
condensates~\cite{kawueda,stampueda}.
The Manakov setting features an SU$(3)$ symmetry of the system. The external potential assumes the same parabolic 
form of $V(x)=\left(1/2\right)\omega^{2}x^{2}$ with $\omega=1$.
Consequently, stationary states can be constructed by assuming
\begin{equation}
\psi_{j}(x,t)=\psi^{0}_{j}(x)e^{-i\mu_{j}t}
\label{three_comp_ansatz}
\end{equation}
which transform Eqs.~\eqref{three_comp_GPE_orig_1}-\eqref{three_comp_GPE_orig_3} into the steady-state system:
\begin{subequations}
\begin{align}
-\frac{1}{2}\frac{d^{2}\psi_{1}^{0}}{d x^{2}} %
+\left(g_{11}|\psi_{1}^{0}|^{2}+g_{12}|\psi_{2}^{0}|^{2}+g_{13}|\psi_{3}^{0}|^{2}\right)\psi_{1}^{0} %
+V(x)\psi_{1}^{0}-\mu_{1}\psi_{1}^{0} &= 0, \label{three_comp_GPE_stead_1} \\
-\frac{1}{2}\frac{d^{2}\psi_{2}^{0}}{d x^{2}} %
+\left(g_{12}|\psi_{1}^{0}|^{2}+g_{22}|\psi_{2}^{0}|^{2}+g_{23}|\psi_{3}^{0}|^{2}\right)\psi_{2}^{0} %
+V(x)\psi_{2}^{0}-\mu_{2}\psi_{2}^{0} &= 0,\label{three_comp_GPE_stead_2}  \\
-\frac{1}{2}\frac{d^{2}\psi_{3}^{0}}{d x^{2}} %
+\left(g_{13}|\psi_{1}^{0}|^{2}+g_{23}|\psi_{2}^{0}|^{2}+g_{33}|\psi_{3}^{0}|^{2}\right)\psi_{3}^{0} %
+V(x)\psi_{3}^{0}-\mu_{3}\psi_{3}^{0} &= 0, \label{three_comp_GPE_stead_3}
\end{align}
\end{subequations}
that we solve numerically. The computational set up is similar to the two-component case, tracing states from their linear limits to a Thomas-Fermi regime. 
For completeness, the BdG stability analysis is presented in the Appendix. It is also relevant to mention that restricting this matrix to its $4 \times 4$ submatrix of the top
left elements and setting $\psi_3=0$, one retrieves naturally as a
special case the corresponding 2-component BdG stability matrix.
Since our goal is to illustrate the generality of our method, we will only examine in detail here two prototypical 
examples, namely the low-lying \state{210} and \state{310} states.

\begin{figure}[pt]
\begin{center}
\includegraphics[height=.19\textheight, angle =0]{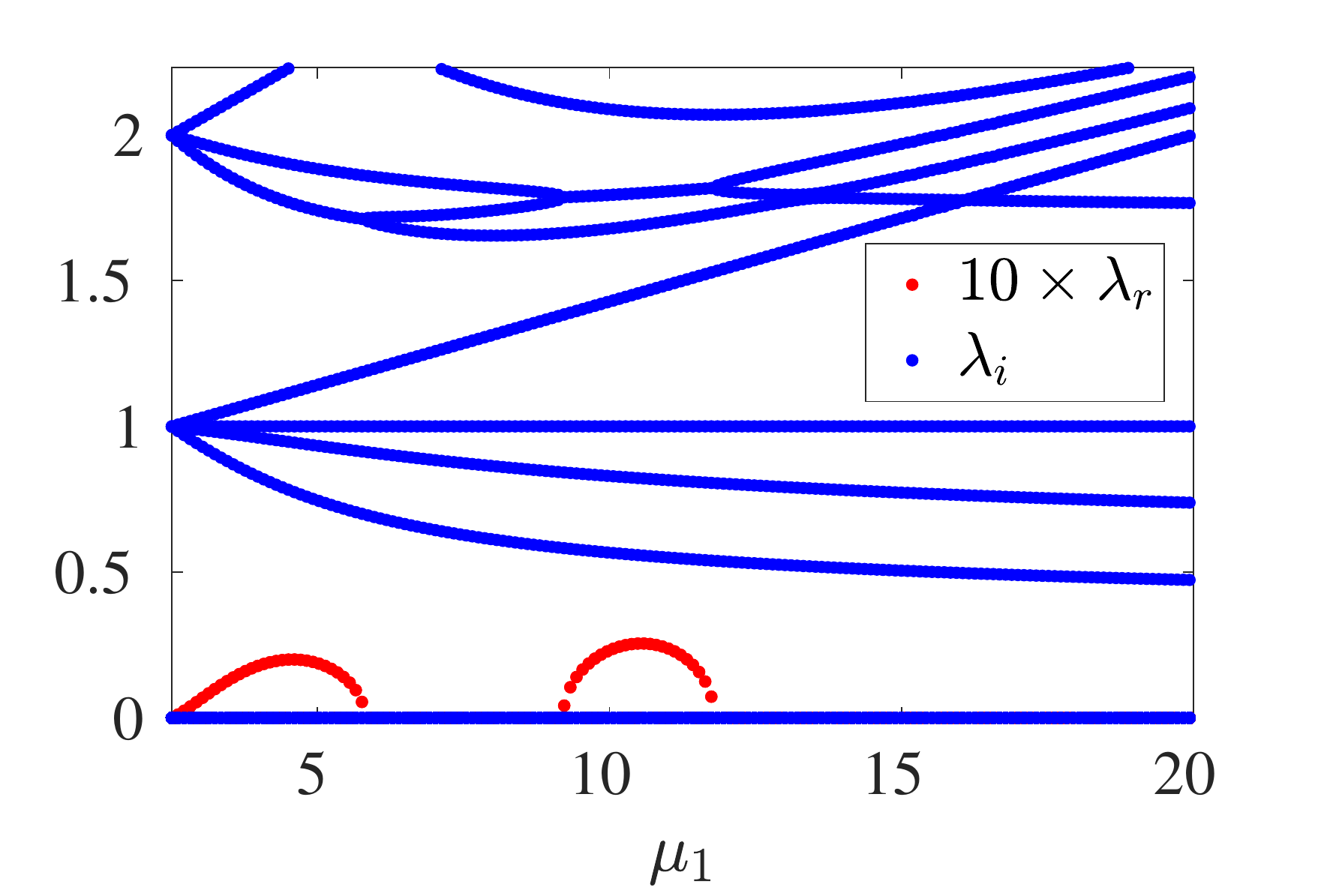}
\includegraphics[height=.19\textheight, angle =0]{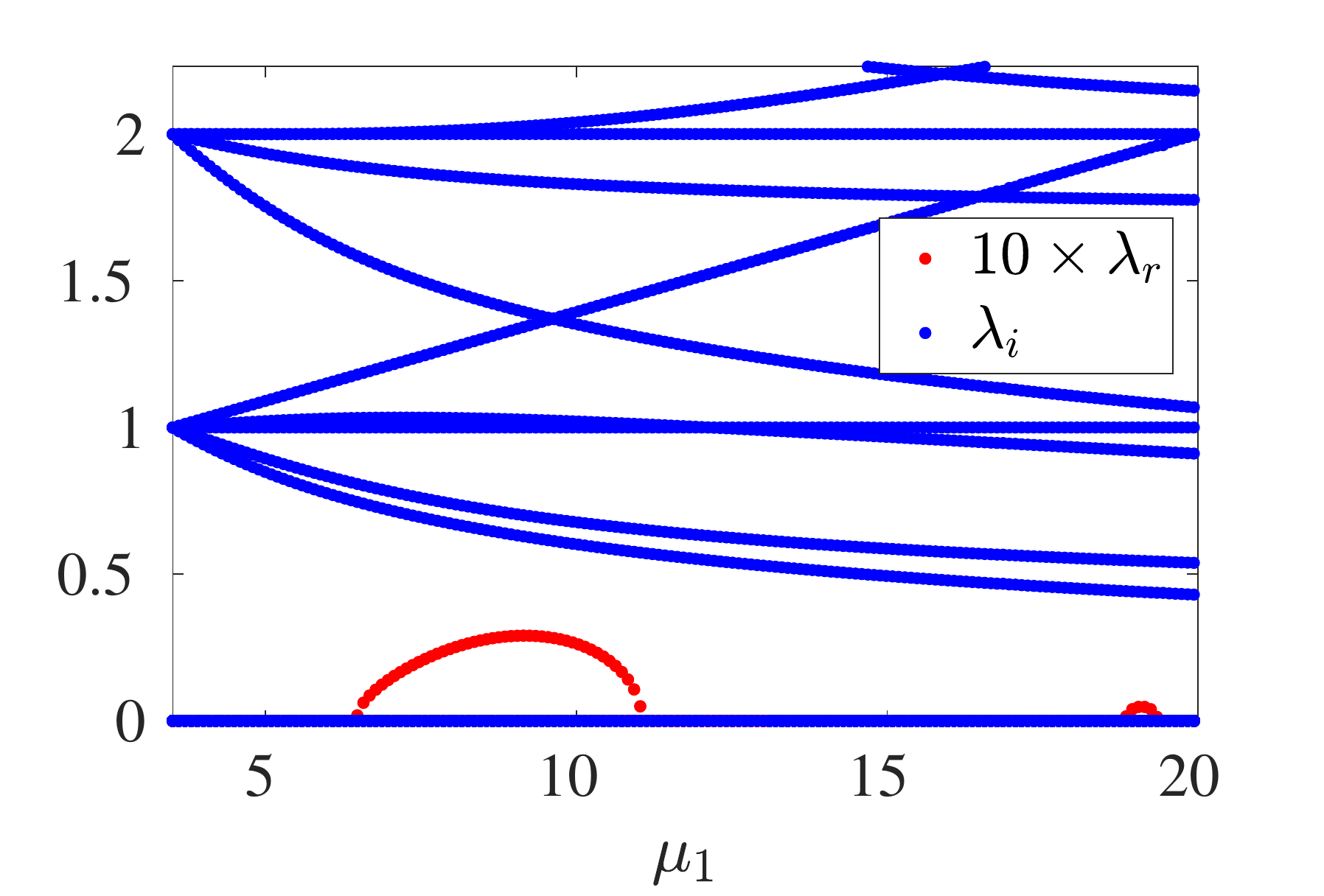}
\end{center}
\caption{
(Color online) %
The BdG spectra ($\lambda=\lambda_{r}+i\lambda_{i}$) for the states \state{210} 
(left) and \state{310} (right) as a function of $\mu_{1}$. Here, the states emanate
from the linear limits and our continuation terminates close to a TF regime with 
$(\mu_{1},\mu_{2},\mu_{3})=(20, 18, 16)$. Note that both states feature wide intervals 
of stability. The real parts in both panels are multiplied by a factor of $10$ 
for visualization purposes.}
\label{Spectra3C}
\end{figure}

\begin{figure}[pt]
\begin{center}
\includegraphics[height=.17\textheight, angle =0]{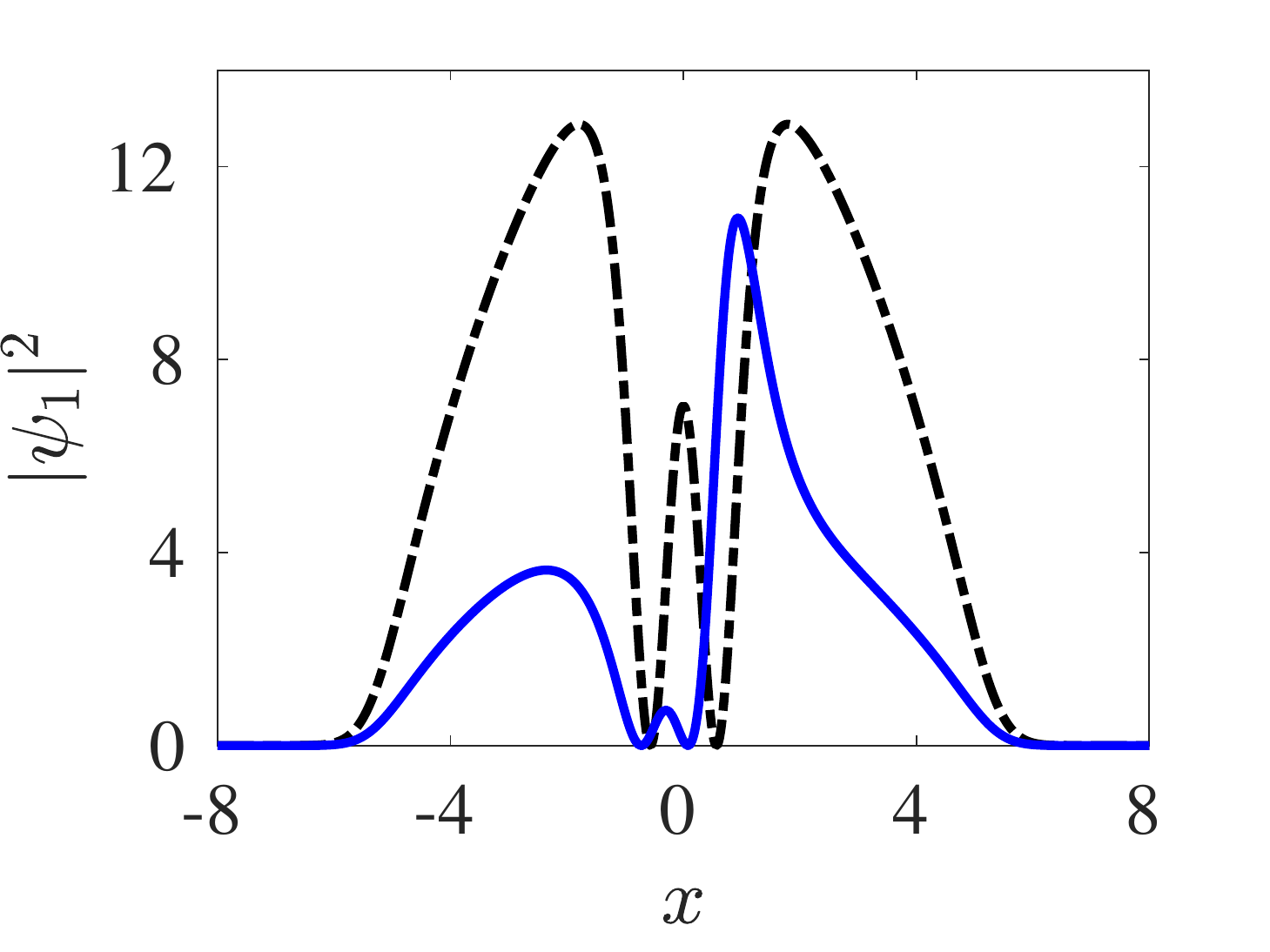}
\includegraphics[height=.17\textheight, angle =0]{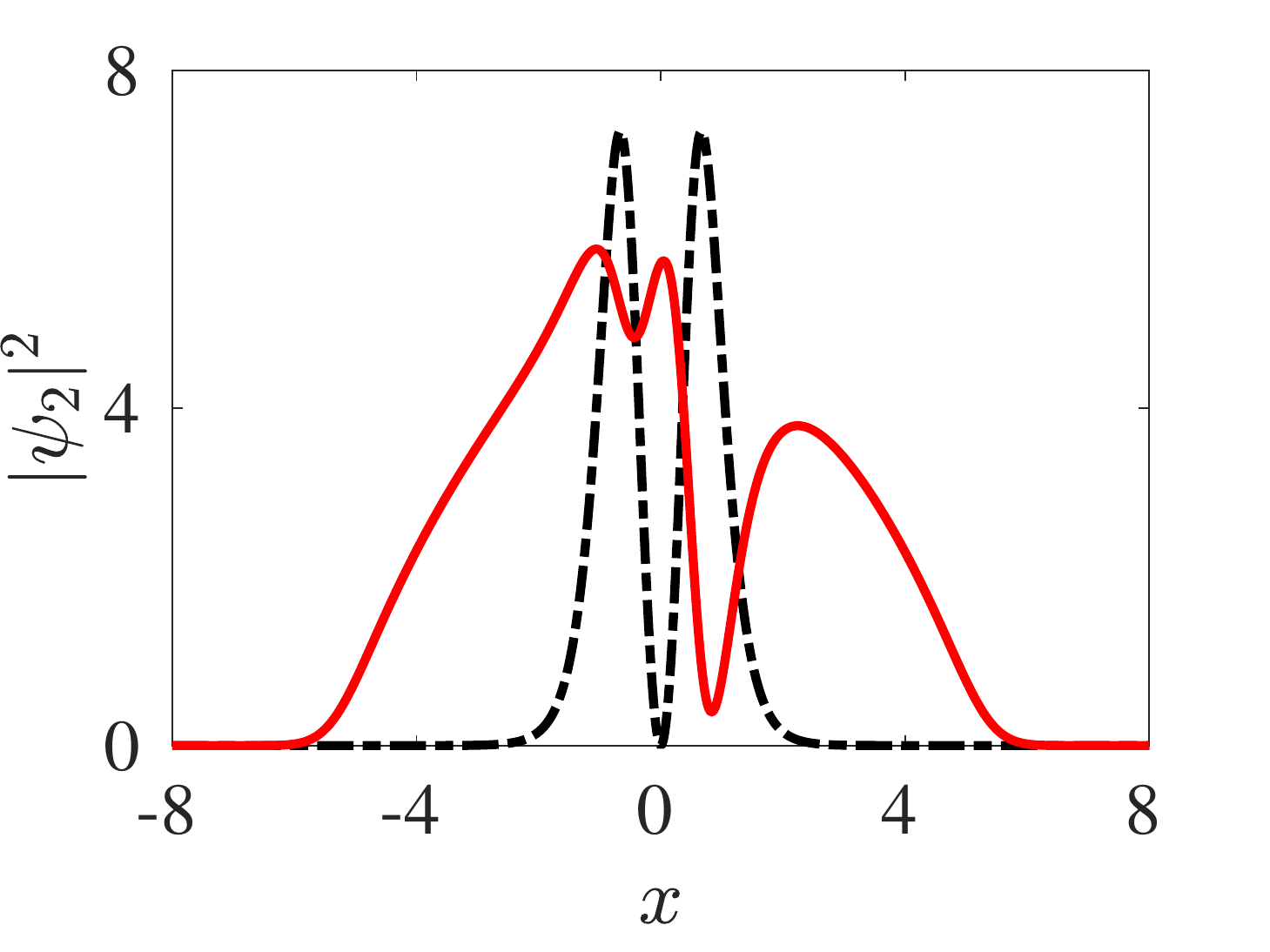}
\includegraphics[height=.17\textheight, angle =0]{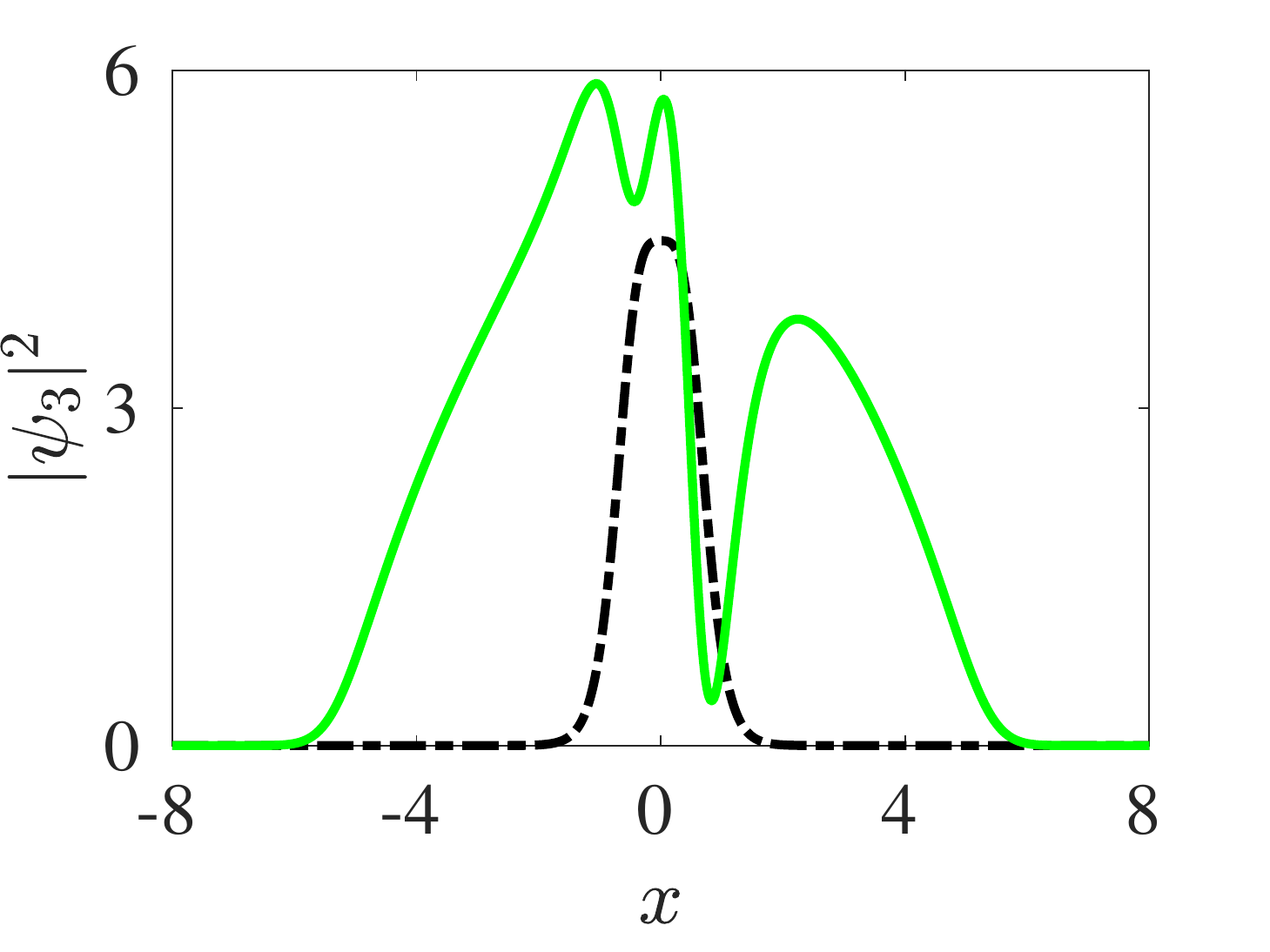}
\includegraphics[height=.17\textheight, angle =0]{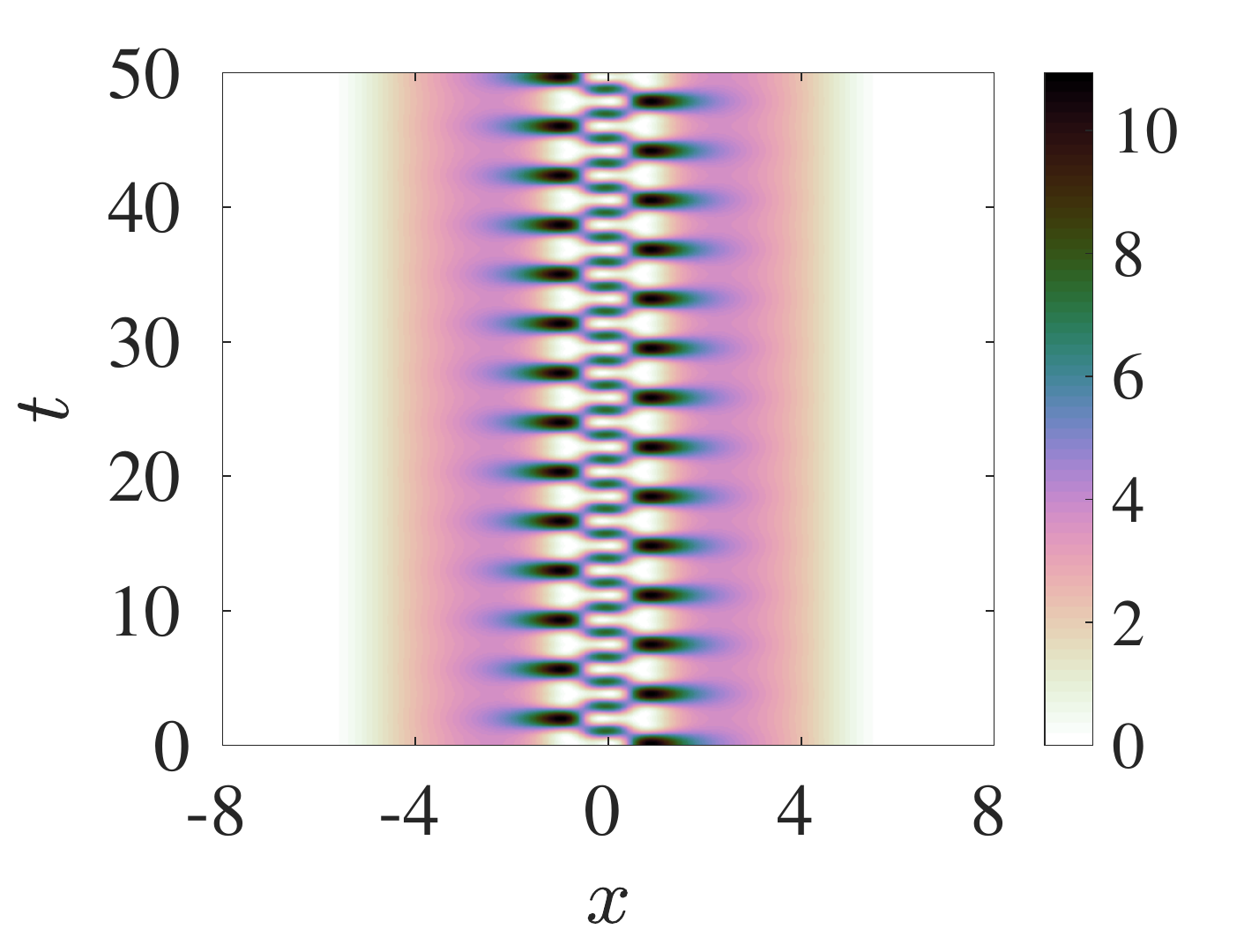}
\includegraphics[height=.17\textheight, angle =0]{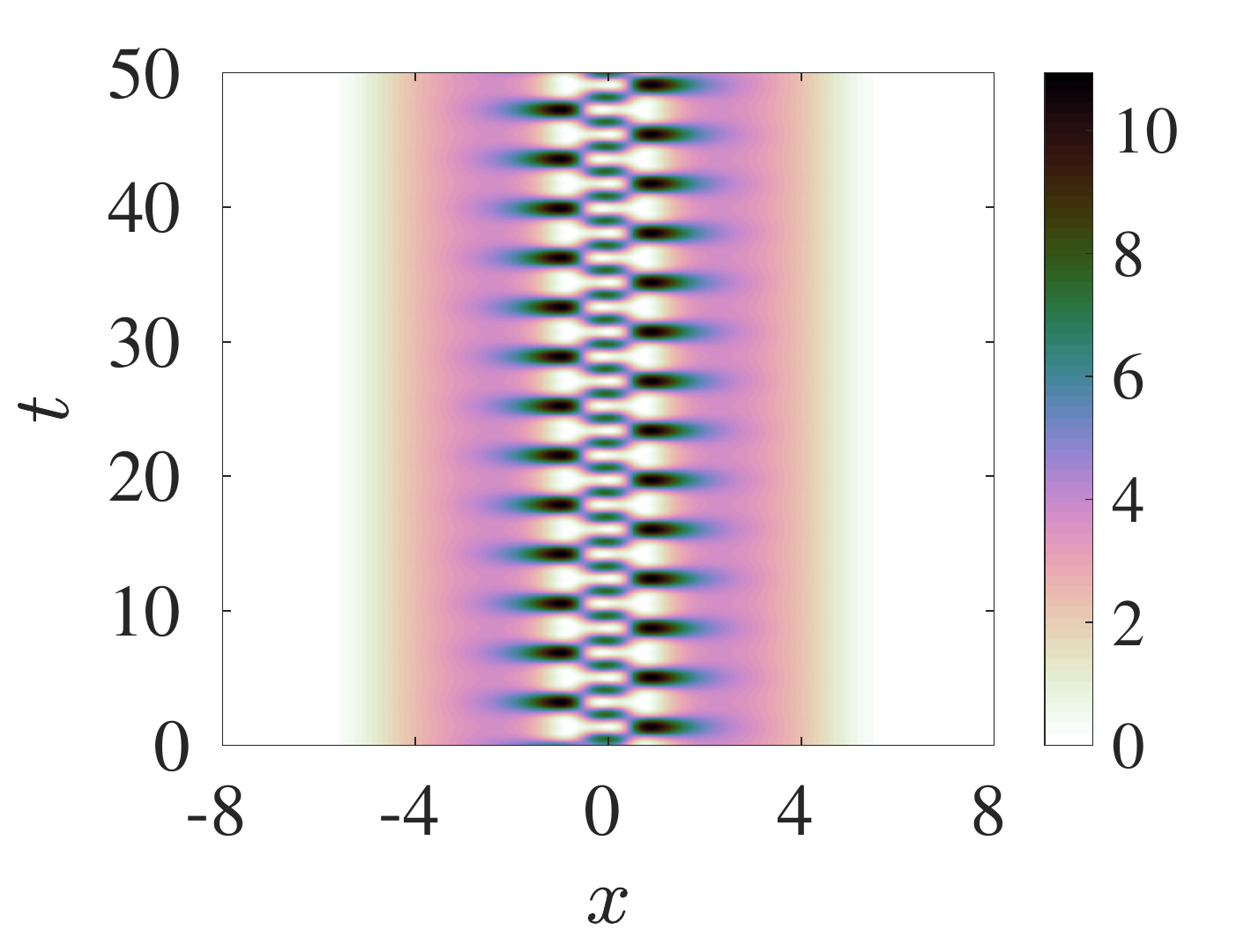}
\includegraphics[height=.17\textheight, angle =0]{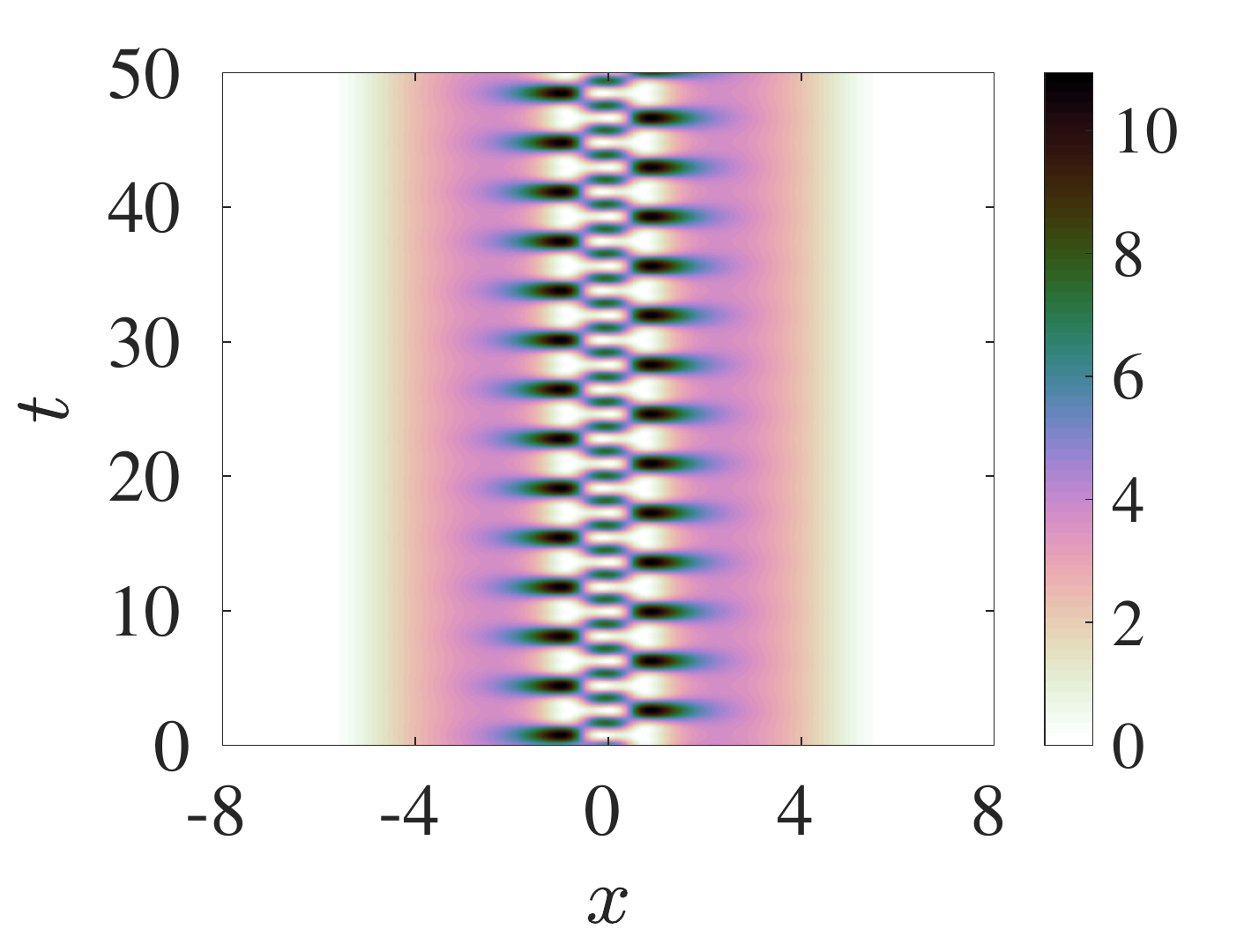}
\end{center}
\caption{
(Color online) %
Summary of results for the \state{210} state. \textit{Top row}:
The densities $|\psi_{j}|^{2}$ ($j=1,2,3$) of the steady-state profiles for 
$\mu_{1}=14.998146$, $\mu_{2}=13.284432$, and $\mu_{3}=11.570432$ are shown with
dashed-dotted black lines, and the SU$(3)$ rotated versions of them are shown
with solid blue, red, and green lines, respectively. \textit{Bottom row}: 
Spatio-temporal evolutions of the densities $|\psi_{1}(x,t)|^{2}$ (left panel),
$|\psi_{2}(x,t)|^{2}$ (middle panel), and $|\psi_{3}(x,t)|^{2}$ (right panel)
are shown where the initial states employed are the SU$(3)$ rotated states
of the top row.
}
\label{S210}
\end{figure}

\begin{figure}[pt]
\begin{center}
\includegraphics[height=.17\textheight, angle =0]{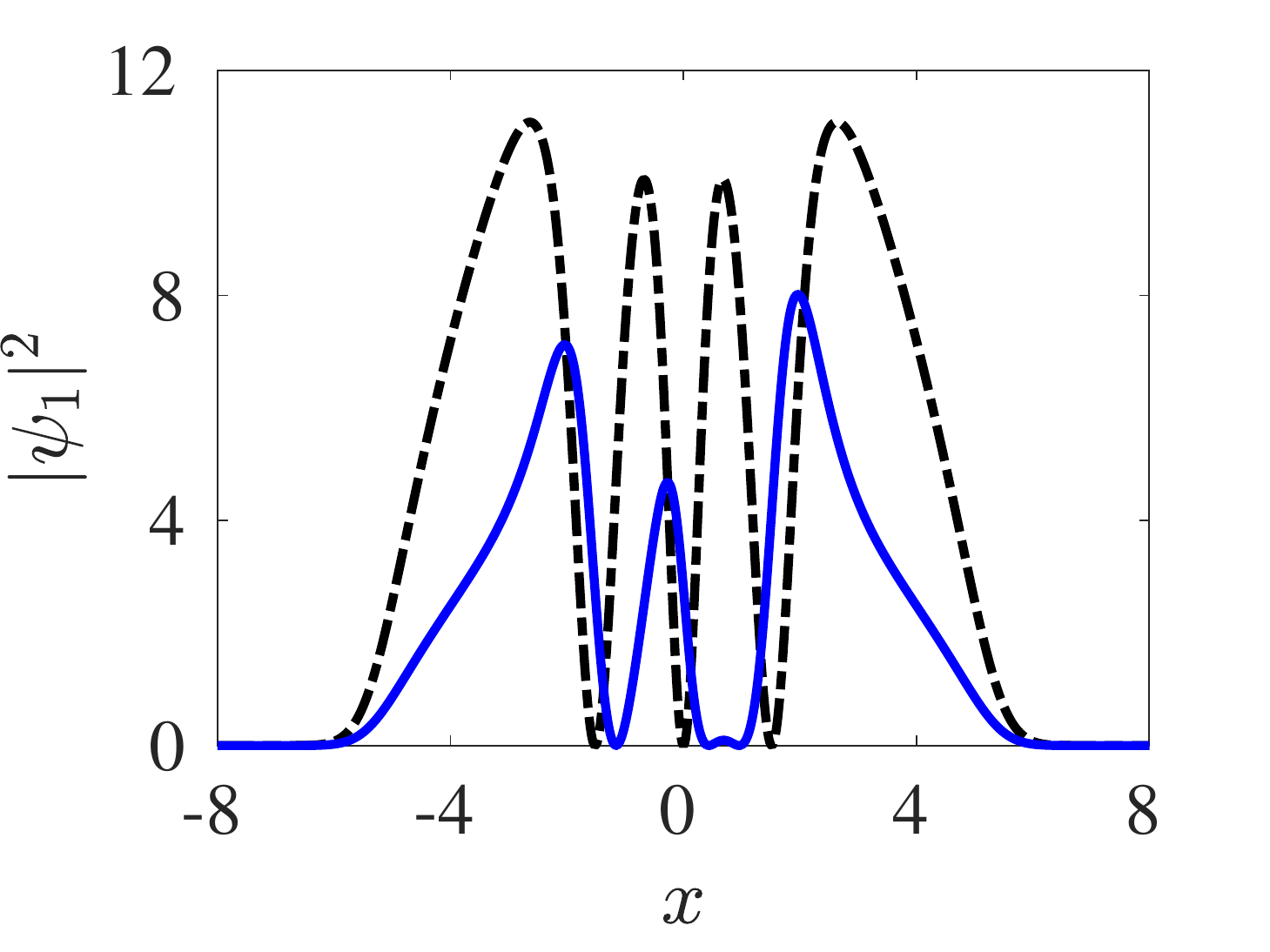}
\includegraphics[height=.17\textheight, angle =0]{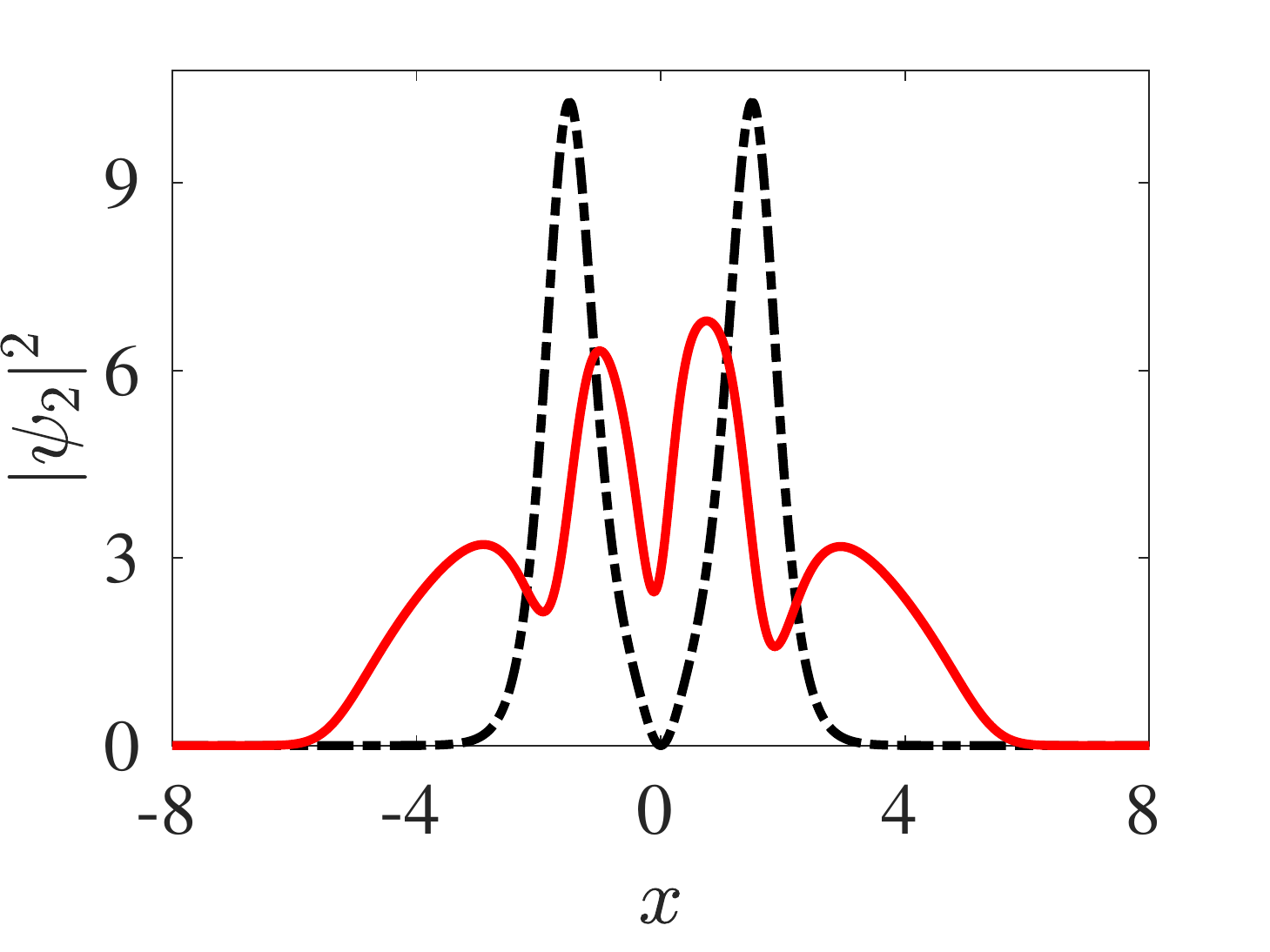}
\includegraphics[height=.17\textheight, angle =0]{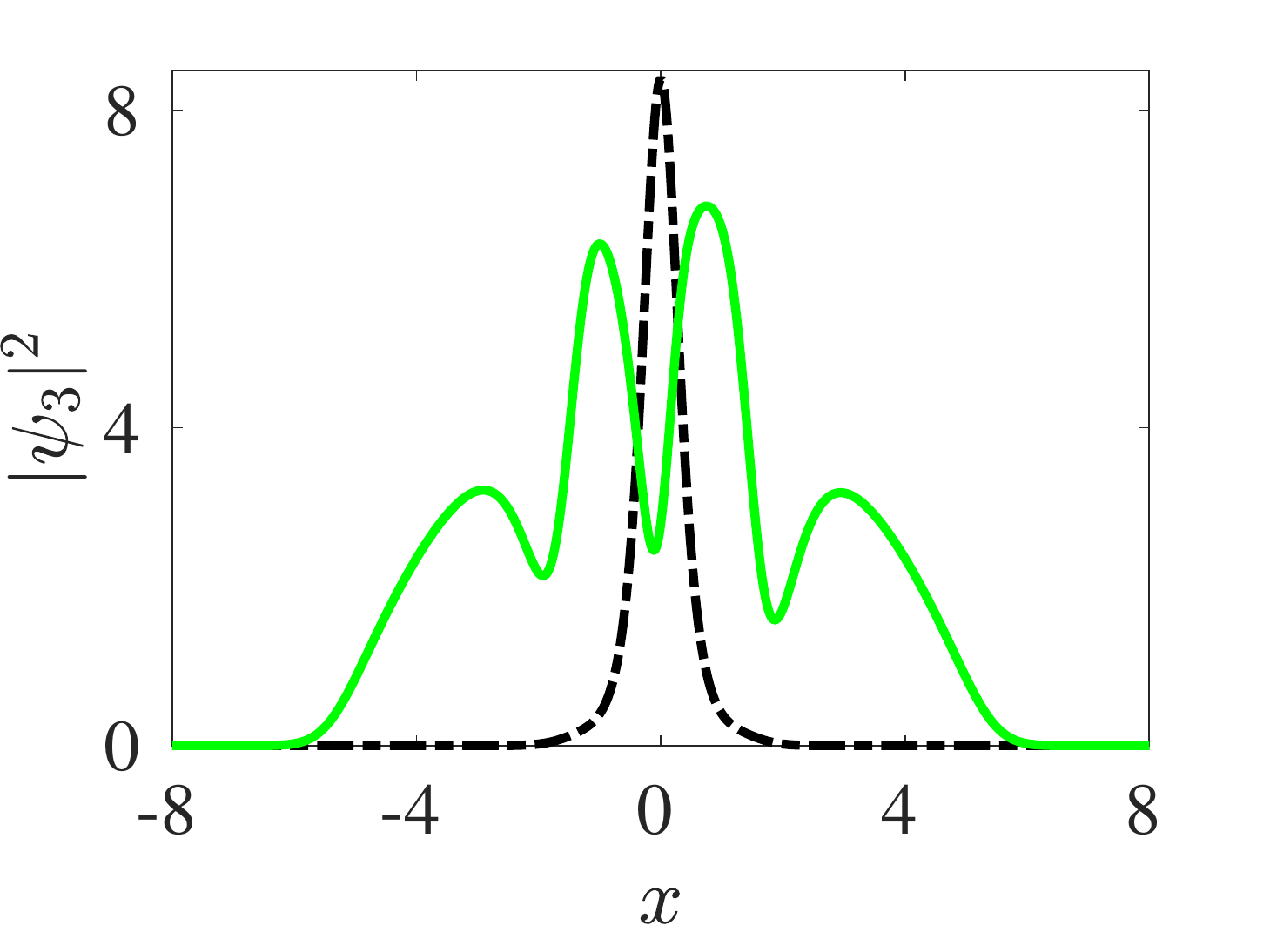}
\includegraphics[height=.17\textheight, angle =0]{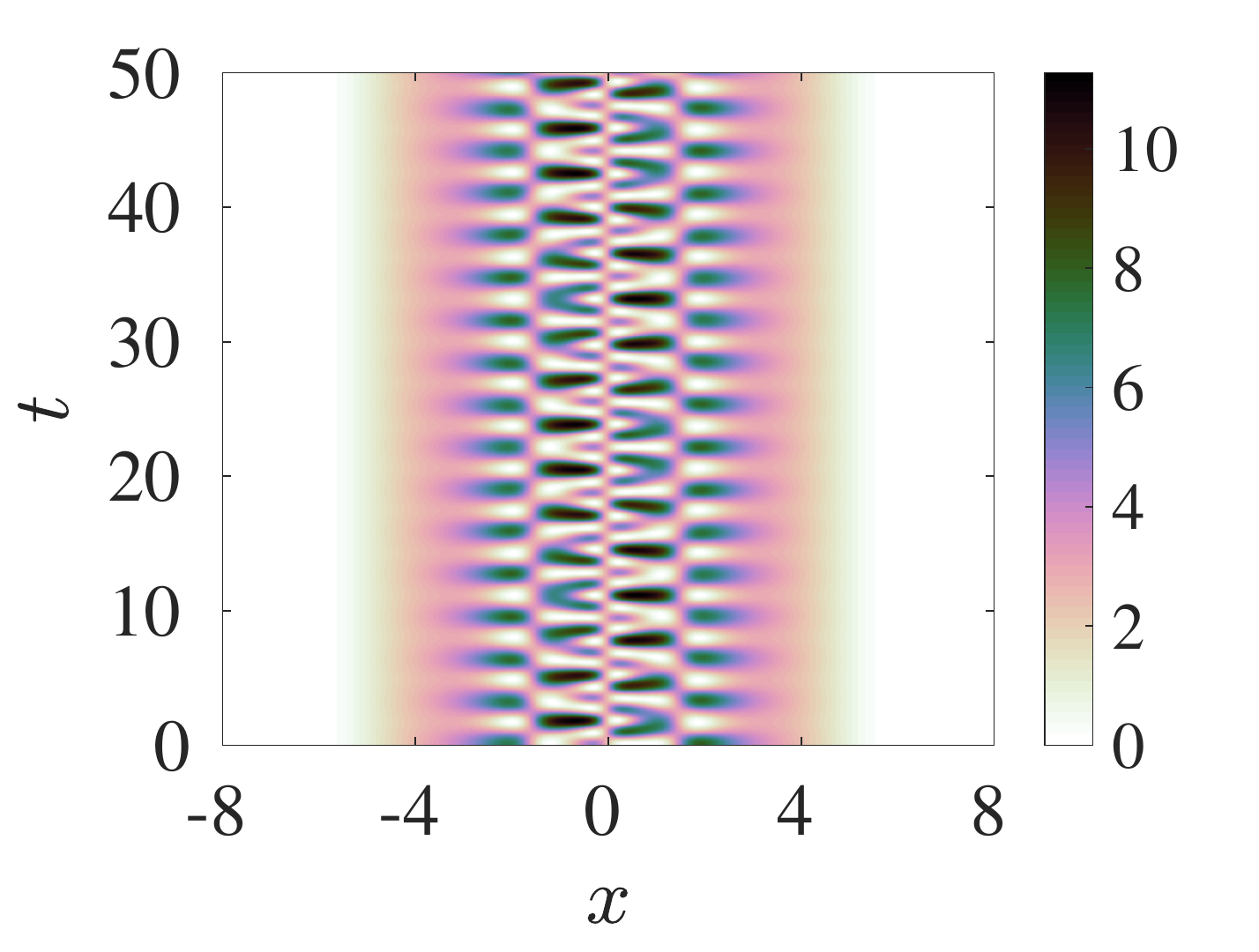}
\includegraphics[height=.17\textheight, angle =0]{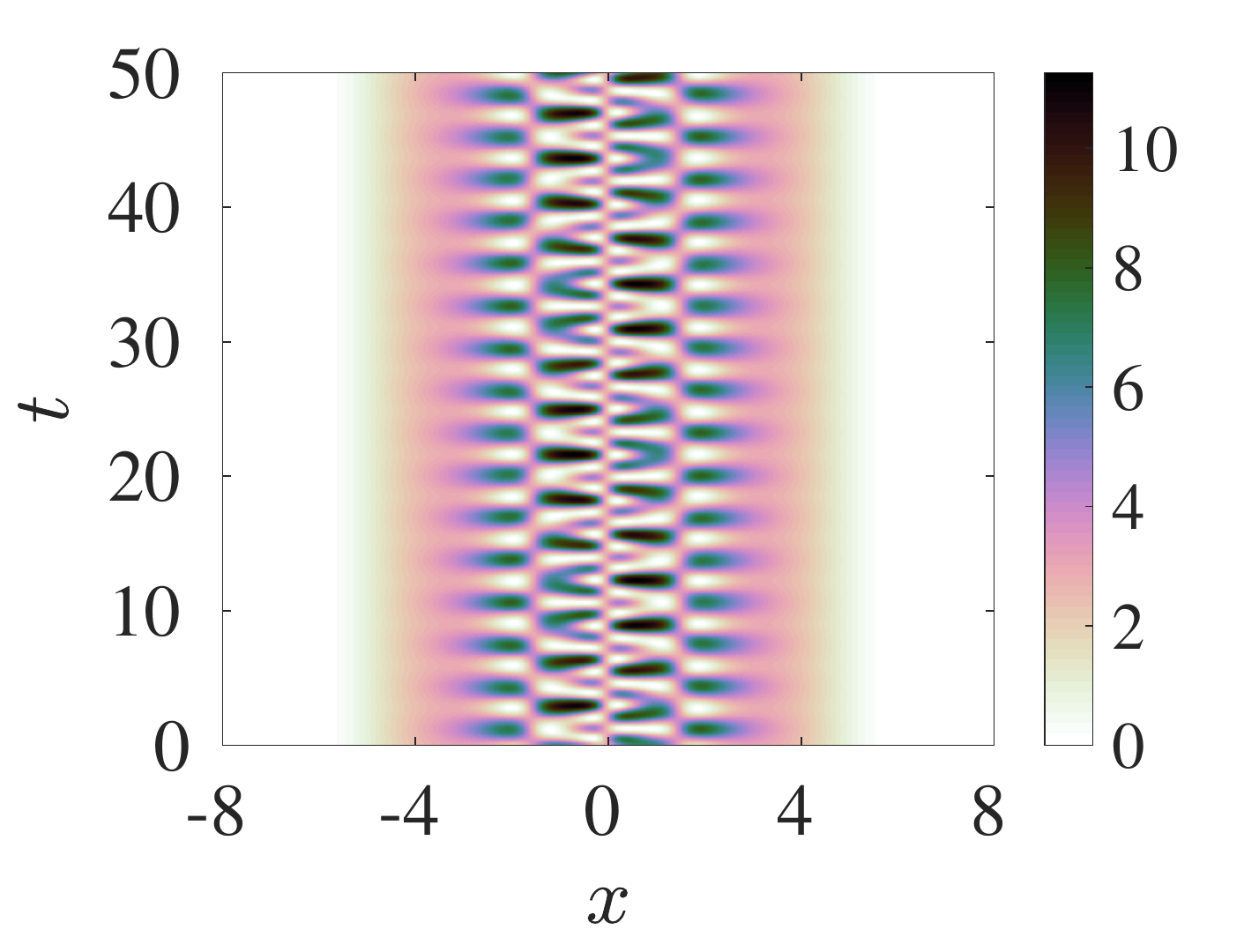}
\includegraphics[height=.17\textheight, angle =0]{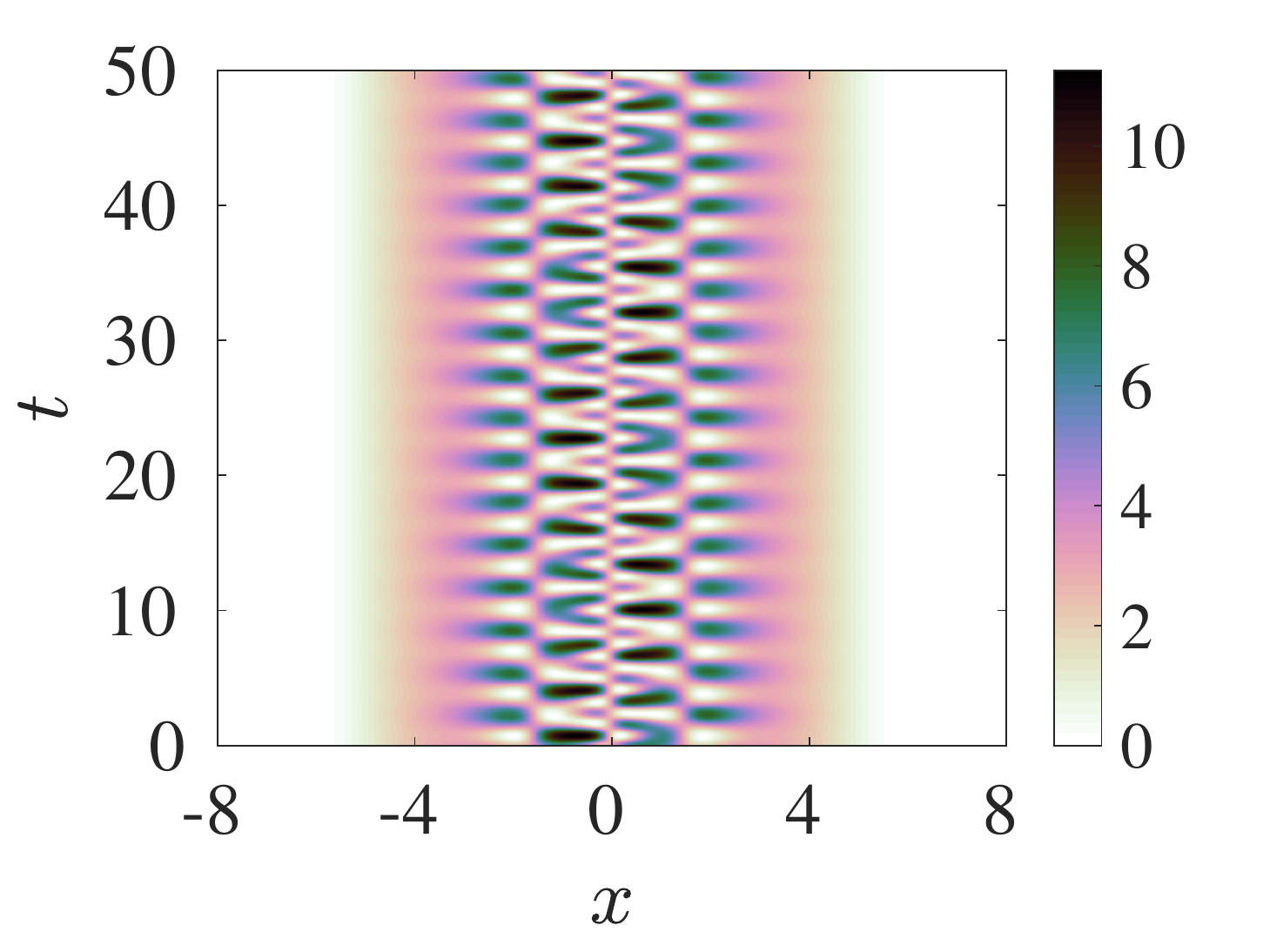}
\end{center}
\caption{(Color online) Same as Fig.~\ref{S210} but for the \state{310} state with chemical potentials $\mu_{1}=15.283574$, $\mu_{2}=13.283574$, and $\mu_{3}=11.569574$. 
Note that there are genuinely two frequencies involved in this dark-dark-dark 
breathing dynamics.
}
\label{S310}
\end{figure}

We perform the continuation of both states \state{210} and \state{310} over the 
chemical potentials from the associated linear limits to a TF regime of $(\mu_{1},\mu_{2},\mu_{3})=(20, 18, 16)$.
The left and right panels of Fig.~\ref{Spectra3C} correspond to the BdG spectra
of \state{210} and \state{310}, respectively. It should be noted that although 
both branches have intervals of instabilities (see the (red colored)
``instability bubbles'' in 
the pertinent panels), there exist \textit{wide} intervals of stability where the
solutions are expected to be long lived. Upon selecting stable steady-states (according
to our spectral stability analysis results), we SU$(3)$-rotate them in order to explore
the possibility of forming breathing yet robust patterns in the
three-component case.
Generally, this can be done by means of a unitary matrix $U=e^{i H \theta}$,
where
$H$ is a linear combination of the so-called Gell-Mann matrices~\cite{gellmann}.
To be more specific, we focus on a symmetric rotation dictated by the
following
unitary matrix~\cite{Lichen:DD}:
\begin{equation}
U=\frac{1}{\sqrt{3}}
\left(
\begin{array}{ccc}
1 & 1 & 1 \\
1 & -\exp(i\pi/3) & \exp(i2\pi/3) \\
1 & \exp(i2\pi/3) & -\exp(i\pi/3) \\
\end{array}
\right).
\end{equation}

%
Figures~\ref{S210} and~\ref{S310} summarize our results for the \state{210} and \state{310},
respectively. In particular, the top panels therein showcase the spatial distribution of the 
densities of the respective components for the unrotated
(dashed-dotted lines) and rotated (solid lines) solutions. The bottom panels in the 
figures offer the spatio-temporal evolution of the subsequent dark-dark-dark beating patterns
($|\psi_{1}(x,t)|^2$, $|\psi_{2}(x,t)|^2$, and $|\psi_{3}(x,t)|^2$ are shown from left to right).
Naturally, there are two dark solitons in each component for the rotated \state{210} whereas there exist 
three for the rotated \state{310}. The beating dynamics of \state{210}
is robustly
periodic, featuring a single-period internal vibration of the state. 
Upon examination, this is a coincidence resulting
from our chosen parameters, where $\mu_1-\mu_2 = \mu_2-\mu_3$, 
yielding only one period, i.e., $2\pi/(\mu_1-\mu_2) \approx 3.6664$. This agrees very well with the 
results of our simulations. On the other hand, two frequencies are genuinely present for the \state{310} beating dynamics, as the above relation has not been selected in our initial
data.
Beating patterns for the \state{210} state in the 
homogeneous setting, i.e., without an external trapping potential, were studied analytically for 
particular solutions~\cite{Lichen:DD}. However, in the present work, we demonstrate that this 
state exists and is, in fact, robust over a wide range of parameters.

\section{Conclusions and Future Challenges}
\label{conclusion}
The present work offered a systematic study of distinct SO$(2)$-induced multiple dark-dark breathing patterns 
from stationary and stable dark-bright and dark-dark bound modes. In particular, we studied the existence and 
stability of these structures from their respective linear limits to typical Thomas-Fermi regimes. We found that
for $n$ solitons, there are $n$ distinct coherent patterns that stem
from the linear limit and which range from fully in-phase to fully out-of-phase ones. 
Analytical results in the homogeneous setting are also discussed: here
the rotation typically involves the breathing of dynamically
non-stationary configurations.
Moreover, we presented a generalization of our approach
to the three-component GPE system to illustrate prototypical case
examples showcasing the generality of the considerations discussed herein.

Motivated by this work, there are multiple avenues for future study that we plan to pursue. One natural extension 
of our work is to generalize considerations
to higher dimensions. In 2D, vortex clusters and/or dark soliton filaments filled 
with bright components of various relative phases are possible, generating, e.g., various vortex cluster-vortex 
cluster breathing patterns. In 3D, vortex filaments and/or dark soliton surfaces filled with bright components 
of various relative phases are relevant for future studies.
Importantly, recent experimental progress, including, e.g., the configurations
reported in~\cite{engels20}, suggests that such states can be accessed
as initial conditions in state-of-the-art experiments and hence the
corresponding vibrational dynamics should, in principle, be
experimentally tractable. 
It is not readily obvious that one can find a systematic way to construct all the topologically distinct states and breathing patterns as in 1D since new 
states can bifurcate away from the linear limit~\cite{Panos:DC2}.
In addition, there are also different possible combinations of linear
eigenmodes.
For instance, $|1,0\rangle$ (where the separated by comma indices
denote here the linear eigenstates in the different spatial
dimensions) produces a dark soliton stripe, 
while $(|1,0\rangle+i|0,1\rangle)/\sqrt{2}$ produces a single vortex,
starting from
essentially the same basis. An 
additional challenge concerns the (in)stability properties of these structures.
It is relevant to seek suitable potential 
configurations to stabilize, e.g., some dark soliton filaments and surfaces against their transverse instabilities 
by adding external pinning potentials. It is also challenging to stabilize certain multiple vortical filament structures, 
and preliminary data suggest that even the double vortex rings filled by either in-phase or out-of-phase bright 
components are extremely difficult to stabilize, at least in a spherical trap. In this situation, one can either 
further increase the chemical potentials or explore instead other trap settings. 

Finally, systematic studies of the three-component system and even
beyond that are also interesting;
notice, in that vein, both the $F=1$ and $F=2$ spinor systems are
presently
experimentally accessible in atomic
condensates~\cite{kawueda,stampueda}. In 
our work, we have studied a few select examples of three-component
structures,
focusing, in particular, on the Manakov case. Yet, more complex structures should 
be accessible under physically realistic (spinor)
perturbations, but
also in higher dimensions. 
Research work along these lines are currently in progress, and will be reported in future publications.

\begin{acknowledgments}
This work is supported by National Natural Science Foundation of China (Contact No. 11775176), 
Basic Research Program of Natural Science of Shaanxi Province (Grant No. 2018KJXX-094), The Key 
Innovative Research Team of Quantum Many-Body Theory and Quantum Control in Shaanxi Province 
(Grant No. 2017KCT-12), and the Major Basic Research Program of Natural Science of Shaanxi 
Province (Grant No. 2017ZDJC-32). W.W.~acknowledges support from the Fundamental Research Funds
for the Central Universities, China. P.G.K. acknowledges support from the US National Science Foundation
under Grants No. PHY-1602994 and DMS-1809074. He also acknowledges support from the Leverhulme 
Trust via a Visiting Fellowship and the Mathematical Institute of the University of Oxford for 
its hospitality during part of this work. We thank the Emei cluster at Sichuan university for providing HPC resources.
\end{acknowledgments}

\section*{Appendix: Linear stability analysis of the three-component GPE}
In this Appendix, we discuss about the setup of the stability analysis problem. 
To this end, we perform a BdG stability analysis of a steady-state solution 
$\psi_{j}^{0}(x)$ by introducing the following perturbation Ans\"atze:
\begin{align}
\widetilde{\psi}_{j}(x,t)=e^{-i\mu_{j}t}\Big\lbrace \psi_{j}^{0}(x) + \varepsilon %
\left(a_{j}(x)e^{\lambda t}+b_{j}^{\ast}(x)e^{\lambda^{\ast}t}\right)\Big\rbrace, %
\quad j=1,2,3,
\label{perturb_ansatz}
\end{align}
where $\varepsilon\ll 1$. Upon substituting Eq.~\eqref{perturb_ansatz} into 
Eqs.~\eqref{three_comp_GPE_orig_1}-\eqref{three_comp_GPE_orig_3}, we obtain 
at order $O(\varepsilon)$ an eigenvalue problem of the form:
\begin{equation}
\rho
\begin{pmatrix}
a_{1} \\
b_{1} \\
a_{2} \\
b_{2} \\
a_{3} \\
b_{3}%
\end{pmatrix}
=%
\begin{pmatrix}
    A_{11}      & A_{12}  &      A_{13}     &     A_{14}      & A_{15}        & A_{16}        \\
-A_{12}^{\ast}  & -A_{11} & -A_{14}^{\ast}  & -A_{13}^{\ast}  &-A_{16}^{\ast} &-A_{15}^{\ast} \\
 A_{13}^{\ast}  & A_{14}  &      A_{33}     &     A_{34}      & A_{35}        & A_{36}        \\
-A_{14}^{\ast}  & -A_{13} & -A_{34}^{\ast}  &    -A_{33}      &-A_{36}^{\ast} &-A_{35}^{\ast} \\
 A_{15}^{\ast}  & A_{16}  &  A_{35}^{\ast}  &     A_{36}      & A_{55}        & A_{56}        \\
-A_{16}^{\ast}  & -A_{15} & -A_{36}^{\ast}  &    -A_{35}      &-A_{56}^{\ast} &-A_{55}
\end{pmatrix}
\begin{pmatrix}
a_{1} \\
b_{1} \\
a_{2} \\
b_{2} \\
a_{3} \\
b_{3}%
\end{pmatrix},
\label{three_GPE_eig_prob}
\end{equation}%
where the distinct matrix elements are given by:
\begin{subequations}
\begin{align}
A_{11}&=-\frac{1}{2}\frac{d^{2}}{dx^{2}}+2g_{11}|\psi_{1}^{0}|^{2}+g_{12}|\psi_{2}^{0}|^{2}%
+g_{13}|\psi_{3}^{0}|^{2}+V(x)-\mu_{1},\nonumber \\
A_{12}&=g_{11}\left(\psi^{0}_{1}\right)^{2},\nonumber\\
A_{13}&=g_{12}\psi_{1}^{0}\left(\psi_{2}^{0}\right)^{\ast},\nonumber\\
A_{14}&=g_{12}\psi_{1}^{0}\psi_{2}^{0},\nonumber\\
A_{15}&=g_{13}\psi_{1}^{0}\left(\psi_{3}^{0}\right)^{\ast},\nonumber\\
A_{16}&=g_{13}\psi_{1}^{0}\psi_{3}^{0},\nonumber\\
A_{33}&=-\frac{1}{2}\frac{d^{2}}{dx^{2}}+g_{12}|\psi_{1}^{0}|^{2}+2g_{22}|\psi_{2}^{0}|^{2}%
+g_{23}|\psi_{3}^{0}|^{2}+V(x)-\mu_{2},\nonumber\\
A_{34}&=g_{22}\left(\psi_{2}^{0}\right)^{2},\nonumber\\
A_{35}&=g_{23}\psi_{2}^{0}\left(\psi_{3}^{0}\right)^{\ast},\nonumber\\
A_{36}&=g_{23}\psi_{2}^{0}\psi_{3}^{0},\nonumber\\
A_{55}&=-\frac{1}{2}\frac{d^{2}}{dx^{2}}+g_{13}|\psi_{1}^{0}|^{2}+g_{23}|\psi_{2}^{0}|^{2}%
+2g_{33}|\psi_{3}^{0}|^{2}+V(x)-\mu_{3},\nonumber\\
A_{56}&=g_{33}\left(\psi_{3}^{0}\right)^{2}.\nonumber
\end{align}
\end{subequations}
Here, $\rho=i\lambda$ is the eigenvalue with the associated eigenvector:
\begin{equation}
\mathbf{W}(x)=\left(a_{1}(x) \, b_{1}(x) \, a_{2}(x) \, b_{2}(x) %
\, a_{3}(x) \, b_{3}(x) \right)^{T}.\nonumber 
\end{equation}
The eigenvalue computations for the three-component case were performed by using the 
FEAST eigenvalue solver~\cite{FEAST:nonHem} where (usually) ~$100$ eigenvalues were
computed with $10^{-8}$ tolerance on the residuals. 

\bibliography{Refs}

\begin{thebibliography}{49}
\expandafter\ifx\csname natexlab\endcsname\relax\def\natexlab#1{#1}\fi
\expandafter\ifx\csname bibnamefont\endcsname\relax
  \def\bibnamefont#1{#1}\fi
\expandafter\ifx\csname bibfnamefont\endcsname\relax
  \def\bibfnamefont#1{#1}\fi
\expandafter\ifx\csname citenamefont\endcsname\relax
  \def\citenamefont#1{#1}\fi
\expandafter\ifx\csname url\endcsname\relax
  \def\url#1{\texttt{#1}}\fi
\expandafter\ifx\csname urlprefix\endcsname\relax\def\urlprefix{URL }\fi
\providecommand{\bibinfo}[2]{#2}
\providecommand{\eprint}[2][]{\url{#2}}

\bibitem[{\citenamefont{Pitaevskii and Stringari}(2003)}]{becbook1}
\bibinfo{author}{\bibfnamefont{L.}~\bibnamefont{Pitaevskii}} \bibnamefont{and}
  \bibinfo{author}{\bibfnamefont{S.}~\bibnamefont{Stringari}},
  \emph{\bibinfo{title}{{B}ose--{E}instein Condensation}}
  (\bibinfo{publisher}{Oxford University Press}, \bibinfo{address}{Oxford, UK},
  \bibinfo{year}{2003}).

\bibitem[{\citenamefont{Pethick and Smith}(2002)}]{becbook2}
\bibinfo{author}{\bibfnamefont{C.}~\bibnamefont{Pethick}} \bibnamefont{and}
  \bibinfo{author}{\bibfnamefont{H.}~\bibnamefont{Smith}},
  \emph{\bibinfo{title}{{B}ose--{E}instein Condensation in Dilute Gases}}
  (\bibinfo{publisher}{Cambridge University Press},
  \bibinfo{address}{Cambridge, UK}, \bibinfo{year}{2002}).

\bibitem[{\citenamefont{Kevrekidis et~al.}(2015)\citenamefont{Kevrekidis,
  Frantzeskakis, and Carretero-Gonz\'{a}lez}}]{Panos:book}
\bibinfo{author}{\bibfnamefont{P.~G.} \bibnamefont{Kevrekidis}},
  \bibinfo{author}{\bibfnamefont{D.~J.} \bibnamefont{Frantzeskakis}},
  \bibnamefont{and}
  \bibinfo{author}{\bibfnamefont{R.}~\bibnamefont{Carretero-Gonz\'{a}lez}},
  \emph{\bibinfo{title}{The Defocusing Nonlinear Schr\"{o}dinger Equation: From
  Dark Solitons to Vortices and Vortex Rings}} (\bibinfo{publisher}{SIAM,
  Philadelphia}, \bibinfo{year}{2015}).

\bibitem[{\citenamefont{Kivshar and Luther-Davies}(1998)}]{DSoptics}
\bibinfo{author}{\bibfnamefont{Y.~S.} \bibnamefont{Kivshar}} \bibnamefont{and}
  \bibinfo{author}{\bibfnamefont{B.}~\bibnamefont{Luther-Davies}},
  \emph{\bibinfo{title}{Dark optical solitons: physics and applications}},
  \bibinfo{journal}{Physics Reports} \textbf{\bibinfo{volume}{298}},
  \bibinfo{pages}{81 } (\bibinfo{year}{1998}), ISSN \bibinfo{issn}{0370-1573}.

\bibitem[{\citenamefont{Abdullaev et~al.}(2005)\citenamefont{Abdullaev, Gammal,
  Kamchatnov, and Tomio}}]{tomio}
\bibinfo{author}{\bibfnamefont{F.}~\bibnamefont{Abdullaev}},
  \bibinfo{author}{\bibfnamefont{A.}~\bibnamefont{Gammal}},
  \bibinfo{author}{\bibfnamefont{A.}~\bibnamefont{Kamchatnov}},
  \bibnamefont{and} \bibinfo{author}{\bibfnamefont{L.}~\bibnamefont{Tomio}},
  \emph{\bibinfo{title}{{Dynamics of bright matter wave solitons in a
  Bose-Einstein condensate}}}, \bibinfo{journal}{Int. J. Mod. Phys. B}
  \textbf{\bibinfo{volume}{19}}, \bibinfo{pages}{3415} (\bibinfo{year}{2005}).

\bibitem[{\citenamefont{Frantzeskakis}(2010)}]{Dimitri:DS}
\bibinfo{author}{\bibfnamefont{D.~J.} \bibnamefont{Frantzeskakis}},
  \emph{\bibinfo{title}{{Dark solitons in atomic Bose{\textendash}Einstein
  condensates: from theory to experiments}}}, \bibinfo{journal}{Journal of
  Physics A: Mathematical and Theoretical} \textbf{\bibinfo{volume}{43}},
  \bibinfo{pages}{213001} (\bibinfo{year}{2010}).

\bibitem[{\citenamefont{Fetter and Svidzinsky}(2001)}]{Alexander2001}
\bibinfo{author}{\bibfnamefont{A.~L.} \bibnamefont{Fetter}} \bibnamefont{and}
  \bibinfo{author}{\bibfnamefont{A.~A.} \bibnamefont{Svidzinsky}},
  \emph{\bibinfo{title}{{Vortices in a trapped dilute Bose-Einstein
  condensate}}}, \bibinfo{journal}{Journal of Physics: Condensed Matter}
  \textbf{\bibinfo{volume}{13}}, \bibinfo{pages}{R135} (\bibinfo{year}{2001}).

\bibitem[{\citenamefont{Fetter}(2009)}]{fetter2}
\bibinfo{author}{\bibfnamefont{A.~L.} \bibnamefont{Fetter}},
  \emph{\bibinfo{title}{{Rotating trapped Bose-Einstein condensates}}},
  \bibinfo{journal}{Rev. Mod. Phys.} \textbf{\bibinfo{volume}{81}},
  \bibinfo{pages}{647} (\bibinfo{year}{2009}).

\bibitem[{\citenamefont{Komineas}(2007)}]{komineas}
\bibinfo{author}{\bibfnamefont{S.}~\bibnamefont{Komineas}},
  \emph{\bibinfo{title}{{Vortex rings and solitary waves in trapped
  {B}ose--{E}instein condensates}}}, \bibinfo{journal}{The European Physical
  Journal Special Topics} \textbf{\bibinfo{volume}{147}}, \bibinfo{pages}{133}
  (\bibinfo{year}{2007}).

\bibitem[{\citenamefont{Proment et~al.}(2012)\citenamefont{Proment, Onorato,
  and Barenghi}}]{PhysRevE.85.036306}
\bibinfo{author}{\bibfnamefont{D.}~\bibnamefont{Proment}},
  \bibinfo{author}{\bibfnamefont{M.}~\bibnamefont{Onorato}}, \bibnamefont{and}
  \bibinfo{author}{\bibfnamefont{C.~F.} \bibnamefont{Barenghi}},
  \emph{\bibinfo{title}{{Vortex knots in a Bose-Einstein condensate}}},
  \bibinfo{journal}{Phys. Rev. E} \textbf{\bibinfo{volume}{85}},
  \bibinfo{pages}{036306} (\bibinfo{year}{2012}).

\bibitem[{\citenamefont{Busch and Anglin}(2001)}]{DBS1}
\bibinfo{author}{\bibfnamefont{T.}~\bibnamefont{Busch}} \bibnamefont{and}
  \bibinfo{author}{\bibfnamefont{J.~R.} \bibnamefont{Anglin}},
  \emph{\bibinfo{title}{{Dark-Bright Solitons in Inhomogeneous Bose-Einstein
  Condensates}}}, \bibinfo{journal}{Phys. Rev. Lett.}
  \textbf{\bibinfo{volume}{87}}, \bibinfo{pages}{010401}
  (\bibinfo{year}{2001}).

\bibitem[{\citenamefont{Becker et~al.}(2008)\citenamefont{Becker, Stellmer,
  Soltan-Panahi, Dörscher, Baumert, Richter, Kronjäger, Bongs, and
  Sengstock}}]{DBS2}
\bibinfo{author}{\bibfnamefont{C.}~\bibnamefont{Becker}},
  \bibinfo{author}{\bibfnamefont{S.}~\bibnamefont{Stellmer}},
  \bibinfo{author}{\bibfnamefont{P.}~\bibnamefont{Soltan-Panahi}},
  \bibinfo{author}{\bibfnamefont{S.}~\bibnamefont{Dörscher}},
  \bibinfo{author}{\bibfnamefont{M.}~\bibnamefont{Baumert}},
  \bibinfo{author}{\bibfnamefont{E.-M.} \bibnamefont{Richter}},
  \bibinfo{author}{\bibfnamefont{J.}~\bibnamefont{Kronjäger}},
  \bibinfo{author}{\bibfnamefont{K.}~\bibnamefont{Bongs}}, \bibnamefont{and}
  \bibinfo{author}{\bibfnamefont{K.}~\bibnamefont{Sengstock}},
  \emph{\bibinfo{title}{{Oscillations and interactions of dark and dark-bright
  solitons in Bose-Einstein condensates}}}, \bibinfo{journal}{Nature Physics}
  \textbf{\bibinfo{volume}{4}}, \bibinfo{pages}{496} (\bibinfo{year}{2008}).

\bibitem[{\citenamefont{Romero-Ros et~al.}(2019)\citenamefont{Romero-Ros,
  Katsimiga, Kevrekidis, and Schmelcher}}]{PS:DB}
\bibinfo{author}{\bibfnamefont{A.}~\bibnamefont{Romero-Ros}},
  \bibinfo{author}{\bibfnamefont{G.~C.} \bibnamefont{Katsimiga}},
  \bibinfo{author}{\bibfnamefont{P.~G.} \bibnamefont{Kevrekidis}},
  \bibnamefont{and}
  \bibinfo{author}{\bibfnamefont{P.}~\bibnamefont{Schmelcher}},
  \emph{\bibinfo{title}{{Controlled generation of dark-bright soliton complexes
  in two-component and spinor Bose-Einstein condensates}}},
  \bibinfo{journal}{Phys. Rev. A} \textbf{\bibinfo{volume}{100}},
  \bibinfo{pages}{013626} (\bibinfo{year}{2019}).

\bibitem[{\citenamefont{Kiehn et~al.}(2019)\citenamefont{Kiehn, Mistakidis,
  Katsimiga, and Schmelcher}}]{PS:DBDD}
\bibinfo{author}{\bibfnamefont{H.}~\bibnamefont{Kiehn}},
  \bibinfo{author}{\bibfnamefont{S.~I.} \bibnamefont{Mistakidis}},
  \bibinfo{author}{\bibfnamefont{G.~C.} \bibnamefont{Katsimiga}},
  \bibnamefont{and}
  \bibinfo{author}{\bibfnamefont{P.}~\bibnamefont{Schmelcher}},
  \emph{\bibinfo{title}{Spontaneous generation of dark-bright and dark-antidark
  solitons upon quenching a particle-imbalanced bosonic mixture}},
  \bibinfo{journal}{Phys. Rev. A} \textbf{\bibinfo{volume}{100}},
  \bibinfo{pages}{023613} (\bibinfo{year}{2019}).

\bibitem[{\citenamefont{Yan et~al.}(2011)\citenamefont{Yan, Chang, Hamner,
  Kevrekidis, Engels, Achilleos, Frantzeskakis, Carretero-Gonz\'alez, and
  Schmelcher}}]{Dong:MDB}
\bibinfo{author}{\bibfnamefont{D.}~\bibnamefont{Yan}},
  \bibinfo{author}{\bibfnamefont{J.~J.} \bibnamefont{Chang}},
  \bibinfo{author}{\bibfnamefont{C.}~\bibnamefont{Hamner}},
  \bibinfo{author}{\bibfnamefont{P.~G.} \bibnamefont{Kevrekidis}},
  \bibinfo{author}{\bibfnamefont{P.}~\bibnamefont{Engels}},
  \bibinfo{author}{\bibfnamefont{V.}~\bibnamefont{Achilleos}},
  \bibinfo{author}{\bibfnamefont{D.~J.} \bibnamefont{Frantzeskakis}},
  \bibinfo{author}{\bibfnamefont{R.}~\bibnamefont{Carretero-Gonz\'alez}},
  \bibnamefont{and}
  \bibinfo{author}{\bibfnamefont{P.}~\bibnamefont{Schmelcher}},
  \emph{\bibinfo{title}{{Multiple dark-bright solitons in atomic Bose-Einstein
  condensates}}}, \bibinfo{journal}{Phys. Rev. A}
  \textbf{\bibinfo{volume}{84}}, \bibinfo{pages}{053630}
  (\bibinfo{year}{2011}).

\bibitem[{\citenamefont{Kevrekidis and Frantzeskakis}(2016)}]{revip}
\bibinfo{author}{\bibfnamefont{P.}~\bibnamefont{Kevrekidis}} \bibnamefont{and}
  \bibinfo{author}{\bibfnamefont{D.}~\bibnamefont{Frantzeskakis}},
  \emph{\bibinfo{title}{{Solitons in coupled nonlinear Schr\"odinger models: A
  survey of recent developments}}}, \bibinfo{journal}{Reviews in Physics}
  \textbf{\bibinfo{volume}{1}}, \bibinfo{pages}{140 } (\bibinfo{year}{2016}),
  ISSN \bibinfo{issn}{2405-4283}.

\bibitem[{\citenamefont{Wang and Kevrekidis}(2017)}]{Wang:DBS}
\bibinfo{author}{\bibfnamefont{W.}~\bibnamefont{Wang}} \bibnamefont{and}
  \bibinfo{author}{\bibfnamefont{P.~G.} \bibnamefont{Kevrekidis}},
  \emph{\bibinfo{title}{{Two-component dark-bright solitons in
  three-dimensional atomic Bose-Einstein condensates}}},
  \bibinfo{journal}{Phys. Rev. E} \textbf{\bibinfo{volume}{95}},
  \bibinfo{pages}{032201} (\bibinfo{year}{2017}).

\bibitem[{\citenamefont{Kevrekidis et~al.}(2018)\citenamefont{Kevrekidis, Wang,
  Carretero-Gonz\'alez, and Frantzeskakis}}]{Wang:AI2}
\bibinfo{author}{\bibfnamefont{P.~G.} \bibnamefont{Kevrekidis}},
  \bibinfo{author}{\bibfnamefont{W.}~\bibnamefont{Wang}},
  \bibinfo{author}{\bibfnamefont{R.}~\bibnamefont{Carretero-Gonz\'alez}},
  \bibnamefont{and} \bibinfo{author}{\bibfnamefont{D.~J.}
  \bibnamefont{Frantzeskakis}}, \emph{\bibinfo{title}{{Adiabatic invariant
  analysis of dark and dark-bright soliton stripes in two-dimensional
  Bose-Einstein condensates}}}, \bibinfo{journal}{Phys. Rev. A}
  \textbf{\bibinfo{volume}{97}}, \bibinfo{pages}{063604}
  (\bibinfo{year}{2018}).

\bibitem[{\citenamefont{Wang et~al.}(2019)\citenamefont{Wang, Kevrekidis, and
  Babaev}}]{Wang:RDS}
\bibinfo{author}{\bibfnamefont{W.}~\bibnamefont{Wang}},
  \bibinfo{author}{\bibfnamefont{P.~G.} \bibnamefont{Kevrekidis}},
  \bibnamefont{and} \bibinfo{author}{\bibfnamefont{E.}~\bibnamefont{Babaev}},
  \emph{\bibinfo{title}{{Ring dark solitons in three-dimensional Bose-Einstein
  condensates}}}, \bibinfo{journal}{Phys. Rev. A}
  \textbf{\bibinfo{volume}{100}}, \bibinfo{pages}{053621}
  (\bibinfo{year}{2019}).

\bibitem[{\citenamefont{Bersano et~al.}(2018)\citenamefont{Bersano, Gokhroo,
  Khamehchi, D'Ambroise, Frantzeskakis, Engels, and Kevrekidis}}]{engels18}
\bibinfo{author}{\bibfnamefont{T.~M.} \bibnamefont{Bersano}},
  \bibinfo{author}{\bibfnamefont{V.}~\bibnamefont{Gokhroo}},
  \bibinfo{author}{\bibfnamefont{M.~A.} \bibnamefont{Khamehchi}},
  \bibinfo{author}{\bibfnamefont{J.}~\bibnamefont{D'Ambroise}},
  \bibinfo{author}{\bibfnamefont{D.~J.} \bibnamefont{Frantzeskakis}},
  \bibinfo{author}{\bibfnamefont{P.}~\bibnamefont{Engels}}, \bibnamefont{and}
  \bibinfo{author}{\bibfnamefont{P.~G.} \bibnamefont{Kevrekidis}},
  \emph{\bibinfo{title}{{Three-Component Soliton States in Spinor $F=1$
  Bose-Einstein Condensates}}}, \bibinfo{journal}{Phys. Rev. Lett.}
  \textbf{\bibinfo{volume}{120}}, \bibinfo{pages}{063202}
  (\bibinfo{year}{2018}).

\bibitem[{\citenamefont{Qu et~al.}(2016)\citenamefont{Qu, Pitaevskii, and
  Stringari}}]{string}
\bibinfo{author}{\bibfnamefont{C.}~\bibnamefont{Qu}},
  \bibinfo{author}{\bibfnamefont{L.~P.} \bibnamefont{Pitaevskii}},
  \bibnamefont{and}
  \bibinfo{author}{\bibfnamefont{S.}~\bibnamefont{Stringari}},
  \emph{\bibinfo{title}{{Magnetic Solitons in a Binary Bose-Einstein
  Condensate}}}, \bibinfo{journal}{Phys. Rev. Lett.}
  \textbf{\bibinfo{volume}{116}}, \bibinfo{pages}{160402}
  (\bibinfo{year}{2016}).

\bibitem[{\citenamefont{Chai et~al.}(2019)\citenamefont{Chai, Lao, Fujimoto,
  Hamazakil, Ueda, and Raman}}]{rueda}
\bibinfo{author}{\bibfnamefont{X.}~\bibnamefont{Chai}},
  \bibinfo{author}{\bibfnamefont{D.}~\bibnamefont{Lao}},
  \bibinfo{author}{\bibfnamefont{K.}~\bibnamefont{Fujimoto}},
  \bibinfo{author}{\bibfnamefont{R.}~\bibnamefont{Hamazakil}},
  \bibinfo{author}{\bibfnamefont{M.}~\bibnamefont{Ueda}}, \bibnamefont{and}
  \bibinfo{author}{\bibfnamefont{C.}~\bibnamefont{Raman}},
  \emph{\bibinfo{title}{{Magnetic solitons in a spin-1 Bose-Einstein
  condensate}}}, \bibinfo{journal}{arXiv preprint arXiv:1912.06672}
  (\bibinfo{year}{2019}).

\bibitem[{\citenamefont{Katsimiga et~al.}(2020)\citenamefont{Katsimiga,
  Mistakidis, Bersano, Ohme, Mossman, Mukherjee, Schmelcher, Engels, and
  Kevrekidis}}]{engels20}
\bibinfo{author}{\bibfnamefont{G.~C.} \bibnamefont{Katsimiga}},
  \bibinfo{author}{\bibfnamefont{S.}~\bibnamefont{Mistakidis}},
  \bibinfo{author}{\bibfnamefont{T.~M.} \bibnamefont{Bersano}},
  \bibinfo{author}{\bibfnamefont{M.~K.~H.} \bibnamefont{Ohme}},
  \bibinfo{author}{\bibfnamefont{S.}~\bibnamefont{Mossman}},
  \bibinfo{author}{\bibfnamefont{K.}~\bibnamefont{Mukherjee}},
  \bibinfo{author}{\bibfnamefont{P.}~\bibnamefont{Schmelcher}},
  \bibinfo{author}{\bibfnamefont{P.}~\bibnamefont{Engels}}, \bibnamefont{and}
  \bibinfo{author}{\bibfnamefont{P.~G.} \bibnamefont{Kevrekidis}},
  \emph{\bibinfo{title}{{Observation and Analysis of Multiple Dark-Antidark
  Solitons in Two-Component Bose-Einstein Condensates}}},
  \bibinfo{journal}{arXiv preprint arXiv:2003.00259}  (\bibinfo{year}{2020}).

\bibitem[{\citenamefont{Park and Shin}(2000)}]{Park:Opticalsolitons}
\bibinfo{author}{\bibfnamefont{Q.-H.} \bibnamefont{Park}} \bibnamefont{and}
  \bibinfo{author}{\bibfnamefont{H.~J.} \bibnamefont{Shin}},
  \emph{\bibinfo{title}{Systematic construction of multicomponent optical
  solitons}}, \bibinfo{journal}{Phys. Rev. E} \textbf{\bibinfo{volume}{61}},
  \bibinfo{pages}{3093} (\bibinfo{year}{2000}).

\bibitem[{\citenamefont{Yan et~al.}(2012)\citenamefont{Yan, Chang, Hamner,
  Hoefer, Kevrekidis, Engels, Achilleos, Frantzeskakis, and Cuevas}}]{Yan:DD}
\bibinfo{author}{\bibfnamefont{D.}~\bibnamefont{Yan}},
  \bibinfo{author}{\bibfnamefont{J.~J.} \bibnamefont{Chang}},
  \bibinfo{author}{\bibfnamefont{C.}~\bibnamefont{Hamner}},
  \bibinfo{author}{\bibfnamefont{M.}~\bibnamefont{Hoefer}},
  \bibinfo{author}{\bibfnamefont{P.~G.} \bibnamefont{Kevrekidis}},
  \bibinfo{author}{\bibfnamefont{P.}~\bibnamefont{Engels}},
  \bibinfo{author}{\bibfnamefont{V.}~\bibnamefont{Achilleos}},
  \bibinfo{author}{\bibfnamefont{D.~J.} \bibnamefont{Frantzeskakis}},
  \bibnamefont{and} \bibinfo{author}{\bibfnamefont{J.}~\bibnamefont{Cuevas}},
  \emph{\bibinfo{title}{{Beating dark{\textendash}dark solitons in
  Bose{\textendash}Einstein condensates}}}, \bibinfo{journal}{Journal of
  Physics B: Atomic, Molecular and Optical Physics}
  \textbf{\bibinfo{volume}{45}}, \bibinfo{pages}{115301}
  (\bibinfo{year}{2012}).

\bibitem[{\citenamefont{Charalampidis et~al.}(2016)\citenamefont{Charalampidis,
  Wang, Kevrekidis, Frantzeskakis, and Cuevas-Maraver}}]{Wang:SO2}
\bibinfo{author}{\bibfnamefont{E.~G.} \bibnamefont{Charalampidis}},
  \bibinfo{author}{\bibfnamefont{W.}~\bibnamefont{Wang}},
  \bibinfo{author}{\bibfnamefont{P.~G.} \bibnamefont{Kevrekidis}},
  \bibinfo{author}{\bibfnamefont{D.~J.} \bibnamefont{Frantzeskakis}},
  \bibnamefont{and}
  \bibinfo{author}{\bibfnamefont{J.}~\bibnamefont{Cuevas-Maraver}},
  \emph{\bibinfo{title}{{SO(2)-induced breathing patterns in multicomponent
  Bose-Einstein condensates}}}, \bibinfo{journal}{Phys. Rev. A}
  \textbf{\bibinfo{volume}{93}}, \bibinfo{pages}{063623}
  (\bibinfo{year}{2016}).

\bibitem[{\citenamefont{Zhao}(2018)}]{Lichen:DD}
\bibinfo{author}{\bibfnamefont{L.-C.} \bibnamefont{Zhao}},
  \emph{\bibinfo{title}{{Beating effects of vector solitons in Bose-Einstein
  condensates}}}, \bibinfo{journal}{Phys. Rev. E}
  \textbf{\bibinfo{volume}{97}}, \bibinfo{pages}{062201}
  (\bibinfo{year}{2018}).

\bibitem[{\citenamefont{Wang and Kevrekidis}(2015)}]{Wang:OD}
\bibinfo{author}{\bibfnamefont{W.}~\bibnamefont{Wang}} \bibnamefont{and}
  \bibinfo{author}{\bibfnamefont{P.~G.} \bibnamefont{Kevrekidis}},
  \emph{\bibinfo{title}{Transitions from order to disorder in multiple dark and
  multiple dark-bright soliton atomic clouds}}, \bibinfo{journal}{Phys. Rev. E}
  \textbf{\bibinfo{volume}{91}}, \bibinfo{pages}{032905}
  (\bibinfo{year}{2015}).

\bibitem[{\citenamefont{Atkinson}(1989)}]{Numanalysis}
\bibinfo{author}{\bibfnamefont{K.~E.} \bibnamefont{Atkinson}},
  \emph{\bibinfo{title}{{An Introduction to Numerical Analysis}}}
  (\bibinfo{publisher}{John Wiley \& Sons}, \bibinfo{year}{1989}).

\bibitem[{\citenamefont{Gaunt et~al.}(2013)\citenamefont{Gaunt, Schmidutz,
  Gotlibovych, Smith, and Hadzibabic}}]{BoxP}
\bibinfo{author}{\bibfnamefont{A.~L.} \bibnamefont{Gaunt}},
  \bibinfo{author}{\bibfnamefont{T.~F.} \bibnamefont{Schmidutz}},
  \bibinfo{author}{\bibfnamefont{I.}~\bibnamefont{Gotlibovych}},
  \bibinfo{author}{\bibfnamefont{R.~P.} \bibnamefont{Smith}}, \bibnamefont{and}
  \bibinfo{author}{\bibfnamefont{Z.}~\bibnamefont{Hadzibabic}},
  \emph{\bibinfo{title}{{Bose-Einstein Condensation of Atoms in a Uniform
  Potential}}}, \bibinfo{journal}{Phys. Rev. Lett.}
  \textbf{\bibinfo{volume}{110}}, \bibinfo{pages}{200406}
  (\bibinfo{year}{2013}).

\bibitem[{\citenamefont{Kawaguchi and Ueda}(2012)}]{kawueda}
\bibinfo{author}{\bibfnamefont{Y.}~\bibnamefont{Kawaguchi}} \bibnamefont{and}
  \bibinfo{author}{\bibfnamefont{M.}~\bibnamefont{Ueda}},
  \emph{\bibinfo{title}{{Spinor Bose-Einstein condensates}}},
  \bibinfo{journal}{Physics Reports} \textbf{\bibinfo{volume}{520}},
  \bibinfo{pages}{253 } (\bibinfo{year}{2012}).

\bibitem[{\citenamefont{Stamper-Kurn and Ueda}(2013)}]{stampueda}
\bibinfo{author}{\bibfnamefont{D.~M.} \bibnamefont{Stamper-Kurn}}
  \bibnamefont{and} \bibinfo{author}{\bibfnamefont{M.}~\bibnamefont{Ueda}},
  \emph{\bibinfo{title}{{Spinor Bose gases: Symmetries, magnetism, and quantum
  dynamics}}}, \bibinfo{journal}{Rev. Mod. Phys.}
  \textbf{\bibinfo{volume}{85}}, \bibinfo{pages}{1191} (\bibinfo{year}{2013}).

\bibitem[{\citenamefont{Ostrovskaya et~al.}(1999)\citenamefont{Ostrovskaya,
  Kivshar, Chen, and Segev}}]{OS99}
\bibinfo{author}{\bibfnamefont{E.~A.} \bibnamefont{Ostrovskaya}},
  \bibinfo{author}{\bibfnamefont{Y.~S.} \bibnamefont{Kivshar}},
  \bibinfo{author}{\bibfnamefont{Z.}~\bibnamefont{Chen}}, \bibnamefont{and}
  \bibinfo{author}{\bibfnamefont{M.}~\bibnamefont{Segev}},
  \emph{\bibinfo{title}{Interaction between vector solitons and solitonic
  gluons}}, \bibinfo{journal}{Opt. Lett.} \textbf{\bibinfo{volume}{24}},
  \bibinfo{pages}{327} (\bibinfo{year}{1999}).

\bibitem[{\citenamefont{Mertes et~al.}(2007)\citenamefont{Mertes, Merrill,
  Carretero-Gonz\'alez, Frantzeskakis, Kevrekidis, and Hall}}]{Hall:BEC2}
\bibinfo{author}{\bibfnamefont{K.~M.} \bibnamefont{Mertes}},
  \bibinfo{author}{\bibfnamefont{J.~W.} \bibnamefont{Merrill}},
  \bibinfo{author}{\bibfnamefont{R.}~\bibnamefont{Carretero-Gonz\'alez}},
  \bibinfo{author}{\bibfnamefont{D.~J.} \bibnamefont{Frantzeskakis}},
  \bibinfo{author}{\bibfnamefont{P.~G.} \bibnamefont{Kevrekidis}},
  \bibnamefont{and} \bibinfo{author}{\bibfnamefont{D.~S.} \bibnamefont{Hall}},
  \emph{\bibinfo{title}{{Nonequilibrium Dynamics and Superfluid Ring
  Excitations in Binary Bose-Einstein Condensates}}}, \bibinfo{journal}{Phys.
  Rev. Lett.} \textbf{\bibinfo{volume}{99}}, \bibinfo{pages}{190402}
  (\bibinfo{year}{2007}).

\bibitem[{\citenamefont{{Manakov}}(1974)}]{Manokov74}
\bibinfo{author}{\bibfnamefont{S.~V.} \bibnamefont{{Manakov}}},
  \emph{\bibinfo{title}{{On the theory of two-dimensional stationary
  self-focusing of electromagnetic waves}}}, \bibinfo{journal}{Soviet Journal
  of Experimental and Theoretical Physics} \textbf{\bibinfo{volume}{38}},
  \bibinfo{pages}{248} (\bibinfo{year}{1974}).

\bibitem[{\citenamefont{{V. B. Matveev and M. A. Salle}}(1991)}]{DT}
\bibinfo{author}{\bibnamefont{{V. B. Matveev and M. A. Salle}}},
  \emph{\bibinfo{title}{{Darboux Transformation and Solitons}}}
  (\bibinfo{publisher}{Springer-Verlag}, \bibinfo{address}{Berlin},
  \bibinfo{year}{1991}).

\bibitem[{\citenamefont{{E. V. Doktorov and S. B.
  Leble}}(2007)}]{Dressingmethod}
\bibinfo{author}{\bibnamefont{{E. V. Doktorov and S. B. Leble}}},
  \emph{\bibinfo{title}{{A Dressing Method in Mathematical Physics}}}
  (\bibinfo{publisher}{Springer-Verlag}, \bibinfo{address}{Berlin},
  \bibinfo{year}{2007}).

\bibitem[{\citenamefont{Hirota}(2004)}]{Hirota}
\bibinfo{author}{\bibfnamefont{R.}~\bibnamefont{Hirota}},
  \emph{\bibinfo{title}{{The Direct Method in Soliton Theory}}}
  (\bibinfo{publisher}{Cambridge University Press},
  \bibinfo{address}{Cambridge, UK}, \bibinfo{year}{2004}).

\bibitem[{\citenamefont{Kanna and Lakshmanan}(2001)}]{Lakshman}
\bibinfo{author}{\bibfnamefont{T.}~\bibnamefont{Kanna}} \bibnamefont{and}
  \bibinfo{author}{\bibfnamefont{M.}~\bibnamefont{Lakshmanan}},
  \emph{\bibinfo{title}{{Exact Soliton Solutions, Shape Changing Collisions,
  and Partially Coherent Solitons in Coupled Nonlinear Schr\"odinger
  Equations}}}, \bibinfo{journal}{Phys. Rev. Lett.}
  \textbf{\bibinfo{volume}{86}}, \bibinfo{pages}{5043} (\bibinfo{year}{2001}).

\bibitem[{\citenamefont{Ling et~al.}(2015)\citenamefont{Ling, Zhao, and
  Guo}}]{Lichen:DT}
\bibinfo{author}{\bibfnamefont{L.}~\bibnamefont{Ling}},
  \bibinfo{author}{\bibfnamefont{L.-C.} \bibnamefont{Zhao}}, \bibnamefont{and}
  \bibinfo{author}{\bibfnamefont{B.}~\bibnamefont{Guo}},
  \emph{\bibinfo{title}{{Darboux transformation and multi-dark soliton for
  {N}-component nonlinear Schr\"{o}dinger equations}}},
  \bibinfo{journal}{Nonlinearity} \textbf{\bibinfo{volume}{28}},
  \bibinfo{pages}{3243} (\bibinfo{year}{2015}).

\bibitem[{\citenamefont{Rajendran et~al.}(2009)\citenamefont{Rajendran,
  Muruganandam, and Lakshmanan}}]{Rajendran_2009}
\bibinfo{author}{\bibfnamefont{S.}~\bibnamefont{Rajendran}},
  \bibinfo{author}{\bibfnamefont{P.}~\bibnamefont{Muruganandam}},
  \bibnamefont{and}
  \bibinfo{author}{\bibfnamefont{M.}~\bibnamefont{Lakshmanan}},
  \emph{\bibinfo{title}{{Interaction of dark{\textendash}bright solitons in
  two-component Bose{\textendash}Einstein condensates}}},
  \bibinfo{journal}{Journal of Physics B: Atomic, Molecular and Optical
  Physics} \textbf{\bibinfo{volume}{42}}, \bibinfo{pages}{145307}
  (\bibinfo{year}{2009}).

\bibitem[{\citenamefont{Dean et~al.}(2013)\citenamefont{Dean, Klotz, Prinari,
  and Vitale}}]{DDDB}
\bibinfo{author}{\bibfnamefont{G.}~\bibnamefont{Dean}},
  \bibinfo{author}{\bibfnamefont{T.}~\bibnamefont{Klotz}},
  \bibinfo{author}{\bibfnamefont{B.}~\bibnamefont{Prinari}}, \bibnamefont{and}
  \bibinfo{author}{\bibfnamefont{F.}~\bibnamefont{Vitale}},
  \emph{\bibinfo{title}{{Dark-dark and dark-bright soliton interactions in the
  two-component defocusing nonlinear Schr\"odinger equation}}},
  \bibinfo{journal}{Applicable Analysis} \textbf{\bibinfo{volume}{92}},
  \bibinfo{pages}{379} (\bibinfo{year}{2013}).

\bibitem[{\citenamefont{Hamner et~al.}(2011)\citenamefont{Hamner, Chang,
  Engels, and Hoefer}}]{DBcounterflow}
\bibinfo{author}{\bibfnamefont{C.}~\bibnamefont{Hamner}},
  \bibinfo{author}{\bibfnamefont{J.~J.} \bibnamefont{Chang}},
  \bibinfo{author}{\bibfnamefont{P.}~\bibnamefont{Engels}}, \bibnamefont{and}
  \bibinfo{author}{\bibfnamefont{M.~A.} \bibnamefont{Hoefer}},
  \emph{\bibinfo{title}{{Generation of Dark-Bright Soliton Trains in
  Superfluid-Superfluid Counterflow}}}, \bibinfo{journal}{Phys. Rev. Lett.}
  \textbf{\bibinfo{volume}{106}}, \bibinfo{pages}{065302}
  (\bibinfo{year}{2011}).

\bibitem[{\citenamefont{Karamatskos et~al.}(2015)\citenamefont{Karamatskos,
  Stockhofe, Kevrekidis, and Schmelcher}}]{DBtunneling}
\bibinfo{author}{\bibfnamefont{E.~T.} \bibnamefont{Karamatskos}},
  \bibinfo{author}{\bibfnamefont{J.}~\bibnamefont{Stockhofe}},
  \bibinfo{author}{\bibfnamefont{P.~G.} \bibnamefont{Kevrekidis}},
  \bibnamefont{and}
  \bibinfo{author}{\bibfnamefont{P.}~\bibnamefont{Schmelcher}},
  \emph{\bibinfo{title}{Stability and tunneling dynamics of a dark-bright
  soliton pair in a harmonic trap}}, \bibinfo{journal}{Phys. Rev. A}
  \textbf{\bibinfo{volume}{91}}, \bibinfo{pages}{043637}
  (\bibinfo{year}{2015}).

\bibitem[{\citenamefont{Katsimiga et~al.}(2018)\citenamefont{Katsimiga,
  Kevrekidis, Prinari, Biondini, and Schmelcher}}]{DBcollisions}
\bibinfo{author}{\bibfnamefont{G.~C.} \bibnamefont{Katsimiga}},
  \bibinfo{author}{\bibfnamefont{P.~G.} \bibnamefont{Kevrekidis}},
  \bibinfo{author}{\bibfnamefont{B.}~\bibnamefont{Prinari}},
  \bibinfo{author}{\bibfnamefont{G.}~\bibnamefont{Biondini}}, \bibnamefont{and}
  \bibinfo{author}{\bibfnamefont{P.}~\bibnamefont{Schmelcher}},
  \emph{\bibinfo{title}{{Dark-bright soliton pairs: Bifurcations and
  collisions}}}, \bibinfo{journal}{Phys. Rev. A} \textbf{\bibinfo{volume}{97}},
  \bibinfo{pages}{043623} (\bibinfo{year}{2018}).

\bibitem[{\citenamefont{Zhao et~al.}(2019)\citenamefont{Zhao, Duan, Gao, and
  Yang}}]{Lichen:MI2}
\bibinfo{author}{\bibfnamefont{L.-C.} \bibnamefont{Zhao}},
  \bibinfo{author}{\bibfnamefont{L.}~\bibnamefont{Duan}},
  \bibinfo{author}{\bibfnamefont{P.}~\bibnamefont{Gao}}, \bibnamefont{and}
  \bibinfo{author}{\bibfnamefont{Z.-Y.} \bibnamefont{Yang}},
  \emph{\bibinfo{title}{{Vector rogue waves on a double-plane wave
  background}}}, \bibinfo{journal}{{EPL} (Europhysics Letters)}
  \textbf{\bibinfo{volume}{125}}, \bibinfo{pages}{40003}
  (\bibinfo{year}{2019}).

\bibitem[{\citenamefont{Gell-Mann}(1962)}]{gellmann}
\bibinfo{author}{\bibfnamefont{M.}~\bibnamefont{Gell-Mann}},
  \emph{\bibinfo{title}{Symmetries of baryons and mesons}},
  \bibinfo{journal}{Phys. Rev.} \textbf{\bibinfo{volume}{125}},
  \bibinfo{pages}{1067} (\bibinfo{year}{1962}).

\bibitem[{\citenamefont{{E. G. Charalampidis, N. Boull\'{e}, P. E. Farrell, and
  P. G. Kevrekidis}}(2019)}]{Panos:DC2}
\bibinfo{author}{\bibnamefont{{E. G. Charalampidis, N. Boull\'{e}, P. E.
  Farrell, and P. G. Kevrekidis}}}, \emph{\bibinfo{title}{{Bifurcation analysis
  of stationary solutions of two-dimensional coupled Gross-Pitaevskii equations
  using deflated continuation}}}, \bibinfo{journal}{arXiv:1912.00023}
  (\bibinfo{year}{2019}).

\bibitem[{\citenamefont{Kestyn et~al.}(2016)\citenamefont{Kestyn, Polizzi, and
  Tang}}]{FEAST:nonHem}
\bibinfo{author}{\bibfnamefont{J.}~\bibnamefont{Kestyn}},
  \bibinfo{author}{\bibfnamefont{E.}~\bibnamefont{Polizzi}}, \bibnamefont{and}
  \bibinfo{author}{\bibfnamefont{T.~P.} \bibnamefont{Tang}},
  \emph{\bibinfo{title}{{Feast Eigensolver for Non-Hermitian Problems}}},
  \bibinfo{journal}{SIAM J. Sci. Comput.} \textbf{\bibinfo{volume}{38}},
  \bibinfo{pages}{S772} (\bibinfo{year}{2016}).

\end{thebibliography}

\end{document}